# *Optimality in goal-dependent analysis of Sharing*

GIANLUCA AMATO and FRANCESCA SCOZZARI
*Dipartimento di Scienze*
*Università di Chieti-Pescara*
(*e-mail:* {amato,scozzari}@sci.unich.it)

**Abstract**

We face the problems of correctness, optimality and precision for the static analysis of logic programs, using the theory of abstract interpretation. We propose a framework with a denotational, goal-dependent semantics equipped with two unification operators for forward unification (calling a procedure) and backward unification (returning from a procedure). The latter is implemented through a matching operation. Our proposal clarifies and unifies many different frameworks and ideas on static analysis of logic programming in a single, formal setting. On the abstract side, we focus on the domain `Sharing` by Jacobs and Langen and provide the best correct approximation of all the primitive semantic operators, namely, projection, renaming, forward and backward unification. We show that the abstract unification operators are strictly more precise than those in the literature defined over the same abstract domain. In some cases, our operators are more precise than those developed for more complex domains involving linearity and freeness.

## 1 Introduction

Abstract interpretation (Cousot and Cousot 1992) is a general theory for static analysis of programs. The basic idea of abstract interpretation is to use the formal semantics of languages to analyze and verify program properties. An abstract interpretation is specified by:

- a concrete domain and a concrete semantics, inductively defined on the syntax of programs from a set of primitive concrete operators;
- an abstract domain, whose elements describe the program properties we want to observe;
- the primitive abstract operators on the abstract domain, which mimic the behavior of the corresponding concrete operators. The abstract semantics is defined from the concrete one by replacing each concrete operator with its abstract counterpart.

Abstract interpretation has been widely used to design static analysis of logic programs. In the literature, we find many proposals for the concrete domain, the concrete semantics, the abstract domain and the abstract operators. For instance, Hans and Winkler (1992) focus on the abstract domains, Howe and King (2003) on the abstract operators, King and Longley (1995) on improving existing analysis using a more refined concrete semantics, while Cortesi et al. (1996) propose a complete framework, combination of particular concrete semantics and abstract domains. In many cases, the correctness of the analysis is taken for granted, since the concrete semantics is not completely specified. However, when applying several of these improvements to a single analysis framework, the improved analysis may significantly differ from the original proposal, and a new proof of correctness is needed for the overall analysis. This is especially true for logic programming, whose basic computational mechanism, unification, is intrinsically more complex than assignment or matching, used in other programming paradigms.

The aim of this article is mainly to clarify and unify several different proposals for the goal-dependent analysis of logic programs. Inspired by the work of Cortesi et al. (1996), we propose a new denotational framework which combines and improves many different ideas appeared in the literature. Later, we focus on the abstract domain `Sharing` by Jacobs and Langen (1992), and we develop an analysis which is strictly more precise than the others in the literature. We formally prove correctness of the overall analysis and optimality of all the involved abstract operators.

When designing a new analysis, one needs to choose a concrete domain and semantics, an abstract domain and abstract operators. Although these choices are related, in the following we will introduce them separately, showing available alternatives, possible improvements and the contributions of this paper.

### *Concrete domain*

Typically, concrete semantics of logic programs are defined over substitutions. However, substitutions are often too informative. For example, consider the one-clause program $\mathtt{p(x,x)}$ and the goal $p(x, y)$. All of $\{x/y\}$, $\{y/x\}$, $\{x/u, y/u\}$, $\{x/v, y/v\}$ are computed answers, corresponding to different choices of most general unifiers and renamed clauses. Often, especially in the case of static analysis, we are not interested in making any distinction among them. Thus, it would be more natural to adopt a domain of equivalence classes of substitutions. Many frameworks for abstract interpretation of logic programs (Jacobs and Langen 1992; Marriott et al. 1994; Levi and Spoto 2003) have adopted similar solutions for avoiding redundancy and causality when choosing computed answers.

Nevertheless, the standard semantics of logic programs, namely SLD resolution, is based on substitutions and unification. Thus, any framework for logic programming should relate, in some way, to standard substitutions, in order to prove that the semantics reflects the underlying operational behavior. However, none of the above frameworks formally states the correspondence between the proposed concrete domain and standard substitutions. Although this correspondence is clear from an intuitive point of view, we think that substitutions are tricky objects, where intuition often fails.

*Our contribution.* We propose a new concrete domain of classes of substitutions, called *existential substitutions*, equipped with a set of primitive operators for projection, renaming and unification. We formally state the correspondence between substitutions and existential substitutions, and in particular between the corresponding unification operators. Moreover, we show the relationship between our proposal and the domain *ESubst* by Jacobs and Langen (1992).

## *Concrete semantics*

We are interested in goal-driven analysis of logic programs. Therefore we need a goal-dependent semantics which is well suited for static analysis, i.e., a collecting semantics over computed answer substitutions. Unfortunately, using a collecting goal-dependent semantics may lead to a loss of precision already at the concrete level, as shown by Marriott et al. (1994). Basically, in any goal-dependent semantics, the unification operator is used twice:

- For performing parameter passing by unifying the given goal and the *call substitution* with the head of the chosen clause. The result is a new goal and an *entry substitution.* This operation is called *forward unification.*
- For propagating back to the initial goal the *exit substitution* (that is, the result of the sub-computation), so obtaining the *answer substitution* for the initial goal. This operation is called *backward unification*[1]

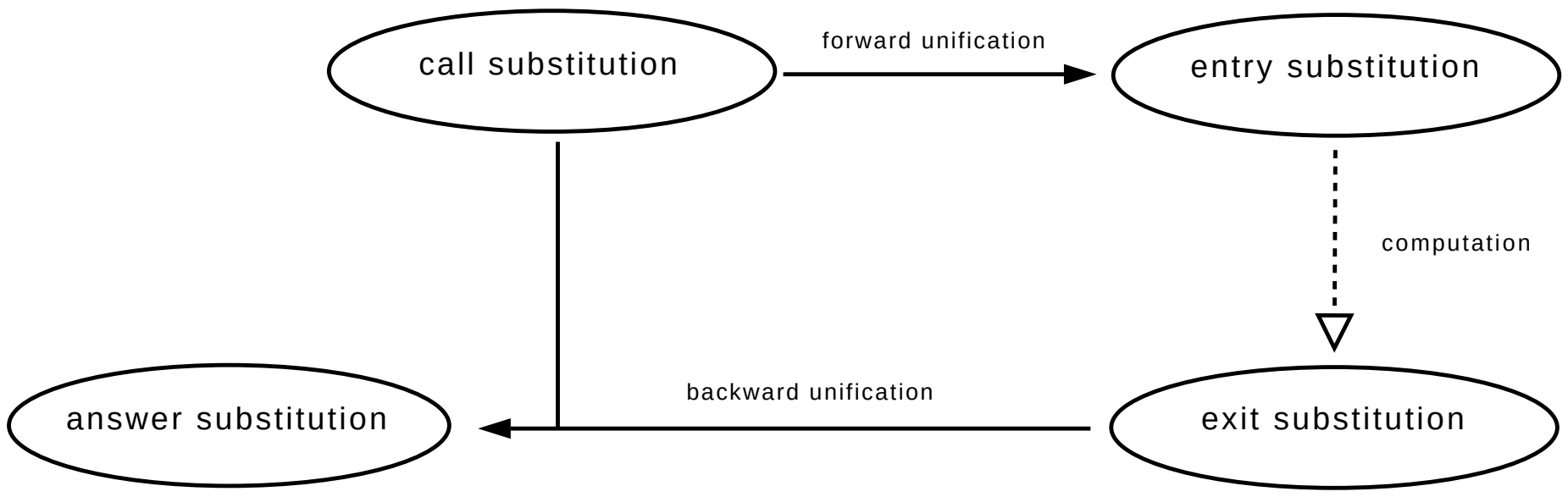

For instance, given the initial goal $p(x)$ and the call substitution $\{x/f(y)\}$, we unify with the head of the clause $\mathtt{p(z)} \leftarrow \mathtt{q(z)}$ by computing the most general unifier $\{x/f(y), z/f(y)\}$, which, projected on the variables of the clause, is simply $\{z/f(y)\}$. Projection is needed in order to avoid an unbounded growing of the set of variables in the entry substitution. This is acceptable at the concrete level, but not at the abstract level, where it may lead to non-terminating analysis. The new goal and entry substitution become $q(z)$ and $\{z/f(y)\}$. Once we have obtained an exit substitution for the goal $q(z)$, for instance $\{z/f(a)\}$, we have to relate this result to the original goal $p(x)$. Thus we need a so-called backward unification, which allows us to conclude that $\{x/f(a)\}$ is an answer for $p(x)$ with call substitution $\{x/f(y)\}$.

[1] We follow Cortesi et al. (1996) and call these operators forward and backward unification. Bruynooghe (1991) and Hans and Winkler (1992) use *procedure entry* and *procedure exit.* Muthukumar and Hermenegildo (1991) use *call_to_entry* and *exit_to_success.*

The backward unification operator introduces a loss of precision, due to the fact that we deal with a set of call substitutions, from which we possibly obtain a set of exit substitutions. Now, when we go backward to obtain the answer substitutions, we may unify a call substitution with an exit substitution which does not pertain to the same computational path (Marriott et al. 1994).

It is possible to reduce the impact of this problem by using two different operators for forward and backward unification (Bruynooghe 1991; Le Charlier et al. 1991). In this way, backward unification can be realized using the operation of matching between substitutions.

*Our contribution.* We propose a denotational goal-dependent semantics equipped with two different forward and backward unification operators. Backward unification uses matching, exploiting the property that the exit substitution is more instantiated than the call substitution. We prove that the concrete semantics is correct and show that the new semantics is strictly more precise than semantics which do not use matching.

### *Abstract domain*

One of the most interesting (and studied) property for logic programs is sharing. The goal of (set) sharing analysis is to detect sets of variables which share a common variable. For instance, in the substitution $\{x/f(z,a), y/g(z)\}$ the variables $x$ and $y$ share the common variable $z$. Typical applications of sharing analysis are in optimization of unification (Søndergaard 1986) and parallelization of logic programs (Hermenegildo and Rossi 1995).

The basic domain for set sharing analysis is `Sharing`, introduced in (Langen 1990; Jacobs and Langen 1992). It is widely recognized that `Sharing` is not very precise, so that it is often combined with other domains for freeness, linearity, groundness or structural information (see Bagnara et al. (2005) for a comparative evaluation). Since this paper does not address the problem to find the best possible domain for set-sharing analysis, we will focus on the domain `Sharing`.

### *Abstract operators*

Once the concrete semantics and the abstract domain have been fixed, the next step is to find suitable abstract operators which mimic the behavior of the concrete ones. The theory of abstract interpretation ensures the existence of the optimal (best correct) abstract operator for each concrete operator. Although the optimal abstract operator enjoys a constructive characterization, this is not amenable to a direct implementation. Therefore, finding an algorithm to compute optimal abstract operators is one of the main difficulties in any abstract interpretation project.

We think that there are several reasons to look for the optimal operator, instead of just using a correct one. In fact, one may argue that a well-design correct operator may be much faster then the optimal one, and does not lose much precision in real programs. Although we agree with this point, we think that knowing the optimal abstract operator, even if we do not plan to implement it, is useful to understand the potentiality and limits of the abstract domain in use, and to guide the search for a more precise (or more efficient) domain. Moreover, at least in the case of sharing analysis, the more precise the abstract operators are, the smaller are the abstract objects computed during the analysis. Therefore, it may be worth spending more time in computing the abstract operators, in order to keep the abstract objects simpler (and the analysis more precise).

The primitive concrete operators used in the semantics of logic programs are renaming, projection, unification and matching. Renaming and projections are not problematic at all: it is generally immediate to find their optimal abstract counterparts, which most of the time are also complete, i.e., they do not lose precision w.r.t. the corresponding concrete operators (Cousot and Cousot 1979; Giacobazzi et al. 2000).

Things are different for unification, which is a very complex operator. In fact, despite several works in this field, the best correct abstraction of unification for the domain `SFL` (King and Longley 1995), which combines sharing, freeness and linearity, is still unknown. For the domain `Sharing`, Cortesi and Filé (1999) have shown that abstract unification defined in Jacobs and Langen (1992) is optimal. However, this result has been obtained for a concrete semantics which uses the same unification operator to compute both forward and backward unification.

We have already said that a specialized backward unification operator may improve precision at the concrete level. In turn, the improvement in precision is reflected at the abstract level, if the abstract backward unification operator is designed to mimic matching instead of standard unification. This idea is implemented in real abstract interpreters such as GAIA (Le Charlier et al. 1991) and PLAI (Muthukumar and Hermenegildo 1992). However, none of the papers which are based on a specialized backward unification operator with matching (Bruynooghe 1991; Le Charlier and Van Hentenryck 1994; Hans and Winkler 1992; Muthukumar and Hermenegildo 1992; King and Longley 1995) has ever proved optimality of the proposed abstract operators. As we will show later, those abstract operators which involve set-sharing information (Hans and Winkler 1992; Muthukumar and Hermenegildo 1992; King and Longley 1995) are not optimal.

In addition, the abstract forward unification operator can be specialized in order to exploit the peculiarity of this process: the variables which occur in the clause head are always renamed apart w.r.t. the goal and the calling substitutions, hence they are free and independent. However, this idea has never been applied before in the general case, but only for abstract domains which explicitly contain freeness and linearity information.

*Our contribution.* We provide abstract operators for renaming, projection, forward unification and backward unification. We prove that all our operators are optimal and that renaming and projection are also complete. We show that abstract forward unification is able to exploit freeness and linearity information. The new backward and forward unification operators strictly improve over previous proposals for the domain `Sharing`.

Although freeness and linearity information are exploited by the forward abstract unification operator, this information is not encoded in the abstract domain, but is just used in the internal steps of the abstract unification algorithm. This means that the algorithm cannot be immediately extended to work with more complex domains, such as `SFL` (King and Longley 1995), retaining optimality. Nonetheless, the abstract unification is able to exploit freeness and linearity better than other algorithms and could be used to improve the unification operation in more complex domains.

### *Plan of the paper*

The next section recalls some basic definitions and the notations about abstract interpretation and substitutions. In Section 3 we define the domain of existentially quantified substitutions and its operators. In Sections 4 and 5 we define the concrete and abstract semantics. Finally, in Sections 6 and 7 we give the algorithms for computing the forward and backward abstract unification and show their correctness and optimality. In Section 8 we compare our framework with related work.

The article is a substantial expansion of (Amato and Scozzari 2002), which introduces preliminary results using standard substitutions. A partial presentation of existential substitutions appeared in Amato and Scozzari (2003).

## 2 Notations

Given a set $A$, let $\wp(A)$ be the powerset of $A$ and $\wp_f(A)$ be the set of finite subsets of $A$. Given two posets $(A, \leq_A)$ and $(B, \leq_B)$, we denote by $A \xrightarrow{m} B$ ($A \xrightarrow{c} B$) the space of monotonic (continuous) functions from $A$ to $B$ ordered pointwise. When an order for $A$ or $B$ is not specified, we assume the least informative order ($x \leq y \iff x = y$). We also use $A \uplus B$ to denote disjoint union and $|A|$ for the cardinality of the set $A$.

Given complete lattices $A, C$, a *Galois connection* (Cousot and Cousot 1979) $\langle \alpha, \gamma \rangle : C \leftrightarrows A$ is given by a pair of maps $\alpha : C \xrightarrow{m} A$, $\gamma : A \xrightarrow{m} C$ such that $\alpha(c) \leq_A a \iff c \leq_C \gamma(a)$. A Galois connection is a *Galois insertion* when $\alpha$ is onto (or equivalently, $\gamma$ is injective). We say that an abstract operator $f^\alpha : A \xrightarrow{m} A$ is *correct* w.r.t. a concrete operator $f : C \xrightarrow{m} C$ when $\forall c \in C.\ (\alpha \circ f)(c) \leq_A (f^\alpha \circ \alpha)(c)$, which is equivalent to $\forall a \in A.\ (f \circ \gamma)(a) \leq_C (\gamma \circ f^\alpha)(a)$ and to $\forall a \in A.\ (\alpha \circ f \circ \gamma)(a) \leq_A f^\alpha(a)$. The abstract operator is *optimal* when $f^\alpha = \alpha \circ f \circ \gamma$. In this case $f^\alpha$ is called the *best correct approximation* of $f$. When $\alpha \circ f = f^\alpha \circ \alpha$ then $f^\alpha$ is said to be *complete*, while if $f \circ \gamma = \gamma \circ f^\alpha$ then $f^\alpha$ is $\gamma$*-complete*.

In the following, we fix a first order signature $(\Sigma, \Pi)$ and an infinite set of variables $\mathcal{V}$. We assume that there are a constant symbol and a function symbol of arity at least two[2]. We use $\mathsf{Terms}$ and $\mathsf{Atoms}$ to denote the sets of terms and atomic formulas (atoms) respectively. Moreover, we call *body* or *goal* a finite sequence of atomic formulas, *clause* an object $H \leftarrow B$ where $H$ is an atom and $B$ is a body, *program* a set of clauses. We use $\Box$ for the empty body and we write $H$ as a short form for $H \leftarrow \Box$. We denote with $\mathsf{Bodies}$, $\mathsf{Clauses}$ and $\mathsf{Progs}$ the set of bodies, clauses and programs respectively. Given a term $t$, we denote by $\mathrm{vars}(t)$ the set of variables occurring in $t$ and by $\mathrm{uvars}(t)$ the subset of $\mathrm{vars}(t)$ whose elements appear once in $t$ (e.g., $\mathrm{uvars}(f(x,y) = f(y,z)) = \{x, z\}$). We apply vars and uvars to any syntactic object, with the obvious meaning. We abuse the notation and write a syntactic object $o$ instead of the set of variables $\mathrm{vars}(o)$, when it is clear from the context (e.g., if $t$ is a term and $x \in \mathcal{V}$, then $x \in t$ should be read as $x \in \mathrm{vars}(t)$).

We denote with $\epsilon$ the empty substitution and by $\{x_1/t_1, \ldots, x_n/t_n\}$ a substitution $\theta$ with $\theta(x_i) = t_i \neq x_i$. Let $\mathrm{dom}(\theta)$ be the set $\{x_1, \ldots, x_n\}$ and $\mathrm{rng}(\theta)$ be the set $\mathrm{vars}(\{t_1, \ldots, t_n\})$. Thus we have that $\mathrm{vars}(\theta) = \mathrm{dom}(\theta) \cup \mathrm{rng}(\theta)$. Given $U \in \wp_f(\mathcal{V})$, let $\theta_{|U}$ be the projection of $\theta$ on $U$, i.e., the unique substitution such that $\theta_{|U}(x) = \theta(x)$ if $x \in U$ and $\theta_{|U}(x) = x$ otherwise. We also write $\theta_{|-U}$ to denote the restriction of $\theta$ over all variables but those in $U$, i.e., $\theta_{|-U} = \theta_{|\mathrm{dom}(\theta)\setminus U}$. Given $\theta_1$ and $\theta_2$ two substitutions with disjoint domains, we denote by $\theta_1 \uplus \theta_2$ the substitution $\theta$ such that $\mathrm{dom}(\theta) = \mathrm{dom}(\theta_1) \cup \mathrm{dom}(\theta_2)$ and $\theta(x) = \theta_i(x)$ if $x \in \mathrm{dom}(\theta_i)$, for each $i \in \{1, 2\}$. The application of a substitution $\theta$ to a term $t$ is written as $t\theta$ or $\theta(t)$. Given two substitutions $\theta$ and $\delta$, their composition, denoted by $\theta \circ \delta$, is given by $(\theta \circ \delta)(x) = \theta(\delta(x))$. A substitution $\rho$ is called renaming if it is a bijection from $\mathcal{V}$ to $\mathcal{V}$ (this is equivalent to say that there exists a substitution $\rho^{-1}$ such that $\rho \circ \rho^{-1} = \rho^{-1} \circ \rho = \epsilon$). A substitution $\theta$ is idempotent when $\mathrm{dom}(\theta) \cap \mathrm{rng}(\theta) = \emptyset$. Instantiation induces a preorder on substitutions: $\theta$ is more general than $\delta$, denoted by $\delta \leq \theta$, if there exists $\sigma$ such that $\sigma \circ \theta = \delta$. If $\approx$ is the equivalence relation induced by $\leq$, we say that $\sigma$ and $\theta$ are equal up to renaming when $\sigma \approx \theta$. The set of substitutions, idempotent substitutions and renamings are denoted by *Subst*, *ISubst* and *Ren* respectively.

Given a set of equations $E$, we write $\sigma = \mathrm{mgu}(E)$ to denote that $\sigma$ is a most general unifier of $E$ such that $\mathrm{vars}(\sigma) \subseteq \mathrm{vars}(E)$. Since $\sigma$ is defined up to renamings, we use this notation only in cases where the choice of the actual unifier does not matter. Any idempotent substitution $\sigma$ is a most general unifier of the corresponding set of equations $\mathrm{Eq}(\sigma) = \{x = \sigma(x) \mid x \in \mathrm{dom}(\sigma)\}$. In the following, we will abuse the notation and denote by $\mathrm{mgu}(\sigma_1, \ldots, \sigma_n)$, when it exists, the substitution $\mathrm{mgu}(\mathrm{Eq}(\sigma_1) \cup \ldots \cup \mathrm{Eq}(\sigma_n))$.

In the rest of the paper, we use: $U$, $V$, $W$ to denote finite sets of variables; $h, k, u, v, w, x, y, z$ for variables; $c, s, t$ for term symbols or terms; $a, b$ for constants; $cl$ for clauses; $\eta, \theta, \sigma, \delta$ for substitutions; $\rho$ for renamings. All these symbols can be subscripted or superscripted.

[2] Otherwise every term has at most one variable and the structure of terms is trivial. We need this assumption in Section 8.1 and in the proofs of optimality of unification and matching.

## 3 Domains of Existentially Quantified Substitutions

The first question when analyzing the behavior of logic programs is what kind of observable we are interested in. Undoubtedly, computed answers have played a prominent role, since they are the result of the process of SLD-resolution. Moreover, they have several nice properties: and-compositionality, condensing and a bottom-up $T_P$-like characterization (van Emden and Kowalski 1976; Bossi et al. 1994). Standard semantics for logic programs, e.g., the $s$-semantics in (Bossi et al. 1994), are defined over equivalence classes of atoms modulo renaming. For example, consider the one-clause program $\mathtt{p(x,x)}$ and the goal $p(x, y)$. All of $p(x, x)$, $p(y, y)$, $p(u, u)$ and $p(v, v)$ are computed instances, corresponding to different choices of most general unifiers and renamed clauses, but we are not interested in making any distinction among them.

However, when we consider a denotational semantics suitable for program analysis, computed answer *substitutions* are much more useful than computed instances, since most of the domains are expressed as abstraction of sets of substitutions. As before, we are not really interested in the substitutions, but in their quotient-set w.r.t. a suitable equivalence relation. But in this case we cannot take renaming as the relevant equivalence relation. Let us consider the substitutions corresponding to the computed instances in the previous example: we obtain $\theta_1 = \{y/x\}$, $\theta_2 = \{x/y\}$, $\theta_3 = \{x/u, y/u\}$ and $\theta_4 = \{x/v, y/v\}$. Although $\theta_1$ and $\theta_2$ are equal up to renaming, the same does not hold for $\theta_3$ and $\theta_4$. Nonetheless, they essentially represent the same answer, since $u$ and $v$ are just two different variables we chose when renaming apart the clause $\mathtt{p(x,x)}$ from the goal $p(x, y)$, and therefore are not relevant. On the other side, if $\theta_3$ and $\theta_4$ were computed answers for the goal $q(x, y, u)$, they would correspond to computed instances $q(u, u, u)$ and $q(v, v, u)$ and therefore would be definitively different. As a consequence, the equivalence relation we need to consider must be coarser than renaming, and must take into account the set of variables of interest, i.e., the set of variables which appear in the goal.

A semantics which takes into account classes of substitutions may follow three possible directions:

1. it may compute only a subset of the computed answer substitutions, provided that the result contains at least one substitution for each equivalence class, e.g., (Cortesi et al. 1996);
2. it may compute all the computed answer substitutions, e.g., (Le Charlier et al. 1991);
3. it may be defined using a quotient domain of substitutions, e.g., (Marriott et al. 1994).

The problem with the first two solutions is that they work by directly manipulating substitutions. It is common knowledge that this is quite tedious and error prone (Shepherdson 1994). This happens because substitutions are too much related to syntax, so that the intuition of what should happen is often betrayed by the reality, when we need to handle problems such as variable clashes and renamings. Actually, at least one framework of the first kind, namely the widely used one in (Cortesi and Filé 1999), has a small flaw due to an unsound treatment of variable clashes (this will be discussed in details in Section 8.2).

Moreover, the first approach is generally pursued by choosing a particular most general unifier and a fixed way of renaming apart terms and substitutions. The semantics is then parametric with respect to these choices. As stated by Jacobs and Langen (1992), this makes difficult to compare different semantics, since each of them may use different conventions for mgu and renaming. We would like to add that this also makes difficult to state properties of a given semantics (such as compositionality properties), since they only hold up to suitable equivalence relations.

For these reasons, we think that the best solution is to move towards a domain of equivalence classes of substitutions. This does not mean we can avoid to work with substitutions altogether, but all the difficulties which arise, such as renaming apart and variables clashes, may be dealt with once and for all at the domain level, reducing the opportunities for subtle mistakes to appear.

### 3.1 Yet another Domain of Existentially Quantified Substitutions

In the literature there are several domains of equivalence classes of substitutions: *ESubst* (Jacobs and Langen 1992), ex-equations (Marriott et al. 1994) and existential Herbrand constraints (Levi and Spoto 2003). For all of them, the basic idea is that some variables, in a substitution or equation, are existentially quantified, so that their names become irrelevant. However, all these proposals depart from the standard notion of substitution. As a result, the relationship between what they compute and the standard set of computed answers for a goal has never been proved. We would like to reconcile these approaches with the standard concept of substitution: in particular, we want to prove that these domains are quotient sets of substitutions, w.r.t. suitable equivalence relations.

We begin by introducing a new equivalence relation $\sim$ over substitutions, which captures the extended notion of renaming which is needed to work with computed answers. Inspired by the seminal paper of Palamidessi (1990), we introduce a new domain $\mathit{Subst}_{\sim}$ of classes of substitutions modulo $\sim$, which will be used in the rest of the paper[3].

Given $\theta_1, \theta_2 \in \mathit{Subst}$ and $U \in \wp_f(\mathcal{V})$, we define the preorder:

$$\theta_1 \preceq_U \theta_2 \iff \exists \delta \in \mathit{Subst}. \forall u \in U.\ \theta_1(u) = \delta(\theta_2(u)) \ . \tag{1}$$

Intuitively, if $\theta_1 \preceq_U \theta_2$, then $\theta_1$ is an instance of $\theta_2$, provided we are only interested in the variables in $U$.

*Example 3.1*

It is easy to check that $\{x/a, y/u\} \preceq_{\{x,y\}} \{y/v\}$, since we may choose $\delta = \{x/a, v/u\}$ in (1). Note that the same does not happen if we consider the standard ordering on substitutions, i.e., $\{x/a, y/u\} \not\preceq \{y/v\}$. Moreover, if we enlarge the set $U$ of variables of interest, we obtain $\{x/a, y/u\} \not\preceq_{\{x,y,v\}} \{y/v\}$. □

[3] In Section 8.1, we will prove that $\mathit{Subst}_{\sim}$ and the domain *ESubst* (Jacobs and Langen 1992) are isomorphic.

Note that, in Equation (1), it is important that $\delta$ is a generic substitution. If we restrict $\delta$ to be idempotent, some equivalences do not hold anymore. For example, $\{x/t(u), y/t(v)\} \preceq_{\{x,y\}} \{x/v, y/u\}$ and this is what we intuitively want, since the names of the variables $u$ and $v$ are not relevant. However, to prove this relation, we choose $\delta = \{u/t(v), v/t(u)\}$ in (1), and it is not an idempotent substitution.

*Proposition 3.2*
$\preceq_U$ is a preorder for any $U \in \wp_f(\mathcal{V})$.

*Proof*
Let $U \in \wp_f(\mathcal{V})$. By definition, $\theta \preceq_U \theta \iff \exists \delta \in \mathit{Subst}.\forall v \in U.\ \theta(v) = \delta(\theta(v))$, which is a tautology by choosing as $\delta$ the empty substitution. Now assume $\theta_1 \preceq_U \theta_2$ and $\theta_2 \preceq_U \theta_3$. Therefore, there exist $\delta_1$ and $\delta_2$ such that, $\forall v \in U$, $\theta_1(v) = \delta_1(\theta_2(v))$ and $\theta_2(v) = \delta_2(\theta_3(v))$. Therefore, $\forall v \in U$, it holds $\theta_1(v) = \delta_1(\theta_2(v)) = \delta_1(\delta_2(\theta_3(v)))$. Therefore, by choosing as $\delta$ the composition $\delta_1 \circ \delta_2$ we have that $\theta_1 \preceq_U \theta_3$. □

The next step is to define the relation:

$$\theta_1 \sim_U \theta_2 \iff \exists \rho \in \mathit{Ren}.\forall v \in U.\ \theta_1(v) = \rho(\theta_2(v)) \ , \tag{2}$$

which will be proved to be the equivalence relation induced by the preorder $\preceq_U$.

*Example 3.3*
It is easy to check that $\{x/v, y/u\} \sim_{\{x,y\}} \epsilon$ by choosing $\rho = \{x/v, v/x, y/u, u/y\}$. Note that $\sim_U$ is coarser than the standard equivalence relation $\approx$: there is no renaming $\rho$ such that $\epsilon = \rho \circ \{x/v, y/u\}$. As it happens for $\preceq$, if we enlarge the set of variables of interest, not all equivalences between substitutions are preserved: for instance, $\{x/v, y/u\} \not\sim_{\{x,y,v\}} \epsilon$. □

*Lemma 3.4*
Let $\theta : V \to \mathcal{V}$ an injective map of variables. Then there exists $\rho \in \mathit{Ren}$ such that $\rho(x) = \theta(x)$ for each $x \in V$ and $\mathrm{vars}(\rho) = V \cup \theta(V)$.

*Proof*
Since $\theta$ is injective, $|V| = |\theta(V)|$, from which it follows that $|V \setminus \theta(V)| = |\theta(V) \setminus V|$. Let $f$ be any bijective map from $\theta(V)\setminus V$ to $V\setminus\theta(V)$, and let us define a substitution $\rho$ as follows:

$$\rho(v) = \begin{cases} \theta(v) & \text{if } v \in V \\ f(v) & \text{if } v \in \theta(V) \setminus V \\ v & \text{otherwise.} \end{cases}$$

Note that, if $x \in V$, $\rho(x) = \theta(x)$ by definition. Moreover, it is easy to check that $\rho$ is bijective, therefore, it is a renaming. Finally, $\mathrm{vars}(\rho) = \mathrm{dom}(\rho) = V \cup (\theta(V) \setminus V) = V \cup \theta(V)$. □

*Proposition 3.5*
The relation $\sim_U$ is the equivalence relation induced by $\preceq_U$.

*Proof*
If $\theta_1 \sim_U \theta_2$ there exists $\rho \in \mathit{Ren}$ such that $\forall v \in U.\ \theta_1(v) = \rho(\theta_2(v))$. By definition of $\preceq_U$, we have that $\theta_1 \preceq_U \theta_2$ by choosing as $\delta$ in (1) the renaming $\rho$. Symmetrically, by choosing as $\delta$ the renaming $\rho^{-1}$ (the inverse of $\rho$), it follows that $\theta_2 \preceq_U \theta_1$.

Now assume that $\theta_1 \preceq_U \theta_2$ and $\theta_2 \preceq_U \theta_1$. Therefore there exist $\delta, \delta' \in \mathit{Subst}$ such that $\theta_2(x) = \delta'(\theta_1(x))$ and $\theta_1(x) = \delta(\theta_2(x))$, thus $\theta_2(x) = \delta'(\delta(\theta_2(x)))$ for each $x \in U$. In general, $\delta$ and $\delta'$ might not be renamings. Our goal is to build a renaming $\rho$, obtained by modifying $\delta$, such that $\theta_1(x) = \rho(\theta_2(x))$, for each $x \in U$. Let $V = \mathrm{vars}(\theta_2(U))$. Since each $v \in V$ belongs to $\mathrm{vars}(\theta_2(x))$ for some $x \in U$, it follows that $(\delta' \circ \delta)(v) = v$ for all $v \in V$. Therefore, $\delta_{|V}$ may be viewed as an injective map from $V$ to $\mathcal{V}$. By Lemma 3.4, there exists $\rho \in \mathit{Ren}$ such that $\rho_{|V} = \delta_{|V}$. Therefore, for each $x \in U$, $\rho(\theta_2(x)) = \delta(\theta_2(x)) = \theta_1(x)$, hence $\theta_1 \sim_U \theta_2$. □

It is worth noting that $\preceq_U$ is coarser than $\leq$ and that $\sim_U$ is coarser than renaming, as shown by the following proposition.

*Proposition 3.6*
Given $\theta \in \mathit{Subst}$, $\rho \in \mathit{Ren}$ and $\delta \in \mathit{Subst}$ then $\rho \circ \theta \sim_U \theta$ and $\delta \circ \theta \preceq_U \theta$ for each $U \in \wp_f(\mathcal{V})$.

*Proof*
Simply choose $\rho$ and $\delta$ as the relevant substitutions in (1) and (2). □

Now, let $\mathit{ISubst}_{\sim_U}$ be the quotient set of *ISubst* w.r.t. $\sim_U$. We define a new domain $\mathit{ISubst}_\sim$ of *existential substitutions* as the disjoint union of all the $\mathit{ISubst}_{\sim_U}$ for $U \in \wp_f(\mathcal{V})$, in formulas:

$$\mathit{ISubst}_\sim = \biguplus_{U \in \wp_f(\mathcal{V})} \mathit{ISubst}_{\sim_U} \ . \tag{3}$$

In the following we write $[\theta]_U$ for the equivalence class of $\theta$ w.r.t. $\sim_U$. We call *canonical representatives* of the equivalence class $[\theta]_U \in \mathit{ISubst}_\sim$ the substitutions $\theta' \in \mathit{ISubst}$ such that $\theta' \sim_U \theta$ and $\mathrm{dom}(\theta') = U$. It is immediate to see that every existential substitution has a canonical representative, although it is not unique. For example, two canonical representatives of $[\{y/f(x)\}]_{x,y,z}$ are $\{y/f(h), x/h, z/k\}$ and $\{y/f(u), x/u, z/v\}$. Working with canonical representatives is of great help, especially in the proofs, since we are sure they have no variables of interest in the range.

By definition of $\preceq_U$, when $\theta \preceq_U \theta'$ then, for all $W \subseteq U$, it holds that $\theta \preceq_W \theta'$. This allows us to define a partial order $\preceq$ over $\mathit{ISubst}_\sim$ given by:

$$[\theta]_U \preceq [\theta']_V \iff U \supseteq V \wedge \theta \preceq_V \theta' \ . \tag{4}$$

Intuitively, $[\theta]_U \preceq [\theta']_V$ means that $\theta$ is an instance of $\theta'$ w.r.t. the variables in $V$, provided that they are all variables of interest of $\theta$. It is easy to show that $\preceq$ is well-defined in $\mathit{ISubst}_\sim$, that is it does not depend on the choice of the representatives.

Note that, although we use equivalence classes of idempotent substitutions, we could build an isomorphic domain by working with equivalence classes of the set of all the substitution. In other words, if we define $\mathit{Subst}_{\sim} = \biguplus_{U \in \wp_f(\mathcal{V})} \mathit{Subst}_{\sim_U}$, we obtain the following:

*Proposition 3.7*
The posets $(\mathit{Subst}_{\sim}, \preceq)$ and $(\mathit{ISubst}_{\sim}, \preceq)$ are isomorphic.

*Proof*
It is enough to prove that, for each $U \in \wp_f(\mathcal{V})$ and $\theta \in \mathit{Subst}$, there exists $\theta' \in \mathit{ISubst}$ such that $\theta \sim_U \theta'$. Let $V = \mathrm{rng}(\theta) \cap \mathrm{dom}(\theta)$ and $W \subseteq \mathcal{V}$ such that $W \cap (U \cup \mathrm{vars}(\theta)) = \emptyset$ and $|V| = |W|$. Moreover, we take a renaming $\rho$ such that $\mathrm{vars}(\rho) = V \cup W$ and $\rho(V) = W$. Then, we may define a substitution $\theta'$ such that

$$\theta' = (\rho \circ \theta)_{|U} \ .$$

Note that $\mathrm{dom}(\theta') = (\mathrm{dom}(\theta) \cup W) \cap U \subseteq \mathrm{dom}(\theta)$ and $\mathrm{rng}(\theta') \subseteq \mathrm{rng}(\theta) \setminus V \cup W$. Therefore, $\mathrm{dom}(\theta') \cap \mathrm{rng}(\theta') = \emptyset$, i.e., $\theta' \in \mathit{ISubst}$. Moreover, by definition, $\theta' \sim_U \theta$. □

The isomorphism between $\mathit{Subst}_{\sim}$ and $\mathit{ISubst}_{\sim}$ holds since a variable in $\mathrm{rng}(\theta)$ is considered not of interest if it also occurs in $\mathrm{dom}(\theta)$. Therefore $\{x/y, y/x\} \sim_{\{x,y\}} \{x/u, y/v\}$, since $y$ and $x$ in the range of $\{x/y, y/x\}$ are just names for existential quantified variables. Obviously $\{x/y\} \not\sim_{\{x,y\}} \{x/u\}$ since here $y$ only appears in the range and is therefore considered as a variable of interest.

### *3.2 Operations on the new Domain*

It is now time to define some useful operations over $\mathit{ISubst}_{\sim}$, which will be used as building blocks for the semantics to be defined further away in the paper. They will also give some more insights over the structure of $\mathit{ISubst}_{\sim}$. To ease notation, we often omit braces from the sets of variables of interest when they are given extensionally. So we write $[\theta]_{x,y}$ instead of $[\theta]_{\{x,y\}}$ and $\sim_{x,y,z}$ instead of $\sim_{\{x,y,z\}}$. When the set of variables of interest is clear from the context or it is not relevant, it will be omitted. Finally, we omit the braces which enclose the bindings of a substitution when it occurs inside an equivalence class, i.e., we write $[x/y]_U$ instead of $[\{x/y\}]_U$.

#### *3.2.1 Projection*

We define an operator which projects an element of $\mathit{ISubst}_{\sim}$ on a given set of variables $V$, given by

$$\pi_V([\sigma]_U) = [\sigma]_{U \cap V} \ , \tag{5}$$

which can be easily proved to be well-defined. Moreover, the following properties hold:

1. $\pi_U \circ \pi_V = \pi_{U \cap V}$;
2. $\pi_U([\sigma]_U) = [\sigma]_U$;
3. $\pi_V$ is monotonic w.r.t. $\preceq$.

### *3.2.2 Renaming*

Another useful operation on classes of substitutions is renaming. We first define the application of a renaming $\rho \in \mathit{Ren}$ to a substitution $\theta \in \mathit{Subst}$ as

$$\rho(\theta) = \{\rho(x)/\rho(\theta(x)) \mid x \in \mathrm{dom}(\theta)\} \ . \tag{6}$$

Intuitively, we treat $\theta$ as a syntactic object and apply the renaming to both left and right hand sides. Note that $\rho(\theta)$ can be equivalently defined as $\rho \circ \theta \circ \rho^{-1}$.

*Proposition 3.8*
Given $\rho \in \mathit{Ren}$ and $\theta \in \mathit{Subst}$ it holds that $\rho(\theta) = \rho \circ \theta \circ \rho^{-1}$.

*Proof*
Let $\theta' = \rho(\theta)$. Since $y \neq \theta(y)$ for all $y \in \mathrm{dom}(\theta)$, then $\rho(y) \neq \rho(\theta(y))$ by injectivity of $\rho$. It follows that $\mathrm{dom}(\theta') = \rho(\mathrm{dom}(\theta))$. We now prove that, for each $x \in \mathcal{V}$, $\theta'(x) = \rho(\theta(\rho^{-1}(x)))$. We distinguish two cases.

- If $x \notin \mathrm{dom}(\theta')$, it follows that $x \notin \rho(\mathrm{dom}(\theta))$ and thus $\rho^{-1}(x) \notin \mathrm{dom}(\theta)$. As a consequence, $\rho(\theta(\rho^{-1}(x))) = \rho(\rho^{-1}(x)) = x = \theta'(x)$.
- If $x \in \mathrm{dom}(\theta')$, then $y = \rho^{-1}(x) \in \mathrm{dom}(\theta)$ and $\theta'(x) = \rho(\theta(y))$. Therefore $\rho(\theta(\rho^{-1}(x))) = \rho(\theta(y)) = \theta'(x)$. □

We may lift this definition to classes of substitutions in the standard way as follows:

$$\rho([\sigma]_U) = [\rho(\sigma)]_{\rho(U)} \ . \tag{7}$$

For example, let $\sigma = \{x/k, y/t(z,k)\}$, $U = \{x, y, z\}$ and consider the renaming:

$$\rho = \{x/u, u/x, y/z, z/y, k/h, h/k\} \ .$$

If we apply $\rho$ to $[\sigma]_U$ we obtain $\rho([\sigma]_U) = [\{u/h, z/t(y,h)\}]_{u,y,z}$. Note that we do not need to worry about variable clashes.

*Theorem 3.9*
The renaming operation is well defined.

*Proof*
It is enough to prove monotonicity w.r.t. the preorder $\preceq_U$. Given $\theta_1, \theta_2 \in \mathit{Subst}$ such that $\theta_1 \preceq_U \theta_2$, we prove that $\rho(\theta_1) \preceq_{\rho(U)} \rho(\theta_2)$. By Prop. 3.8, we need to show that $\rho \circ \theta_1 \circ \rho^{-1} \preceq_{\rho(U)} \rho \circ \theta_2 \circ \rho^{-1}$, which is equivalent to $\theta_1 \circ \rho^{-1} \preceq_{\rho(U)} \theta_2 \circ \rho^{-1}$ thanks to Prop. 3.6. By hypothesis, there exists a substitution $\delta \in \mathit{Subst}$ such that $\theta_1(x) = \delta(\theta_2(x))$ for all $x \in U$. Therefore, for all $v \in \rho(U)$, it holds $\theta_1(\rho^{-1}(v)) = \delta(\theta_2(\rho^{-1}(v)))$, which is the thesis. □

Several properties hold for the renaming operation:

1. $(\rho_1 \circ \rho_2)([\theta]_V) = \rho_1(\rho_2([\theta]_V))$;
2. $\rho$ is monotonic w.r.t. $\preceq$;
3. $\rho(\pi_V([\theta]_U)) = \pi_{\rho(V)}(\rho([\theta]_U))$;
4. $\rho_1([\theta]_U) = \rho_2([\theta]_U)$ if $\rho_{1|U} = \rho_{2|U}$.

We just prove the last two, since the first is trivial and the second one immediately follows from the proof of Theorem 3.9. Note that the first point implies that $\rho : \mathit{ISubst}_\sim \to \mathit{ISubst}_\sim$ is invertible.

*Proposition 3.10*
Renaming is a congruence w.r.t. $\pi$, i.e.,

$$\rho(\pi_V([\theta]_U)) = \pi_{\rho(V)}(\rho([\theta]_U)) \ .$$

for $[\theta]_U \in \mathit{ISubst}_\sim$ and $\rho \in \mathit{Ren}$.

*Proof*
By definition $\rho(\pi_V([\theta]_U)) = \rho([\theta]_{U\cap V}) = [\rho(\theta)]_{\rho(U\cap V)}$. Since $\rho$ is bijective, $\rho(U \cap V) = \rho(U) \cap \rho(V)$ and therefore $\rho(\pi_V([\theta]_U)) = \pi_{\rho(V)}([\rho(\theta)]_{\rho(U)}) = \pi_{\rho(V)}(\rho([\theta]_U))$, which concludes the proof. □

*Proposition 3.11*
Renaming only depends from the variables of interest, i.e., if $\rho_1, \rho_2 \in \mathit{Ren}$, $[\theta]_U \in \mathit{ISubst}_\sim$ and $\rho_{1|U} = \rho_{2|U}$, then $\rho_1([\theta]_U) = \rho_2([\theta]_U)$. In particular, if $\rho_1|_U = id$, then $\rho_1([\theta]_U) = [\theta]_U$.

*Proof*
Let us denote $\rho_1(U) = \rho_2(U)$ by $W$. We need to prove that $\rho_1(\theta) \sim_W \rho_2(\theta)$. It is obvious that $\rho_1^{-1}{}_{|W} = \rho_2^{-1}{}_{|W}$. Therefore, given $\rho = \rho_1 \circ \rho_2^{-1}$, we have that for each $x \in W$, $\rho(\rho_2(\theta)(x)) = \rho(\rho_2(\theta(\rho_2^{-1}(x)))) = \rho_1(\theta(\rho_1^{-1}(x)))$. □

### *3.2.3 Unification*

Given $U, V \in \wp_f(\mathcal{V})$, $[\theta_1]_U, [\theta_2]_V \in \mathit{ISubst}_\sim$, we define the most general unifier between these two classes as the mgu of suitably chosen representatives, where variables not of interest are renamed apart. In formulas:

$$\mathrm{mgu}([\theta_1]_U, [\theta_2]_V) = [\mathrm{mgu}(\theta_1', \theta_2')]_{U\cup V} \tag{8}$$

where $\theta_1 \sim_U \theta_1' \in \mathit{ISubst}$, $\theta_2 \sim_V \theta_2' \in \mathit{ISubst}$ and $(U \cup \mathrm{vars}(\theta_1')) \cap (V \cup \mathrm{vars}(\theta_2')) \subseteq U \cap V$. The last condition is needed to avoid variables clashes between the chosen representatives $\theta_1'$ and $\theta_2'$.

*Example 3.12*
Let $\theta_1 = \{x/a, y/t(v_1, v_1, v_2)\}$ and $\theta_2 = \{y/t(a, v_2, v_1), z/b\}$. Then

$$\mathrm{mgu}([\theta_1]_{x,y}, [\theta_2]_{y,z}) = [\{x/a, y/t(a, a, v), z/b\}]_{x,y,z}$$

by choosing $\theta_1' = \theta_1$ and $\theta_2' = \{y/t(a, w, v), z/b\}$. In this case we have

$$\{x/a, y/t(a, a, v), z/b\} \sim_{x,y,z} \mathrm{mgu}(\theta_1', \theta_2') = \{x/a, y/t(a, a, v), z/b, v_1/a, w/a, v_2/v\} \ . \quad \square$$

We may prove that mgu over $\mathit{ISubst}_\sim$ is well defined and that $\mathrm{mgu}([\theta_1]_U, [\theta_2]_V)$ is the greatest lower bound of $[\theta_1]_U$ and $[\theta_2]_V$ w.r.t. $\preceq$.

*Theorem 3.13*
mgu is well-defined.

*Proof*
We begin by proving that, given $\theta_1, \theta'_1, \theta_2 \in \mathit{ISubst}$, if $\theta_1 \sim_U \theta'_1$ with $(U \cup \mathrm{vars}(\theta_1)) \cap (V \cup \mathrm{vars}(\theta_2)) \subseteq U \cap V$ and $(U \cup \mathrm{vars}(\theta'_1)) \cap (V \cup \mathrm{vars}(\theta_2)) \subseteq U \cap V$, then $\mathrm{mgu}(\theta_1, \theta_2) \sim_{U \cup V} \mathrm{mgu}(\theta'_1, \theta_2)$. We have the following equalities:

$$\begin{aligned}
&\mathrm{mgu}(\theta_1, \theta_2) \\
\sim_{U\cup V}\ &\mathrm{mgu}(\theta_1, \theta_2)_{|U\cup V} \\
=\ &\mathrm{mgu}(\theta_{1|U}, \theta_2, \theta_{1|-U})_{|U\cup V} \\
=\ &(\mathrm{mgu}(\theta_{1|U}, \theta_2) \circ \theta_{1|-U})_{|U\cup V} \\
=\ &\mathrm{mgu}(\theta_{1|U}, \theta_2)_{|U\cup V} \ .
\end{aligned}$$

In the last step, we use the fact that $\mathrm{dom}(\theta_{1|-U})$ is disjoint from $\mathrm{vars}(\theta_{1|U})$ by idempotency of $\theta_1$ and it is disjoint from $\mathrm{vars}(\theta_2)$ by the assumptions $(U \cup \mathrm{vars}(\theta_1)) \cap (V \cup \mathrm{vars}(\theta_2)) \subseteq U \cap V$. Since $\theta_1 \sim_U \theta'_1$, there exists $\rho \in \mathit{Ren}$ such that $(\rho \circ \theta'_1)_{|U} = \theta_{1|U}$. The restriction of $\rho$ to $\mathrm{vars}(\theta'_{1|U})$ is an injective map of variables whose range is $\mathrm{vars}(\theta_{1|U})$. By applying Lemma 3.4, it follows that we may choose a $\rho$ such that $\mathrm{vars}(\rho) \subseteq \theta_1(U) \cup \theta'_1(U) \subseteq \mathrm{vars}(\theta_1) \cup \mathrm{vars}(\theta'_1) \cup \mathrm{vars}(U)$. Then $\mathrm{vars}(\rho) \cap V \subseteq U$. We have:

$$\begin{aligned}
&\mathrm{mgu}(\theta_{1|U}, \theta_2)_{|U\cup V} \\
=\ &\mathrm{mgu}((\rho \circ \theta'_1)_{|U}, \theta_2)_{|U\cup V} \\
=\ &(\mathrm{mgu}((\rho \circ \theta'_1)_{|U}, \theta_2) \circ \theta')_{|U\cup V} && \text{[for each } \theta' \text{ s.t. } \mathrm{dom}(\theta') \cap (U \cup V) = \emptyset] \\
=\ &\mathrm{mgu}((\rho \circ \theta'_1)_{|U}, \theta_2, (\rho \circ \theta'_1)_{|-U})_{|U\cup V} && \text{[by choosing } \theta' = (\rho \circ \theta'_1)_{|-U}] \\
=\ &\mathrm{mgu}(\rho \circ \theta'_1, \theta_2)_{|U\cup V} \\
=\ &(\rho' \circ \mathrm{mgu}(\theta'_1, \theta_2))_{|U\cup V} && \text{[by (Palamidessi 1990, Theorem 5.10)]} \\
\sim_{U\cup V}\ &\mathrm{mgu}(\theta'_1, \theta_2)_{|U\cup V} && \text{[by Prop. 3.6]} \\
\sim_{U\cup V}\ &\mathrm{mgu}(\theta'_1, \theta_2) \ .
\end{aligned}$$

which proves the required property. Now, to prove the general theorem, assume there are $\theta_1 \sim_U \theta'_1$, $\theta_2 \sim_V \theta'_2$ with $(U \cup \mathrm{vars}(\theta_1)) \cap (V \cup \mathrm{vars}(\theta_2)) \subseteq U \cap V$ and $(U \cup \mathrm{vars}(\theta'_1)) \cap (V \cup \mathrm{vars}(\theta'_2)) \subseteq U \cap V$. Then, consider a new substitution $\theta''_1 \sim_U \theta'_1$ such that $(U \cup \mathrm{vars}(\theta''_1)) \cap (V \cup \mathrm{vars}(\theta_2)) \subseteq U \cap V$, $(U \cup \mathrm{vars}(\theta''_1)) \cap (V \cup \mathrm{vars}(\theta'_2)) \subseteq U \cap V$ and we repeatedly apply the previous property, obtaining

$$\mathrm{mgu}(\theta_1, \theta_2) \sim_{U\cup V} \mathrm{mgu}(\theta''_1, \theta_2) \sim_{U\cup V} \mathrm{mgu}(\theta''_1, \theta'_2) \sim_{U\cup V} \mathrm{mgu}(\theta'_1, \theta'_2) \ . \quad \square$$

Note that, in the proof, the condition $(U \cup \mathrm{vars}(\theta'_1)) \cap (V \cup \mathrm{vars}(\theta'_2)) \subseteq U \cap V$ implies that $\mathrm{vars}(\theta'_1) \cap V \subseteq U \cap V$ and $\mathrm{vars}(\theta'_2) \cap U \subseteq U \cap V$. If we relax the condition to $\mathrm{vars}(\theta'_1) \cap \mathrm{vars}(\theta'_2) \subseteq U \cap V$ then this property no longer holds and mgu ceases to be well defined. This is actually the origin of the flaw in (Cortesi and Filé 1999) which we will examine in Section 8.2.

*Example 3.14*
Consider $\theta_1 = \{x/a\}$ and $\theta_2 = \{u/b\}$. Assume we have a relaxed definition of mgu as stated above. Then, to compute $\mathrm{mgu}([\theta_1]_x, [\theta_2]_{u,v})$ we may choose $\theta'_1 = \theta_1$ and $\theta'_2 = \theta_2$ to obtain $\{x/a, u/b\}$. But with the relaxed condition we might also choose $\theta'_1 = \{x/a, v/a\}$ and $\theta'_2 = \theta_2$ since it is true that $\mathrm{vars}(\theta'_1) \cap \mathrm{vars}(\theta'_2) = \emptyset$. However $\mathrm{mgu}(\theta'_1, \theta'_2) = \{x/a, v/a, u/b\} \not\sim_{x,u,v} \{x/a, u/b\}$. □

*Theorem 3.15*
mgu is the greatest lower bound of $(\mathit{ISubst}_\sim, \preceq)$.

*Proof*
If $[\theta]_{U\cup V} = \mathrm{mgu}([\theta_1]_U, [\theta_2]_V)$, we may assume, without loss of generality, that $\theta = \mathrm{mgu}(\theta_1, \theta_2)$ and $\theta_1$, $\theta_2$ are canonical representatives. It immediately follows that $\theta \leq \theta_1$ and therefore $\theta \preceq_U \theta_1$. In the same way, $\theta \preceq_V \theta_2$.

Now, assume $[\eta]_{U\cup V} \preceq [\theta_1]_U$ and $[\eta]_{U\cup V} \preceq [\theta_2]_V$. We want to prove that $[\eta]_{U\cup V} \preceq [\theta]_{U\cup V}$. By definition of $\preceq$, there is a $\sigma_1$ such that $\eta(x) = \sigma_1(\theta_1(x))$ for each $x \in U$. We may choose $\sigma_1$ such that $\mathrm{dom}(\sigma_1) \subseteq \mathrm{rng}(\theta_1)$. In the same way, there is $\sigma_2$ such that $\mathrm{dom}(\sigma_2) \subseteq \mathrm{rng}(\theta_2(x))$ and $\eta(x) = \sigma_2(\theta_2(x))$ for each $x \in V$. We may define a new substitution $\sigma$ such that

$$\sigma(x) = \begin{cases} \sigma_1(\theta_1(x)) & \text{if } x \in U \cup \mathrm{dom}(\sigma_1),\\ \sigma_2(\theta_2(x)) & \text{if } x \in V \cup \mathrm{dom}(\sigma_2),\\ x & \text{otherwise.} \end{cases}$$

Note that this definition is correct, since the first two cases may occur simultaneously only if $x \in U \cap V$, which implies $\sigma_1(\theta_1(x)) = \sigma_2(\theta_2(x)) = \eta(x)$. It is easy to check that $\eta \sim_{U\cup V} \sigma$ and $\sigma = \sigma \circ \theta_1 = \sigma \circ \theta_2$. Therefore

$$\eta \sim_{U\cap V} \sigma \leq \mathrm{mgu}(\theta_1, \theta_2) = \theta \ ,$$

i.e., $\eta \preceq_{U\cup V} \theta$, which proves the thesis. □

We now give some properties which relate the mgu with the other operations on $\mathit{ISubst}_\sim$, namely renaming and projection.

*Proposition 3.16*
$\rho$ is a congruence w.r.t. unification. In formulas, if $E$ is a set of equations and $[\theta_1]_{U_1}, [\theta_2]_{U_2} \in \mathit{ISubst}_\sim$ then it holds that:

- $\mathrm{mgu}(\rho(E)) = \rho(\mathrm{mgu}(E))$
- $\rho(\mathrm{mgu}([\theta_1]_{U_1}, [\theta_2]_{U_2})) = \mathrm{mgu}(\rho([\theta_1]_{U_1}), \rho([\theta_2]_{U_2}))$ .

*Proof*
The first property is trivial since the unification algorithm does not depend on the actual name of variables. Therefore, to prove the second property, we only need to check that $\mathrm{mgu}([\theta_1]_{U_1}, [\theta_2]_{U_2}) = [\mathrm{mgu}(\theta'_1, \theta'_2)]_{U_1\cup U_2}$ (according to Eq. 8) implies $\mathrm{mgu}(\rho([\theta_1]_{U_1}), \rho([\theta_2]_{U_2})) = [\mathrm{mgu}(\rho(\theta'_1), \rho(\theta'_2))]_{\rho(U_1)\cup\rho(U_2)}$. First of all, since $\theta'_1 \sim_{U_1} \theta_1$, then $\rho(\theta'_1) \sim_{\rho(U_1)} \rho(\theta_1)$, by Theorem 3.9. With the same reasoning, we obtain that $\rho(\theta'_2) \sim_{\rho(U_2)} \rho(\theta_2)$. Then, we prove that $(\rho(U_1) \cup \mathrm{vars}(\rho(\theta'_1))) \cap (\rho(U_2) \cup \mathrm{vars}(\rho(\theta'_2))) \subseteq \rho(U_1) \cap \rho(U_2)$. It is obvious that $\rho(\mathrm{vars}(\theta)) = \mathrm{vars}(\rho(\theta))$. Therefore, since $\rho$ is bijective,

$$\begin{aligned}&(\rho(U_1) \cup \mathrm{vars}(\rho(\theta'_1))) \cap (\rho(U_2) \cup \mathrm{vars}(\rho(\theta'_2)))\\ &\qquad = \rho((U_1 \cup \mathrm{vars}(\theta'_1)) \cap (U_2 \cup \mathrm{vars}(\theta'_2))) \subseteq \rho(U_1 \cap U_2) = \rho(U_1) \cap \rho(U_2) \ .\end{aligned}$$ □

*Proposition 3.17*
Given a set of variables $V$ and $[\theta_1]_{U_1}, [\theta_2]_{U_2} \in \mathit{ISubst}_\sim$, we have that

$$\pi_V(\mathrm{mgu}(\pi_V([\theta_1]_{U_1}), [\theta_2]_{U_2})) = \mathrm{mgu}(\pi_V([\theta_1]_{U_1}), \pi_V([\theta_2]_{U_2})) \ .$$

*Proof*
First observe that $\pi_V(\mathrm{mgu}(\pi_V([\theta_1]_{U_1}), [\theta_2]_{U_2}))) = [\theta]_{V\cap((V\cap U_1)\cup U_2)} = [\theta]_{V\cap(U_1\cup U_2)}$ where $\theta \in \mathrm{mgu}(\theta'_1, \theta'_2)$, $\theta'_1$ and $\theta'_2$ are canonical representatives of $[\theta_1]_{V\cap U_1}$ and $[\theta_2]_{U_2}$ and $\mathrm{vars}(\theta'_1)\cap\mathrm{vars}(\theta'_2) \subseteq V\cap U_1\cap U_2$. Note that $\theta'_2 \sim_{U_2} \theta_2$ and therefore $\theta'_2 \sim_{V\cap U_2} \theta_2$. Moreover $(\mathrm{vars}(\theta'_1)\cup(V\cap U_1))\cap(\mathrm{vars}(\theta'_2)\cup(V\cap U_2)) \subseteq V\cap U_1\cap U_2$, and therefore $\theta'_1$ and $\theta'_2$ are valid representatives to compute $\mathrm{mgu}(\pi_V([\theta_1]_{U_1}), \pi_V([\theta_2]_{U_2}))$ according to (8). Therefore $[\theta]_{V\cap(U_1\cup U_2)} = \mathrm{mgu}(\pi_V([\theta_1]_{U_1}), \pi_V([\theta_2]_{U_2}))$ and this proves the thesis. □

Thanks to the above properties, the algebraic structure of the domain $\mathit{ISubst}_\sim$ is very similar to (locally finite) cylindric algebras (Henkin et al. 1971). In particular, if the unit element is defined as $[\epsilon]_\emptyset$, the diagonal elements are given by the substitutions $[x/y]_{\{x,y\}}$ and cylindrification is defined as $c_x([\theta]_V) = \pi_{V\setminus\{x\}}([\theta]_V)$, then these operators satisfy the axioms defining a cylindric algebra. The fundamental difference is that the underlying set $\mathit{ISubst}_\sim$ is not a boolean algebra.

It would be possible, as in (Palamidessi 1990), to define a "least common anti-instance" operator which corresponds to the least upper bound in $\mathit{ISubst}_\sim$. However, since it is not used in the semantic framework we are going to describe, we omit to define this operator.

## 4 Concrete Semantics

Since we are interested in goal-dependent analysis of logic programs, we need a goal-dependent semantics which is well suited for static analysis, i.e., a collecting semantics over computed answers. Unfortunately, using a collecting goal-dependent semantics may lead to a loss of precision already at the concrete level, as shown by Marriott et al. (1994). It is possible to reduce the impact of this problem by using two different operators for forward and backward unification. In particular, it turns out that backward unification may be realized using the operation of matching between substitutions (Bruynooghe 1991; Le Charlier et al. 1991). We follow the same approach and define a new denotational framework based on existential substitutions and inspired by (Cortesi et al. 1994).

### *4.1 Concrete Domain*

We start to define the concrete domain for the semantics. A concrete object is essentially a set of existential substitutions with a fixed set of variables of interest. In formulas:

$$\mathtt{Psub} = \{[\Theta, U] \mid \Theta \subseteq \mathit{ISubst}_{\sim_U}, U \in \wp_f(\mathcal{V})\} \cup \{\bot_{\mathrm{Ps}}, \top_{\mathrm{Ps}}\}$$

where $\top_{\mathrm{Ps}}$ and $\bot_{\mathrm{Ps}}$ are the top and bottom elements respectively and

$$[\Theta_1, U_1] \sqsubseteq_{\mathrm{Ps}} [\Theta_2, U_2] \iff U_1 = U_2 \text{ and } \Theta_1 \subseteq \Theta_2 \ .$$

The notation we adopt may appear clumsy, since the set of variables of interest $U$ in $[\Theta, U]$ may be derived from $\Theta$. However, when we move to the abstract domain, we need to explicitly keep track of this set $U$. By using $[\Theta, U]$ in $\mathtt{Psub}$, we want to keep a consistent notation for both concrete and abstract domains.

It turns out that $(\mathtt{Psub}, \sqsubseteq_{\mathrm{Ps}})$ is a complete lattice, and we denote by $\sqcup_{\mathrm{Ps}}$ its least upper bound, which is given by

$$\begin{aligned} \top_{\mathrm{Ps}} \sqcup_{\mathrm{Ps}} \chi = \chi \sqcup_{\mathrm{Ps}} \top_{\mathrm{Ps}} &= \top_{\mathrm{Ps}} \\ \bot_{\mathrm{Ps}} \sqcup_{\mathrm{Ps}} \chi = \chi \sqcup_{\mathrm{Ps}} \bot_{\mathrm{Ps}} &= \chi \\ [\Theta_1, U_1] \sqcup_{\mathrm{Ps}} [\Theta_2, U_2] &= \begin{cases} [\Theta_1 \cup \Theta_2, U_1] & \text{if } U_1 = U_2, \\ \top_{\mathrm{Ps}} & \text{otherwise.} \end{cases} \end{aligned} \tag{9}$$

We now define the main operations over $\mathtt{Psub}$, that is: projection on a set of variables, unification of an object with a single substitution and the operation for matching two objects of $\mathtt{Psub}$. All the operations are strict: when one of the argument is $\bot_{\mathrm{Ps}}$ the result is $\bot_{\mathrm{Ps}}$. If no argument is $\bot_{\mathrm{Ps}}$ and at least one of the argument is $\top_{\mathrm{Ps}}$ the result is $\top_{\mathrm{Ps}}$. Therefore, in the following, we will omit the cases for the objects $\bot_{\mathrm{Ps}}$ and $\top_{\mathrm{Ps}}$.

Given $[\Theta, U] \in \mathtt{Psub}$ and $V \subseteq \mathcal{V}$, we define the projection of $[\Theta, U]$ on the set of variables $V$ as

$$\pi_{\mathrm{Ps}}([\Theta, U], V) = [\{\pi_V([\sigma]_U) \mid [\sigma]_U \in \Theta\}, U \cap V] \ . \tag{10}$$

The concrete unification $\mathsf{unif}_{\mathrm{Ps}} : \mathtt{Psub} \times \mathit{ISubst} \to \mathtt{Psub}$ is given by:

$$\mathsf{unif}_{\mathrm{Ps}}([\Theta, U], \delta) = [\{\mathrm{mgu}([\sigma]_U, [\delta]_{\mathrm{vars}(\delta)}) \mid [\sigma]_U \in \Theta\}, U \cup \mathrm{vars}(\delta)]. \tag{11}$$

The operations $\pi_{\mathrm{Ps}}$ and $\mathsf{unif}_{\mathrm{Ps}}$ are just the pointwise extensions of $\pi$ and mgu. Note that, in $\mathsf{unif}_{\mathrm{Ps}}$, the argument $\delta$ may have variables which do not appear in $U$. This is not always the case in literature. For example, in (Cortesi and Filé 1999; Bagnara et al. 2005) we find a variant of $\mathsf{unif}_{\mathrm{Ps}}$ which only consider the case when $\mathrm{vars}(\delta) \subseteq U$. When this does not happen, the same effect is obtained by first enlarging the set of variables of interest $U$, and then applying unification. Although nothing changes at the concrete level, this gives a loss of precision when we move to the abstract side, since the composition of two optimal abstract operators is generally less precise than the optimal abstract counterpart of the whole $\mathsf{unif}_{\mathrm{Ps}}$ (see Section 6).

Finally, we define the matching operation. The idea is to design an operator which performs unification between two substitutions $[\theta_1]_{U_1}$ and $[\theta_2]_{U_2}$ only if the process of unification does not instantiate the first substitution. In other words, we require that if we compute $\mathrm{mgu}([\theta_1]_{U_1}, [\theta_2]_{U_2})$ and we only observe variables in $U_1$, that is $\pi_{U_1}(\mathrm{mgu}([\theta_1]_{U_1}, [\theta_2]_{U_2}))$, then we obtain exactly $[\theta_1]_{U_1}$. The next proposition shows this is equivalent to require that $\theta_1 \preceq_{U_1 \cap U_2} \theta_2$.

*Proposition 4.1*

Given two existential substitutions $[\theta_1]_{U_1}$ and $[\theta_2]_{U_2}$, we have that $\theta_1 \preceq_{U_1 \cap U_2} \theta_2$ iff $[\theta_1]_{U_1} = \pi_{U_1}(\mathrm{mgu}([\theta_1]_{U_1}, [\theta_2]_{U_2}))$.

*Proof*
By Prop. 3.17 we obtain $\pi_{U_1}(\mathrm{mgu}([\theta_1]_{U_1}, [\theta_2]_{U_2})) = \mathrm{mgu}(\pi_{U_1}([\theta_1]_{U_1}), \pi_{U_1}([\theta_2]_{U_2})) = \mathrm{mgu}([\theta_1]_{U_1}, [\theta_2]_{U_1 \cap U_2})$. Since mgu is the greatest lower bound of $\mathit{ISubst}_\sim$, we have that $[\theta_1]_{U_1} = \mathrm{mgu}([\theta_1]_{U_1}, [\theta_2]_{U_1 \cap U_2})$ iff $[\theta_1]_{U_1} \preceq [\theta_2]_{U_1 \cap U_2}$ which, by definition, is equivalent to $\theta_1 \preceq_{U_1 \cap U_2} \theta_2$. □

We can now define the matching operator $\mathsf{match}_{\mathrm{Ps}} : \mathtt{Psub} \times \mathtt{Psub} \to \mathtt{Psub}$ as follows:

$$\mathsf{match}_{\mathrm{Ps}}([\Theta_1, U_1], [\Theta_2, U_2]) = [\{\mathrm{mgu}([\theta_1]_{U_1}, [\theta_2]_{U_2}) \mid \\ \theta_1 \preceq_{U_1 \cap U_2} \theta_2, [\theta_1]_{U_1} \in \Theta_1, [\theta_2]_{U_2} \in \Theta_2\}, U_1 \cup U_2] \ . \tag{12}$$

The above operator allows us to unify all the pairs of substitutions $[\theta_1]_{U_1} \in \Theta_1$ and $[\theta_2]_{U_2} \in \Theta_2$, under the condition that the common variables in $U_1$ and $U_2$ may not be further instantiated w.r.t. their values in $\theta_1$.

*Example 4.2*
Let $\Theta_1 = \{[x/y]_{x,y}\}$ and $\Theta_2 = \{[u/x]_{u,x}, [x/t(u)]_{u,x}\}$. Then

$$\mathsf{match}_{\mathrm{Ps}}([\Theta_1, \{x, y\}], [\Theta_2, \{u, x\}]) = [\{[x/y, u/y]_{x,y,u}\}, \{x, y, u\}] \ .$$

Note that $[y/t(u), x/t(u)]_{u,x,y}$, obtained by unifying $[x/y]_{x,y}$ with $[x/t(u)]_{u,x}$, is not in the result of matching. This is because $[x/t(u)]_{u,x}$ is strictly more instantiated then $[x/y]_{x,y}$ w.r.t. the variable $x$ and therefore $\{x/y\} \not\preceq_x \{x/t(u)\}$. □

*Proposition 4.3*
The operations $\pi_{\mathrm{Ps}}$, $\mathsf{unif}_{\mathrm{Ps}}$ and $\mathsf{match}_{\mathrm{Ps}}$ are continuous over $\mathtt{Psub}$.

*Proof*
Trivial from their definitions. If we do not consider the element $\top_{\mathrm{Ps}}$, they are actually additive. □

### *4.2 Semantics*

Using the operators defined so far, we introduce a denotational semantics for logic programs. It computes, for a given goal $G$, the set of computed answers for a program w.r.t. $G$ modulo the equivalence relation $\sim_{\mathrm{vars}(G)}$. It is a goal-dependent *collecting semantics* (Cousot and Cousot 1994), in that it works by computing the set of possibly entry and exit substitutions at each point in the program.

We call *denotation* an element in the set of continuous maps:

$$\mathcal{D}en = \mathsf{Atoms} \to \mathtt{Psub} \xrightarrow{c} \mathtt{Psub} \ . \tag{13}$$

We have the following semantic functions:

$$\mathcal{P} : \mathsf{Progs} \to \mathcal{D}en$$
$$\mathcal{C} : \mathsf{Clauses} \to \mathcal{D}en \xrightarrow{c} \mathcal{D}en$$
$$\mathcal{B} : \mathsf{Bodies} \to \mathcal{D}en \xrightarrow{c} \mathtt{Psub} \xrightarrow{c} \mathtt{Psub} \ .$$

The corresponding definitions[4], given $d \in \mathit{Den}$ and $x \in \mathtt{Psub}$, are:

$$\begin{aligned}
&\mathcal{P}[\![P]\!] = \mathit{lfp}\ \lambda d. \left( \bigsqcup_{cl \in P}{}_{\mathrm{Ps}} \mathcal{C}[\![cl]\!] d \right)\\
&\mathcal{C}[\![H \leftarrow B]\!]\ d\ A\ \chi = \mathbf{U}^b_{\mathrm{Ps}}((\mathcal{B}[\![B]\!] d \mathbf{U}^f_{\mathrm{Ps}}(\chi, A, H)), \chi, H, A)\\
&\mathcal{B}[\![\Box]\!]\ d\ \chi = \chi\\
&\mathcal{B}[\![A, B]\!]\ d\ \chi = \mathcal{B}[\![B]\!] d (d A \chi)
\end{aligned}$$

defined by means of the following operators:

$$\begin{aligned}
&\mathbf{U}^f_{\mathrm{Ps}} : \mathtt{Psub} \times \mathsf{Atoms} \times \mathsf{Atoms} \to \mathtt{Psub}\ ,\\
&\mathbf{U}^b_{\mathrm{Ps}} : \mathtt{Psub} \times \mathtt{Psub} \times \mathsf{Atoms} \times \mathsf{Atoms} \to \mathtt{Psub}\ .
\end{aligned}$$

$\mathbf{U}^f_{\mathrm{Ps}}$ and $\mathbf{U}^b_{\mathrm{Ps}}$ are respectively the *forward* and *backward* unification (Muthukumar and Hermenegildo 1992). They are used according to the following pattern:

- the *forward unification*, in order to compute the set of *entry substitutions* $\mathbf{U}^f_{\mathrm{Ps}}(\chi, A, H)$ from the set of *call substitutions* $\chi$;
- the *backward unification*, in order to compute the set of *answer substitutions* $\mathbf{U}^b_{\mathrm{Ps}}((\mathcal{B}[\![B]\!] d \mathbf{U}^f_{\mathrm{Ps}}(\chi, A, H)), \chi, H, A)$ starting from the set of *exit substitutions* $\mathcal{B}[\![B]\!] d \mathbf{U}^f_{\mathrm{Ps}}(\chi, A, H)$.

The formal definitions of $\mathbf{U}^f_{\mathrm{Ps}}$ and $\mathbf{U}^b_{\mathrm{Ps}}$ are the following:

$$\mathbf{U}^f_{\mathrm{Ps}}([\Theta, U], A_1, A_2) = \pi_{\mathrm{Ps}}(\mathsf{unif}_{\mathrm{Ps}}(\rho([\Theta, U]), \mathrm{mgu}(\rho(A_1) = A_2)), \mathrm{vars}(A_2))\ , \tag{14}$$

where $\rho$ is a renaming such that $\rho(U \cup \mathrm{vars}(A_1)) \cap \mathrm{vars}(A_2) = \emptyset$ and $\rho([\Theta, U]) = [\{\rho([\sigma]_U) \mid [\sigma]_U \in \Theta\}, \rho(U)]$ is the obvious lifting of renamings from $\mathit{ISubst}_\sim$ to $\mathtt{Psub}$.

$$\begin{aligned}
&\mathbf{U}^b_{\mathrm{Ps}}([\Theta_1, U_1], [\Theta_2, U_2], A_1, A_2) =\\
&\qquad \pi_{\mathrm{Ps}}(\mathsf{match}_{\mathrm{Ps}}(\rho([\Theta_1, U_1]), \mathsf{unif}_{\mathrm{Ps}}([\Theta_2, U_2], \mathrm{mgu}(\rho(A_1) = A_2))), U_2 \cup \mathrm{vars}(A_2))
\end{aligned} \tag{15}$$

where $\rho$ is a renaming such that $\rho(U_1 \cup \mathrm{vars}(A_1)) \cap (U_2 \cup \mathrm{vars}(A_2)) = \emptyset$. If $\rho(A_1)$ and $A_2$ do not unify, the results for both the operations is assumed to be $\bot_{\mathrm{Ps}}$.

*Example 4.4*
Consider the goal $p(x, y, z)$ with $y = f(x, z)$ and the trivial program $P$ with just one clause

```
p(u,v,w).
```

We first compute the concrete semantics $\mathcal{P}[\![P]\!] = \mathit{lfp}\ \lambda d. \mathcal{C}[\![\mathtt{p(u,v,w)} \leftarrow \Box]\!] d$. According to the semantic definition, we have that:

$$\mathcal{C}[\![\mathtt{p(u,v,w)} \leftarrow \Box]\!] d = \lambda A. \lambda \chi. \mathbf{U}^b_{\mathrm{Ps}}((\mathcal{B}[\![\Box]\!] d \mathbf{U}^f_{\mathrm{Ps}}(\chi, A, p(u, v, w))), \chi, p(u, v, w), A)\ .$$

[4] Here we use the lambda notation, writing $\mathit{lfp}\ \lambda x. E(x)$ to denote the least fixed point of the function $f$ given by $f(x) = E(x)$.

Since $\mathcal{B}[\![\Box]\!]d = \lambda\chi.\chi$, this is equivalent to

$$\lambda A.\lambda\chi.\mathbf{U}^b_{\mathrm{Ps}}(\mathbf{U}^f_{\mathrm{Ps}}(\chi, A, p(u,v,w)), \chi, p(u,v,w), A) \enspace ,$$

from which we immediately obtain the semantics of the program $P$:

$$\mathcal{P}[\![P]\!] = \lambda A.\lambda\chi.\mathbf{U}^b_{\mathrm{Ps}}(\mathbf{U}^f_{\mathrm{Ps}}(\chi, A, p(u,v,w)), \chi, p(u,v,w), A) \enspace .$$

We now compute the semantics of the goal $p(x,y,z)$ with $y = f(x,z)$. In order to improve readability, we will omit subscripts on classes of substitutions.

$$\begin{aligned}
&\mathcal{P}[\![P]\!]p(x,y,z)[\{[y/f(x,z)]\}, \{x,y,z\}] = \\
&\qquad \mathbf{U}^b_{\mathrm{Ps}}(\mathbf{U}^f_{\mathrm{Ps}}([\{[y/f(x,z)]\}, \{x,y,z\}], p(x,y,z), p(u,v,w)), \\
&\qquad\quad [\{[y/f(x,z)]\}, \{x,y,z\}], p(u,v,w), p(x,y,z)) \enspace .
\end{aligned}$$

We first compute the forward unification

$$\begin{aligned}
&\mathbf{U}^f_{\mathrm{Ps}}([\{[y/f(x,z)]\}, \{x,y,z\}], p(x,y,z), p(u,v,w)) = \\
&\qquad\qquad [\{[u/x', v/f(x',z'), w/z']\}, \{u,v,w\}] \enspace ,
\end{aligned}$$

where we have renamed $x$ and $z$ to $x'$ and $z'$ to avoid ambiguities, although it is not needed. Now we can compute the semantics of the goal.

$$\begin{aligned}
&\mathcal{P}[\![P]\!]p(x,y,z)[\{[y/f(x,z)]\}, \{x,y,z\}] \\
&= \mathbf{U}^b_{\mathrm{Ps}}([\{[u/x', v/f(x',z'), w/z']\}, \{u,v,w\}], [\{[y/f(x,z)]\}, \{x,y,z\}], \\
&\qquad p(u,v,w), p(x,y,z)) \\
&= \pi_{\mathrm{Ps}}(\mathsf{match}_{\mathrm{Ps}}([\{[u/x', v/f(x',z'), w/z']\}, \{u,v,w\}], \\
&\qquad [\{[u/x, v/f(x,z), w/z, y/f(x,z)]\}, \{u,v,w,x,y,z\}]), \{x,y,z\}) \\
&= \pi_{\mathrm{Ps}}([\{[u/x, v/f(x,z), w/z, y/f(x,z)]\}, \{u,v,w,x,y,z\}], \{x,y,z\}) \\
&= [\{[y/f(x,z)]\}, \{x,y,z\}]
\end{aligned}$$

Thus, we have only one computed answer substitution for the goal $p(x,y,z)$ with $y = f(x,z)$, which is $\{y/f(x,z)\}$. □

*Theorem 4.5*
$\mathbf{U}^f_{\mathrm{Ps}}$ and $\mathbf{U}^b_{\mathrm{Ps}}$ are well defined, in that they are independent from the choice of $\rho$. Moreover, they are continuous.

*Proof*
Continuity is trivial from their definition, therefore we only need to prove the independence from the choice of the renaming $\rho$. We only consider the case when none of the arguments are $\bot_{\mathrm{Ps}}$ or $\top_{\mathrm{Ps}}$, since otherwise the result is always $\bot_{\mathrm{Ps}}$ or $\top_{\mathrm{Ps}}$. Moreover, note that, given atoms $A_1$ and $A_2$, if $\rho_1$ and $\rho_2$ are renamings such that $\rho_i(\mathrm{vars}(A_1)) \cap \mathrm{vars}(A_2) = \emptyset$ for $i \in \{1,2\}$, then $\rho_1(A_1)$ and $A_2$ unify iff $\rho_2(A_1)$ and $A_2$ unify. Therefore, we can restrict ourselves to the case where the two atoms given as arguments, appropriately renamed, do unify. Otherwise, the result is always $\bot_{\mathrm{Ps}}$.

Observe that, by Prop. 3.16, given $\rho \in \mathit{Ren}$, $[\theta_1]_{U_1}, [\theta_2]_{U_2} \in \mathit{ISubst}_\sim$, we have that $\rho(\mathrm{mgu}([\theta_1]_{U_1}, [\theta_2]_{U_2})) = \mathrm{mgu}(\rho([\theta_1]_{U_1}), \rho([\theta_2]_{U_2}))$. By definition of $\mathsf{unif}_{\mathrm{Ps}}$, it follows that $\rho(\mathsf{unif}_{\mathrm{Ps}}([\Theta, U], \delta)) = \mathsf{unif}_{\mathrm{Ps}}(\rho([\Theta, U]), \rho(\delta))$, since $\mathrm{vars}(\rho(\delta)) = \rho(\mathrm{vars}(\delta))$.

Let $\rho_1, \rho_2$ be renamings. We first show that

$$\pi_{\mathrm{Ps}}(\mathsf{unif}_{\mathrm{Ps}}(\rho_1([\Theta, U]), \mathrm{mgu}(\rho_1(A_1) = A_2)), \mathrm{vars}(A_2)) = \\ \pi_{\mathrm{Ps}}(\mathsf{unif}_{\mathrm{Ps}}(\rho_2([\Theta, U]), \mathrm{mgu}(\rho_2(A_1) = A_2)), \mathrm{vars}(A_2))$$

provided that $\rho_i(U \cup \mathrm{vars}(A_1)) \cap \mathrm{vars}(A_2) = \emptyset$, for $i \in \{1, 2\}$. Let $W = \rho_1(U \cup \mathrm{vars}(A_1))$ and $\delta = (\rho_2 \circ \rho_1^{-1})_{|W}$. Then $\delta$ may be viewed as an injective map from $V$ to $\mathcal{V}$, since it is the composition of injective functions. By Lemma 3.4 there exists a renaming $\rho$ such that $\rho_{|W} = \delta$ and $\mathrm{vars}(\rho) = \mathrm{vars}(\delta) \subseteq W \cup \mathrm{rng}(\delta) \subseteq W \cup \rho_2(U \cup \mathrm{vars}(A_1))$. Observe that $\mathrm{vars}(\rho) \cap \mathrm{vars}(A_2) = \emptyset$ since, by hypothesis, for each $i \in \{1, 2\}$ it is the case that $\rho_i(U \cup \mathrm{vars}(A_1)) \cap \mathrm{vars}(A_2) = \emptyset$. Thus the following equivalences hold:

$$\begin{aligned}
& \pi_{\mathrm{Ps}}(\mathsf{unif}_{\mathrm{Ps}}(\rho_1([\Theta, U]), \mathrm{mgu}(\rho_1(A_1) = A_2)), \mathrm{vars}(A_2)) \\
= \;& \rho(\pi_{\mathrm{Ps}}(\mathsf{unif}_{\mathrm{Ps}}(\rho_1([\Theta, U]), \mathrm{mgu}(\rho_1(A_1) = A_2)), \mathrm{vars}(A_2))) \\
& \quad [\text{since } \rho_{|\mathrm{vars}(A_2)} = id \text{ and by Prop. 3.11}] \\
= \;& \pi_{\mathrm{Ps}}(\rho(\mathsf{unif}_{\mathrm{Ps}}(\rho_1([\Theta, U]), \mathrm{mgu}(\rho_1(A_1) = A_2))), \mathrm{vars}(A_2)) \\
& \quad [\text{since } \rho \text{ is a congruence for } \pi_{\mathrm{Ps}} \text{ by Prop. 3.10}] \\
= \;& \pi_{\mathrm{Ps}}(\mathsf{unif}_{\mathrm{Ps}}(\rho(\rho_1([\Theta, U])), \mathrm{mgu}(\rho(\rho_1(A_1)) = \rho(A_2))), \mathrm{vars}(A_2)) \\
& \quad [\text{since } \rho \text{ is a congruence for } \mathsf{unif}_{\mathrm{Ps}} \text{ by Prop. 3.16}] \\
= \;& \pi_{\mathrm{Ps}}(\mathsf{unif}_{\mathrm{Ps}}(\rho_2([\Theta, U]), \mathrm{mgu}(\rho_2(A_1)) = A_2), \mathrm{vars}(A_2)) \\
& \quad [\text{since } (\rho \circ \rho_1)_{|U \cup \mathrm{vars}(A_1)} = \rho_{2|U \cup \mathrm{vars}(A_1)} \text{ and by Prop. 3.11}] \;.
\end{aligned}$$

We now show that $\mathbf{U}^b_{\mathrm{Ps}}$ is independent from the choice of the renaming. First of all, note that by Prop. 3.16 and Theorem 3.9 the following follows:

$$\rho(\mathsf{match}_{\mathrm{Ps}}([\Theta_1, U_1], [\Theta_2, U_2])) = \mathsf{match}_{\mathrm{Ps}}(\rho([\Theta_1, U_1]), \rho([\Theta_2, U_2])) \;.$$

Assume given $\rho_1, \rho_2 \in \mathit{Ren}$ such that $\rho_i(U_1 \cup \mathrm{vars}(A_1)) \cap (U_2 \cup \mathrm{vars}(A_2)) = \emptyset$, for $i \in \{1, 2\}$. Let $W = \rho_1(U_1 \cup \mathrm{vars}(A_1))$ and $\delta = (\rho_2 \circ \rho_1^{-1})_{|W}$. As shown above, there exists $\rho \in \mathit{Ren}$ such that $\rho_{|W} = \delta$ and $\mathrm{vars}(\rho) = \mathrm{vars}(\delta) \subseteq W \cup \rho_2(U_1 \cup \mathrm{vars}(A_1))$. Observe that $\delta_{|U_2 \cup \mathrm{vars}(A_2)} = id$. Thus the following equivalences hold, where $Z = U_2 \cup \mathrm{vars}(A_2)$:

$$\begin{aligned}
& \pi_{\mathrm{Ps}}(\mathsf{match}_{\mathrm{Ps}}(\rho_1([\Theta_1, U_1]), \mathsf{unif}_{\mathrm{Ps}}([\Theta_2, U_2], \mathrm{mgu}(\rho_1(A_1) = A_2))), Z) \\
= \;& \rho(\pi_{\mathrm{Ps}}(\mathsf{match}_{\mathrm{Ps}}(\rho_1([\Theta_1, U_1]), \mathsf{unif}_{\mathrm{Ps}}([\Theta_2, U_2], \mathrm{mgu}(\rho_1(A_1) = A_2))), Z)) \\
= \;& \pi_{\mathrm{Ps}}(\mathsf{match}_{\mathrm{Ps}}(\rho(\rho_1([\Theta_1, U_1])), \mathsf{unif}_{\mathrm{Ps}}(\rho([\Theta_2, U_2]), \mathrm{mgu}(\rho(\rho_1(A_1)) = \rho(A_2)))), Z) \\
= \;& \pi_{\mathrm{Ps}}(\mathsf{match}_{\mathrm{Ps}}(\rho_2([\Theta_1, U_1]), \mathsf{unif}_{\mathrm{Ps}}([\Theta_2, U_2], \mathrm{mgu}(\rho_2(A_1) = A_2))), Z) \;.
\end{aligned}$$

This concludes the proof of the theorem. □

*Theorem 4.6*

All the semantic functions are well defined and continuous.

*Proof*

The proof is trivial since the semantic functions are obtained by composition, application, projection and tupling of continuous functions. Therefore, they are continuous and compute continuous denotations. Moreover, they do not depend on the choice of $\rho$ in $\mathbf{U}^f_{\mathrm{Ps}}$ and $\mathbf{U}^b_{\mathrm{Ps}}$, as proved in Theorem 4.5. □

Note that several frameworks have been developed for logic programs, and not all of them use the same operators for forward and backward unification. We will discuss the benefits of our choices later, when we introduce the abstract operators, since the relative merits of the different proposals mainly arise when speaking about abstractions.

### *4.3 Correctness and Completeness*

The semantics we have defined in this section is significant only up to the point that, studying its properties, it is possible to derive some conclusions about the properties of the real operational behavior of logic programs. We said before that we considered as the relevant operational observable of our analysis the set of classes of computed answers for a goal. Therefore, the best we can expect from our collecting semantics is that it enables us to recover the set of computed answer for each goal. Our first theorem is a partial positive answer to this question.

*Theorem 4.7*

(Semantic Correctness) Given a program $P$ and an goal $G$, if $\theta$ is a computed answer for the goal $G$, then $\mathcal{B}[\![G]\!](\mathcal{P}[\![P]\!])G[\{\epsilon\}, \mathrm{vars}(G)] \sqsupseteq_{\mathrm{Ps}} [\{[\theta]\}, \mathrm{vars}(G)]$.

*Proof*

The proof, quite long and tedious, may be found in the Appendix A. □

Therefore, we know that all the computed answers may be obtained by our semantics. However, the opposite is not true: the semantics given in this paper, although more precise than a semantics which only uses unification, is not complete w.r.t. computed answers. Actually, Marriott et al. (1994, Section 5.5) give an example where a collecting goal-dependent semantics computes a substitution which is not a computed answer. When matching is used to compute the backward unification, as it is the case in our framework, that example does not work anymore (see Example 7.3).

However, also with the use of matching, the collecting semantics computes substitutions which are not computed answers. Consider the program $P$ given by the following clauses:

```
p(x,y) :- q(x).
q(x).
```

We want to compute $\mathcal{P}[\![P]\!]p(x,y)[\Theta, \{x,y\}]$ where $\Theta = \{[x/y], [x/a]\}$. It is easy to check that

$$\mathcal{P}[\![P]\!]q(x)[\Delta, \{x\}] = [\Delta, \{x\}]$$

for each $[\Delta, \{x\}] \in$ `Psub`. Therefore, this implies that

$$\mathcal{P}[\![P]\!]p(x,y)[\Theta,\{x,y\}] = [\{[x/y],[x/a],[x/a,y/a]\},\{x,y\}] \ .$$

The substitution $[x/a, y/a]$ arises from calling $q(x)$ with the substitution $[x/a]$ and matching the result with $[x/y]$, which is not forbidden by matching. However, there is no substitution in the class of $[\{x/a, y/a\}]_{x,y}$ which is a computed answer for the goal $p(x, y)$ in the program $P$ with entry substitution in $\Theta$.

This loss of precision is not relevant for downward-closed abstract domains, where goal-dependent collecting semantics are more precise than goal-independent ones. This is not the case for upward-closed abstract domain, where goal-independent semantics are more precise than goal-dependent ones. García de la Banda et al. (1998) deal with this topic and show several semantics which combine a goal-dependent and a goal-independent computation to improve precision over all the conditions.

## 5 Abstract Domain and Semantics

Several abstract domains have been used for analyses of sharing and aliasing. We use the domain `Sharing` (Jacobs and Langen 1992; Cortesi and Filé 1999) which computes set-sharing information:

$$\texttt{Sharing} = \{[A,U] \mid A \subseteq \wp(U), (A \neq \emptyset \Rightarrow \emptyset \in A), U \in \wp_f(\mathcal{V})\} \cup \{\top_{\mathrm{Sh}}, \bot_{\mathrm{Sh}}\} \ .$$

Intuitively, an abstract object $[A, U]$ describes the relations between the variables in $U$: if $S \in A$, the variables in $S$ are allowed to share a common variable. For instance, $[\{\{x, y\}, \{z\}, \emptyset\}, \{x, y, z\}]$ represents the (equivalence classes of) substitutions where $x$ and $y$ may possibly share, while $z$ is independent from both $x$ and $y$: $\{x/y\}$ and $\epsilon$ are two of such substitutions while $\{x/z\}$ is not.

The domain is ordered like `Psub`, with $\top_{\mathrm{Sh}}$ and $\bot_{\mathrm{Sh}}$ as the greatest and least element respectively, and $[A_1, U_1] \sqsubseteq_{\mathrm{Sh}} [A_2, U_2]$ iff $U_1 = U_2$ and $A_1 \subseteq A_2$. The least upper bound satisfies the following property:

$$[A_1,U_1] \sqcup_{\mathrm{Sh}} [A_2,U_2] = \begin{cases} [A_1 \cup A_2, U_1] & \text{if } U_1 = U_2, \\ \top_{\mathrm{Sh}} & \text{otherwise.} \end{cases} \tag{16}$$

To design the abstraction from `Psub` to `Sharing`, we first define a map $\alpha_{\mathrm{Sh}}$ : $\mathit{ISubst}_{\sim} \rightarrow$ `Sharing` as

$$\alpha_{\mathrm{Sh}}([\sigma]_V) = [\{\mathit{occ}(\sigma,y) \cap V \mid y \in \mathcal{V}\}, V] \ . \tag{17}$$

where $\mathit{occ}(\sigma, y) = \{z \in \mathcal{V} \mid y \in \mathrm{vars}(\sigma(z))\}$ is the set of variables $z$ such that $y$ occurs in $\sigma(z)$. For instance, $\mathit{occ}(\{x/t(y, z), x'/z, y'/z'\}, z) = \{x, x', z\}$. We call *sharing group* an element of $\wp_f(\mathcal{V})$.

We say that $x$ is *independent* from $y$ in $[\sigma]_V$ when, given $\alpha_{\mathrm{Sh}}([\sigma]_V) = [S, U]$, there is no $X \in S$ such that $\{x, y\} \subseteq X$. Given $U \in \wp(\mathcal{V})$, we say that $x$ is independent from $U$ in $[\sigma]_V$ when it is independent from $y$ for each $y \in U$ different from $x$. Finally, $x$ is independent in $[\sigma]_V$ if it is independent from $V$ in $[\sigma]_V$.

*Proposition 5.1*

The map $\alpha_{\mathrm{Sh}} : \mathit{ISubst}_{\sim} \to \mathtt{Sharing}$ is well defined, i.e., it does not depend on the choice of representatives.

*Proof*
If $\sigma \sim_V \sigma'$, let $\rho \in \mathit{Ren}$ such that $\sigma'(x) = \rho(\sigma(x))$ for each $x \in V$. Then

$$\begin{aligned} \mathit{occ}(\sigma', \rho(y)) \cap V &= \{z \in V \mid \rho(y) \in \mathrm{vars}(\sigma'(z))\} \\ &= \{z \in V \mid y \in \rho^{-1}(\mathrm{vars}(\rho(\sigma(z))))\} \\ &= \{z \in V \mid y \in \mathrm{vars}(\sigma(z))\} \\ &= \mathit{occ}(\sigma, y) \cap V \ . \end{aligned}$$

Therefore, $x \in \mathit{occ}(\sigma, y) \cap V$ iff $x \in \mathit{occ}(\sigma', \rho(y)) \cap V$, which proves the thesis. □

The abstraction map may be lifted pointwise to $\alpha_{\mathrm{Sh}} : \mathtt{Psub} \to \mathtt{Sharing}$ as follows:

$$\alpha_{\mathrm{Sh}}(\bot_{\mathrm{Ps}}) = \bot_{\mathrm{Sh}} \qquad \alpha_{\mathrm{Sh}}(\top_{\mathrm{Ps}}) = \top_{\mathrm{Sh}}$$
$$\alpha_{\mathrm{Sh}}([\Theta, U]) = \bigsqcup_{[\sigma]_U \in \Theta}{}_{\mathrm{Sh}} \ \alpha_{\mathrm{Sh}}([\sigma]_U) \tag{18}$$

To ease the notation, often we will write a sharing group as the sequence of its elements in any order (e.g., `xyz` represents $\{x, y, z\}$) and we omit the empty set when clear from the context. For example:

$$\begin{aligned} \alpha_{\mathrm{Sh}}([\{[\epsilon]\}, \{x, y, z\}]) &= [\{\mathtt{x}, \mathtt{y}, \mathtt{z}\}, \{x, y, z\}] \\ \alpha_{\mathrm{Sh}}([\{[x/y, z/a]\}, \{x, y, z\}]) &= [\{\mathtt{xy}\}, \{x, y, z\}] \\ \alpha_{\mathrm{Sh}}([\{[\epsilon], [x/y, z/a]\}, \{x, y, z\}]) &= [\{\mathtt{xy}, \mathtt{x}, \mathtt{y}, \mathtt{z}\}, \{x, y, z\}] \ . \end{aligned}$$

Since $\alpha_{\mathrm{Sh}}$ is additive, there is an induced concretization function $\gamma_{\mathrm{Sh}}$, the right adjoint of $\alpha_{\mathrm{Sh}}$, which maps each abstract object to the set of substitutions it represents:

$$\gamma_{\mathrm{Sh}}([S, U]) = [\{[\theta]_U \mid \alpha_{\mathrm{Sh}}([\theta]_U) \sqsubseteq_{\mathrm{Sh}} [S, U]\}, U] \ . \tag{19}$$

Note that each abstract object represents the *possible* relations between variables: a substitution in which all the variables in $U$ are ground is always in $\gamma_{\mathrm{Sh}}([A, U])$, independently from $A$.

*Proposition 5.2*
$\langle \alpha_{\mathrm{Sh}}, \gamma_{\mathrm{Sh}} \rangle : \mathtt{Psub} \leftrightarrows \mathtt{Sharing}$ defines a Galois insertion.

*Proof*
That $\langle \alpha_{\mathrm{Sh}}, \gamma_{\mathrm{Sh}} \rangle$ is a Galois connection immediately follows from the fact they are an adjoint pair. Now, we want to prove that $\alpha_{\mathrm{Sh}}$ is onto. Given $[S, V] \in \mathtt{Sharing}$ and $X \in S$, consider the substitution $\theta_X$ defined as

$$\theta_X(x) = \begin{cases} w & \text{if } x \in X \\ a & \text{if } x \in V \setminus X \\ x & \text{otherwise.} \end{cases}$$

where $w$ is a fresh variable not in $V$. It is easy to check that $\alpha_{\mathrm{Sh}}([\theta_X]_V) = [\{X\}, S]$

and therefore $\alpha_{\mathrm{Sh}}([\{[\theta_X]_V \mid X \in S\}, V]) = [S, V]$. Moreover, we have $\alpha_{\mathrm{Sh}}(\bot_{\mathrm{Ps}}) = \bot_{\mathrm{Sh}}$ and $\alpha_{\mathrm{Sh}}(\top_{\mathrm{Ps}}) = \top_{\mathrm{Sh}}$. □

### 5.1 The Abstract Semantics

The abstract semantics is obtained by replacing, in the definition of the concrete semantics in Section 4.2, the concrete domain `Psub` with the abstract domain `Sharing` and the basic operators, namely, least upper bound $\sqcup_{\mathrm{Ps}}$, forward unification $\mathbf{U}^f_{\mathrm{Ps}}$ and backward unification $\mathbf{U}^b_{\mathrm{Ps}}$ with their corresponding abstract counterparts. The abstract least upper bound $\sqcup_{\mathrm{Sh}}$ has been already defined in the previous section. We recall that, on the concrete side, we have defined the forward and backward unification operators in (14), (15) as:

$$\mathbf{U}^f_{\mathrm{Ps}}([\Theta, U], A_1, A_2) = \pi_{\mathrm{Ps}}(\mathsf{unif}_{\mathrm{Ps}}(\rho([\Theta, U]), \mathrm{mgu}(\rho(A_1) = A_2)), \mathrm{vars}(A_2))$$

$$\mathbf{U}^b_{\mathrm{Ps}}([\Theta_1, U_1], [\Theta_2, U_2], A_1, A_2) = \pi_{\mathrm{Ps}}(\mathsf{match}_{\mathrm{Ps}}(\rho([\Theta_1, U_1]), \mathsf{unif}_{\mathrm{Ps}}([\Theta_2, U_2], \mathrm{mgu}(\rho(A_1) = A_2))), U_2 \cup \mathrm{vars}(A_2))$$

The abstract forward and backward unification operators are obtained by replacing, in the above definitions, the primitive operators with their abstract counterparts, namely, abstract projection $\pi_{\mathrm{Sh}}$, abstract renaming $\rho$, abstract unification $\mathsf{unif}_{\mathrm{Sh}}$ and abstract matching $\mathsf{match}_{\mathrm{Sh}}$.

The abstract operators behave exactly as the concrete ones on $\top_{\mathrm{Sh}}$ and $\bot_{\mathrm{Sh}}$. Abstract projection and renaming are defined as:

$$\pi_{\mathrm{Sh}}([A_1, U_1], U_2) = [\{B \cap U_2 \mid B \in A_1\}, U_1 \cap U_2] \ , \tag{20}$$

$$\rho([A, U]) = [\rho(A), \rho(U)] \ . \tag{21}$$

The definition of the abstract versions of matching and unification is the main argument of the rest of this paper. Here we show some properties of completeness for projection and renaming. Since the concrete and abstract operators behave in the same way on top and bottom elements, here and in the following proofs we only consider the case when all the arguments are different from $\bot_{\mathrm{Ps}}/\bot_{\mathrm{Sh}}$ and $\top_{\mathrm{Ps}}/\top_{\mathrm{Sh}}$.

*Theorem 5.3*
$\pi_{\mathrm{Sh}}$ is correct and complete w.r.t. $\pi_{\mathrm{Ps}}$.

*Proof*
Given $[\Theta, V] \in$ `Psub`, we prove that $\alpha_{\mathrm{Sh}}(\pi_{\mathrm{Ps}}([\Theta, V], U)) = \pi_{\mathrm{Sh}}(\alpha_{\mathrm{Sh}}([\Theta, V]), U)$. We first prove that, for each $[\phi]_V \in \mathit{ISubst}_\sim$, it holds that $\pi_{\mathrm{Sh}}(\alpha_{\mathrm{Sh}}([\phi]_V), U) = \alpha_{\mathrm{Sh}}([\phi]_{V \cap U})$. Actually

$$\begin{aligned} \alpha_{\mathrm{Sh}}([\phi]_{V\cap U}) &= [\{\mathit{occ}(\phi, z) \cap V \cap U \mid z \in \mathcal{V}\}, V \cap U] \\ &= \pi_{\mathrm{Sh}}([\{\mathit{occ}(\phi, z) \cap V \mid z \in \mathcal{V}\}, V], U) \\ &= \pi_{\mathrm{Sh}}(\alpha_{\mathrm{Sh}}([\phi]_V), U) \ . \end{aligned}$$

The result for the lifted $\alpha_{\mathrm{Sh}}$ follows trivially. □

*Theorem 5.4*
Abstract renaming is correct, complete and $\gamma$-complete w.r.t. concrete renaming.

*Proof*
First of all, given $\rho \in \mathit{Ren}, y \in \mathcal{V}$ and $\phi \in \mathit{Subst}$, we prove that $\mathit{occ}(\rho(\phi), \rho(y)) = \rho(\mathit{occ}(\phi, y))$. Actually:

$$\begin{aligned}
\mathit{occ}(\rho(\phi), \rho(y)) &= \{z \in \mathcal{V} \mid \rho(y) \in \mathrm{vars}(\rho(\phi(\rho^{-1}(z))))\} \\
&= \{z \in \mathcal{V} \mid y \in \mathrm{vars}(\phi(\rho^{-1}(z)))\} \\
&= \{\rho(k) \mid k \in \mathcal{V}, y \in \mathrm{vars}(\phi(k))\} \qquad \text{[by letting } k = \rho^{-1}(z)] \\
&= \rho(\mathit{occ}(\phi), y) \ .
\end{aligned}$$

Then we prove that, given $[\phi]_V \in \mathtt{Psub}$ and $\rho \in \mathit{Ren}$, $\alpha_{\mathrm{Sh}}(\rho([\phi]_V)) = \rho(\alpha_{\mathrm{Sh}}([\phi]_V))$. Using the fact that $\rho$ as an operation over $\mathit{ISubst}_\sim$ is bijective, we have:

$$\begin{aligned}
\alpha_{\mathrm{Sh}}(\rho([\phi]_V)) &= [\{\mathit{occ}(\rho(\phi), z) \cap \rho(V) \mid z \in \mathcal{V}\}, \rho(V)] \\
&= [\{\rho(\mathit{occ}(\phi, \rho^{-1}(z)) \cap \rho(V) \mid z \in \mathcal{V}\}, \rho(V)] \\
&= \rho([\mathit{occ}(\phi, k) \cap V \mid k \in \mathcal{V}\}, V]) \qquad \text{[by letting } z = \rho(k)] \\
&= \rho(\alpha_{\mathrm{Sh}}([\phi]_V)) \ .
\end{aligned}$$

This property, lifted to $\mathtt{Psub}$, gives the completeness of abstract renaming. Finally, we need to prove that renaming is $\gamma$-complete, i.e., that $\gamma_{\mathrm{Sh}} \circ \rho = \rho \circ \gamma_{\mathrm{Sh}}$.

$$\begin{aligned}
\gamma_{\mathrm{Sh}}(\rho([S, V])) &= \gamma_{\mathrm{Sh}}([\rho(S), \rho(V)]) \\
&= \big[\{[\theta]_V \mid \alpha_{\mathrm{Sh}}([\theta]_V) \sqsubseteq_{\mathrm{Sh}} \rho(S)\}, \rho(V)\big] \\
&= \big[\{\rho([\theta]_V) \mid \alpha_{\mathrm{Sh}}(\rho([\theta]_V)) \sqsubseteq_{\mathrm{Sh}} \rho(S)\}, \rho(V)\big] \\
&= \big[\{\rho([\theta]_V) \mid \rho(\alpha_{\mathrm{Sh}}([\theta]_V)) \sqsubseteq_{\mathrm{Sh}} \rho(S)\}, \rho(V)\big] \\
&= \big[\{\rho([\theta]_V) \mid \alpha_{\mathrm{Sh}}([\theta]_V) \sqsubseteq_{\mathrm{Sh}} S\}, \rho(V)\big] \\
&= \rho(\gamma_{\mathrm{Sh}}([S, V])) \ .
\end{aligned}$$

which concludes the proof of the theorem. □

## 6 Forward Unification

We briefly recall from (Cortesi and Filé 1999; Bagnara et al. 2002) the definition of the standard operator $\mathsf{unif}'_{\mathrm{Sh}}$ for abstract unification on $\mathtt{Sharing}$. The abstract unification is performed between a set of sharing groups $A$ and a single substitution $\delta$, under the assumption that $\mathrm{vars}(\delta) \subseteq U$, and it is defined as follows:

$$\mathsf{unif}'_{\mathrm{Sh}}([A, U], \delta) = [\mathbf{u}_{\mathrm{Sh}}(A, \delta), U] \tag{22}$$

where $\mathbf{u}_{\mathrm{Sh}} : \wp(\wp_f(\mathcal{V})) \times \mathit{ISubst} \to \wp(\wp_f(\mathcal{V}))$ is defined by induction as follows:

$$\begin{aligned}
\mathbf{u}_{\mathrm{Sh}}(A, \epsilon) &= A \\
\mathbf{u}_{\mathrm{Sh}}(A, \{x/t\} \uplus \theta) &= \mathbf{u}_{\mathrm{Sh}}(A \setminus (\mathbf{rel}(A, \{x\}) \cup \mathbf{rel}(A, \mathrm{vars}(t))) \\
&\qquad \cup \mathbf{bin}(\mathbf{rel}(A, \{x\})^*, \mathbf{rel}(A, \mathrm{vars}(t))^*), \theta).
\end{aligned} \tag{23}$$

The auxiliary operators used in the definition of $\mathbf{u}_{\mathrm{Sh}}$ are given by:

- the *closure under union* (or *star union*) $(.)^* : \wp(\wp_f(\mathcal{V})) \to \wp(\wp_f(\mathcal{V}))$

$$A^* = \left\{ \bigcup T \mid \emptyset \neq T \in \wp_f(A) \right\}^5 \ ; \tag{24}$$

- the *extraction of relevant components* $\mathbf{rel} : \wp(\wp_f(\mathcal{V})) \times \wp_f(\mathcal{V}) \to \wp(\wp_f(\mathcal{V}))$:

$$\mathbf{rel}(A, V) = \{T \in A \mid T \cap V \neq \emptyset\} \ ; \tag{25}$$

- the *binary union* $\mathbf{bin} : \wp(\wp_f(\mathcal{V})) \times \wp(\wp_f(\mathcal{V})) \to \wp(\wp_f(\mathcal{V}))$:

$$\mathbf{bin}(A, B) = \{T_1 \cup T_2 \mid T_1 \in A, T_2 \in B\} \ . \tag{26}$$

We recall that we will often abuse the notation and write $\mathbf{rel}(A, o)$ for $\mathbf{rel}(A, \mathrm{vars}(o))$ and $x \in o$ for $x \in \mathrm{vars}(o)$ where $o$ is any syntactic object.

*Example 6.1*
Take $A = \{\mathtt{xy}, \mathtt{xz}, \mathtt{y}\}$, $U = \{w, x, y, z\}$ and $\delta = \{x/t(y,z), w/t(y)\}$. Note that, since $w$ does not appear in $A$, then $w$ is always bound to a ground term in $\gamma_{\mathrm{Sh}}([A, U])$. We have $\mathbf{rel}(A, x) = \{\mathtt{xy}, \mathtt{xz}\}$, $\mathbf{rel}(A, y) = \{\mathtt{xy}, \mathtt{y}\}$, $\mathbf{rel}(A, z) = \{\mathtt{xz}\}$ and therefore

$$\begin{aligned} \mathbf{u}_{\mathrm{Sh}}(A, \{x/t(y,z)\}) =& A \setminus \{\mathtt{xy}, \mathtt{xz}, \mathtt{y}\} \cup \mathbf{bin}(\{\mathtt{xy}, \mathtt{xz}\}^*, \{\mathtt{xy}, \mathtt{xz}, \mathtt{y}\}^*) \\ =& \mathbf{bin}(\{\mathtt{xy}, \mathtt{xz}, \mathtt{xyz}\}, \{\mathtt{xy}, \mathtt{xz}, \mathtt{xyz}, \mathtt{y}\}) \\ =& \{\mathtt{xy}, \mathtt{xz}, \mathtt{xyz}\} \ . \end{aligned}$$

If we take $B = \{\mathtt{xy}, \mathtt{xz}, \mathtt{xyz}\}$, we obtain $\mathbf{rel}(B, w) = \emptyset$, $\mathbf{rel}(B, y) = \{\mathtt{xy}, \mathtt{xyz}\}$ and therefore

$$\begin{aligned} \mathbf{u}_{\mathrm{Sh}}(A, \delta) =& \mathbf{u}_{\mathrm{Sh}}(B, \{w/t(y)\}) \\ =& B \setminus \{\mathtt{xy}, \mathtt{xyz}\} \cup \mathbf{bin}(\emptyset, \{\mathtt{xy}, \mathtt{xyz}\}^*) \\ =& B \setminus \{\mathtt{xy}, \mathtt{xyz}\} \\ =& \{\mathtt{xz}\} \ . \quad \square \end{aligned}$$

It is worth noting that $\mathsf{unif}'_{\mathrm{Sh}}$ is not the abstract counterpart of $\mathsf{unif}_{\mathrm{Ps}}$, because $\mathsf{unif}'_{\mathrm{Sh}}([S, U], \delta)$ is defined only under the condition that $\mathrm{vars}(\delta) \subseteq U$. Since this is not enough to define a goal-dependent semantics, when this solution is adopted, there is the need of an operator to expand the set of variables of interest in a substitution. Let us introduce the following concrete operator:

$$\iota_{\mathrm{Ps}}([\Theta, U], V) = [\{\mathrm{mgu}([\sigma]_U, [\epsilon]_V) \mid [\sigma]_U \in \Theta\}, U \cup V] \ , \tag{27}$$

whose optimal abstract counterpart is simply given by:

$$\iota_{\mathrm{Sh}}([\Theta, U], V) = [\Theta \cup \{\{x\} \mid x \in V \setminus U\}, U \cup V)] \ . \tag{28}$$

By using $\iota_{\mathrm{Ps}}$, the operator $\mathsf{unif}_{\mathrm{Ps}}$ can be equivalently rewritten as:

$$\mathsf{unif}_{\mathrm{Ps}}([\Theta, U], \theta) = \mathsf{unif}_{\mathrm{Ps}}(\iota_{\mathrm{Ps}}([\Theta, U], \mathrm{vars}(\theta)), \theta) \ , \tag{29}$$

[5] Note that, due to the condition $T \neq \emptyset$, the notation $A^+$ would be more appropriate. However, we retain the notation $A^*$ for historical reasons.

and now, in the right hand side, $\iota_{\mathrm{Ps}}([\Theta, U], \mathrm{vars}(\theta))$ is an object of the kind $[\Delta, U \cup \mathrm{vars}(\theta)]$. Therefore, a correct abstract forward unification operator for $\mathbf{U}^{f}_{\mathrm{Ps}}$ may be obtained as

$$\begin{aligned}\mathbf{U'}^{f}_{\mathrm{Sh}}([\Theta, U], A_1, A_2) = \pi_{\mathrm{Sh}}(\mathsf{unif}'_{\mathrm{Sh}}(\iota_{\mathrm{Sh}}(\rho([\Theta, U]), \mathrm{vars}(\rho(A_1)) \cup \mathrm{vars}(A_2)),\\ \mathrm{mgu}(\rho(A_1) = A_2)), \mathrm{vars}(A_2)) \enspace ,\end{aligned} \tag{30}$$

provided that $\rho$ is a renaming such that $\rho(U \cup \mathrm{vars}(A_1)) \cap \mathrm{vars}(A_2) = \emptyset$. However, $\mathbf{U'}^{f}_{\mathrm{Sh}}$ is not optimal w.r.t. $\mathbf{U}^{f}_{\mathrm{Ps}}$.

*Example 6.2*
We keep on Example 4.4 and compute the abstract counterpart of the concrete forward unification

$$\begin{aligned}&\mathbf{U}^{f}_{\mathrm{Ps}}([\{[y/f(x,z)]\}, \{x,y,z\}], p(x,y,z), p(u,v,w)) =\\ &\qquad [\{[u/x, v/f(x,z), w/z]\}, \{u,v,w\}] \enspace .\end{aligned}$$

Since the abstraction of $[\{[y/f(x,z)]\}, \{x,y,z\}]$ is $[\{\mathtt{xy}, \mathtt{yz}\}, \{x,y,z\}]$, we compute:

$$\begin{aligned}&\mathbf{U'}^{f}_{\mathrm{Sh}}([\{\mathtt{xy}, \mathtt{yz}\}, \{x,y,z\}], p(x,y,z), p(u,v,w)) =\\ &\quad \pi_{\mathrm{Sh}}([\mathbf{u}_{\mathrm{Sh}}(\{\mathtt{xy}, \mathtt{yz}, \mathtt{u}, \mathtt{v}, \mathtt{w}\}, \{x/u, y/v, z/w\}), \{x,y,z,u,v,w\}], \{u,v,w\}) =\\ &\quad \pi_{\mathrm{Sh}}([\{\mathtt{xyuv}, \mathtt{yzvw}, \mathtt{xyzuvw}\}, \{x,y,z,u,v,w\}], \{u,v,w\}) =\\ &\quad [\{\mathtt{uv}, \mathtt{vw}, \mathtt{uvw}\}, \{u,v,w\}] \enspace .\end{aligned}$$

There exists a sharing group $\mathtt{uvw}$ computed by the forward unification. However, when computing $\mathsf{unif}_{\mathrm{Ps}}(\gamma_{\mathrm{Sh}}([\{\mathtt{xy}, \mathtt{yz}\}, \{x,y,z\}]), \{x/u, y/v, z/w\})$ we know that $u, v$ and $w$ are free in $\gamma_{\mathrm{Sh}}([\{\mathtt{xy}, \mathtt{yz}\}, \{x,y,z\}]$. Following (Hans and Winkler 1992), we can avoid computing the star unions when considering the binding $y/v$ in $\mathbf{u}_{\mathrm{Sh}}$, obtaining the smaller result $[\{\mathtt{xyuv}, \mathtt{yzvw}\}, \{x,y,z,u,v,w\}]$. If we now compute the projection on the variables $\{u,v,w\}$ we obtain the entry substitution $[\{\mathtt{uv}, \mathtt{vw}\}, \{u,v,w\}]$, with an obvious gain of precision. □

*Example 6.3*
Let us consider the following unification.

$$\begin{aligned}&\mathbf{U'}^{f}_{\mathrm{Sh}}([\{\mathtt{xy}, \mathtt{xz}\}, \{x,y,z\}], p(x,y,z), p(t(u,v),h,k)) =\\ &\qquad \pi_{\mathrm{Sh}}([\mathbf{bin}(\{\mathtt{xyh}, \mathtt{xzk}, \mathtt{xyzhk}\}, \{\mathtt{u}, \mathtt{v}, \mathtt{uv}\}), \{x,y,z,h,k,u\}], \{u,v,h,k\}) \enspace .\end{aligned}$$

Since the term $t(u,v)$ is linear and independent from $x$, following (Hans and Winkler 1992) we can avoid to compute the star union over $\{\mathtt{xy}, \mathtt{xz}\}$, obtaining the abstract object $[\mathbf{bin}(\{\mathtt{xyh}, \mathtt{xzk}\}, \{\mathtt{u}, \mathtt{v}, \mathtt{uv}\}), \{x,y,z,h,k,u\}]$. If we project on $\{h,k,u,v\}$ we obtain $\mathbf{bin}(\{\mathtt{h}, \mathtt{k}\}, \{\mathtt{u}, \mathtt{v}, \mathtt{uv}\})$ against $\mathbf{bin}(\{\mathtt{h}, \mathtt{k}, \mathtt{hk}\}, \{\mathtt{u}, \mathtt{v}, \mathtt{uv}\})$. In this way, we are able to prove the independence of $h$ from $k$. □

These examples show that, when computing forward abstract unification by first enlarging the domain of variables of interest, there is a loss of precision. In fact, such a forward abstract unification operator is not optimal. We now show that it is possible to design an optimal operator for forward unification which is able to exploit linearity and freeness information that stems from the fact that variables in the third argument of $\mathbf{U}^f_{\mathrm{Ps}}$ are fresh. Note that we are not proposing to embed freeness and linearity information inside the domain, but only to use all the information coming from the syntax of the clauses.

### *6.1 The Refined Forward Unification*

We are going to define an abstract operator $\mathsf{unif}_{\mathrm{Sh}}$ which is correct and optimal w.r.t. $\mathsf{unif}_{\mathrm{Ps}}$.

*Definition 6.4*
The *abstract unification* $\mathsf{unif}_{\mathrm{Sh}} : \mathtt{Sharing} \times \mathit{ISubst} \to \mathtt{Sharing}$ is defined as

$$\mathsf{unif}_{\mathrm{Sh}}([S_1, U_1], \theta) = [\mathbf{u}^f_{\mathrm{Sh}}(S_1 \cup \{\{x\} \mid x \in U_2\}, U_2, \theta), U_1 \cup U_2]$$

where $U_2 = \mathrm{vars}(\theta) \setminus U_1$ and $\mathbf{u}^f_{\mathrm{Sh}} : \wp(\wp_f(\mathcal{V})) \times \wp_f(\mathcal{V}) \times \mathit{ISubst} \to \wp(\wp_f(\mathcal{V}))$ is defined as:

$$\begin{aligned}
\mathbf{u}^f_{\mathrm{Sh}}(S, U, \epsilon) &= S \\
\mathbf{u}^f_{\mathrm{Sh}}(S, U, \{x/t\} \uplus \delta) &= \mathbf{u}^f_{\mathrm{Sh}}((S \setminus (\mathbf{rel}(S,t) \cup \mathbf{rel}(S,x))) \cup \\
&\qquad \mathbf{bin}(\mathbf{rel}(S,x), \mathbf{rel}(S,t)), U \setminus \{x\}, \delta) \qquad \text{if } x \in U \\
\mathbf{u}^f_{\mathrm{Sh}}(S, U, \{x/t\} \uplus \delta) &= \mathbf{u}^f_{\mathrm{Sh}}((S \setminus (\mathbf{rel}(S,t) \cup \mathbf{rel}(S,x))) \cup \\
&\qquad \mathbf{bin}(\mathbf{rel}(S,x), \mathbf{rel}(S,Y)^*) \cup \\
&\qquad \mathbf{bin}(\mathbf{rel}(S,x)^*, \mathbf{rel}(S,Z)^*) \cup \\
&\qquad \mathbf{bin}(\mathbf{bin}(\mathbf{rel}(S,x)^*, \mathbf{rel}(S,Z)^*), \mathbf{rel}(S,Y)^*), \\
&\qquad U \setminus \mathrm{vars}(\{x/t\}), \delta) \qquad \text{if } x \notin U
\end{aligned}$$

where $Y = \mathrm{uvars}(t) \cap U$, $Z = \mathrm{vars}(t) \setminus Y$.

The idea is simply to carry on, in the second argument of $\mathbf{u}^f_{\mathrm{Sh}}$, the set of variables which are definitively free and to apply the optimizations for the abstract unification with linear terms and free variables (Hans and Winkler 1992). Actually, while the case for $x \in U$ is standard, the case for $x \notin U$ exploits some optimizations which are not found in the literature. When $Z = \emptyset$, we obtain:

$$(S \setminus (\mathbf{rel}(S,t) \cup \mathbf{rel}(S,x))) \cup \mathbf{bin}(\mathbf{rel}(S,x), \mathbf{rel}(S,Y)^*) \ ,$$

which is the standard result when the term $t$ is linear and independent from $x$. However, when $Z \neq \emptyset$, the standard optimizations which appear, e.g., in (Hans and Winkler 1992), do not apply, since $t$ cannot be proved to be linear and independent from $x$, and we should obtain the following standard result:

$$(S \setminus (\mathbf{rel}(S,t) \cup \mathbf{rel}(S,x))) \cup \mathbf{bin}(\mathbf{rel}(S,x)^*, \mathbf{rel}(S,t)^*) \ .$$

We are able to avoid some star unions by distinguishing the variables in $t$ which are "linear and independent" (the set $Y$) from the others (the set $Z$), and observing that two sharing groups in $\mathbf{rel}(S,x)$ may be merged together only under the effect of the unification with some variable in $Z$. We will come back later to this topic.

We can now define the forward abstract unification $\mathbf{U}^f_{\mathrm{Sh}} : \mathtt{Sharing} \times \wp_f(\mathcal{V}) \times \mathsf{Atoms} \times \mathsf{Atoms} \to \mathtt{Sharing}$. We only need to introduce the necessary renamings and projections, as done for the concrete case:

$$\mathbf{U}^f_{\mathrm{Sh}}([S_1, U_1], A_1, A_2) = \pi_{\mathrm{Sh}}(\mathsf{unif}_{\mathrm{Sh}}(\rho([S_1, U_1]), \mathrm{mgu}(\rho(A_1) = A_2)), \mathrm{vars}(A_2)) \quad (31)$$

with $\rho$ a renaming such that $\rho(U_1 \cup \mathrm{vars}(A_1)) \cap \mathrm{vars}(A_2) = \emptyset$.

*Example 6.5*
We keep on Examples 4.4 and 6.2 and compute the abstract counterpart of the concrete forward unification

$$\mathbf{U}^f_{\mathrm{Ps}}([\{[y/f(x,z)]\}, \{x,y,z\}], p(x,y,z), p(u,v,w)) = \\ [\{[u/x, v/f(x,z), w/z]\}, \{u,v,w\}]$$

using our optimized forward unification operator.

$$\begin{aligned} &\mathbf{U}^f_{\mathrm{Sh}}([\{\mathtt{xy}, \mathtt{yz}\}, \{x,y,z\}], p(x,y,z), p(u,v,w)) = \\ &\quad \pi_{\mathrm{Sh}}(\mathsf{unif}_{\mathrm{Sh}}(\{\mathtt{xy}, \mathtt{yz}\}, \{x/u, y/v, z/w\}), \{u,v,w\}) = \\ &\quad \pi_{\mathrm{Sh}}([\{\mathtt{uvxy}, \mathtt{vwyz}\}, \{u,v,w,x,y,z\}], \{u,v,w\}) = \\ &\quad [\{\mathtt{uv}, \mathtt{vw}\}, \{u,v,w\}] \ . \end{aligned}$$

Thus the optimized operator is able to prove that $u$ and $w$ are independent after the unification. □

### *6.2 Correctness of Forward Unification*

We prove that the unification operator $\mathsf{unif}_{\mathrm{Sh}}$ is correct w.r.t. the concrete operator $\mathsf{unif}_{\mathrm{Ps}}$. We begin to analyze the abstract behavior of unification when the second argument is a substitution with only one binding. Let $\sigma$ and $\{x/t\}$ be the two substitutions we want to unify. In this simple case, the resultant sharing groups can be easily computed by exploiting the substitution $\delta = \mathrm{mgu}(x\sigma = t\sigma)$. We show that, under suitable conditions, any sharing group either belongs to $\alpha_{\mathrm{Sh}}([\sigma]_U])$ or is of the form $occ(\sigma, occ(\delta, v)) \cap U$, where $v \in \mathrm{vars}(x\sigma = t\sigma)$.

*Proposition 6.6*
Let $[\sigma]_U \in \mathit{ISubst}_\sim$ and $\{x/t\} \in \mathit{ISubst}$ such that $\mathrm{vars}(\{x/t\}) \subseteq U$ and $\sigma$ and $\{x/t\}$ unify. If $\alpha_{\mathrm{Sh}}([\sigma]_U]) \sqsubseteq_{\mathrm{Sh}} [S, U]$ and $\delta = \mathrm{mgu}(x\sigma = t\sigma)$, we obtain:

$$\begin{aligned} \alpha_{\mathrm{Sh}}(\mathrm{mgu}([\sigma]_U, [x/t]_U)) \sqsubseteq_{\mathrm{Sh}} & [(S \setminus (\mathbf{rel}(S,x) \cup \mathbf{rel}(S,t))) \\ & \cup \{occ(\sigma, occ(\delta, v)) \cap U \mid v \in \mathrm{vars}(x\sigma = t\sigma)\}, U] \ . \end{aligned}$$

*Proof*
The proof can be found in the Appendix as Prop. B.3 □

This result may be refined by introducing further hypotheses. We have anticipated that our abstract algorithm takes advantage of the fact that some variables are known to be free in order to to produce better results than standard abstract unification. We may be more formal.

*Definition 6.7*
We say that a variable $x \in \mathcal{V}$ is *free* in $[\theta]_V$ when $\theta_{|V}(x) \in \mathcal{V}$.

Note that this definition does not depend on the choice of the representative for $[\theta]_V$. Moreover, if $x$ is free and independent from $V$ in $[\theta]_V$, there exists a representative $\theta' \sim_V \theta$ such that $x \notin \text{vars}(\theta')$. It is enough to take $\theta' = \theta''_{|\{-x\}}$ where $\theta''$ is a canonical representative.

Now, we consider again Prop. 6.6, but we assume $x$ to be free and independent from $U$ in $[\sigma]_U$. A result similar to the following proposition has been already proved in the literature, e.g., (Hans and Winkler 1992). Since our treatment of substitutions is slightly different from the standard one, for the sake of completeness we present the altered proof.

*Proposition 6.8*
Let $[\sigma]_U \in \mathit{ISubst}_\sim$ and $\{x/t\} \in \mathit{ISubst}$ such that $\text{vars}(\{x/t\}) \subseteq U$ and $\sigma$ and $\{x/t\}$ unify. If $\alpha_{\text{Sh}}([\sigma]_U) \sqsubseteq_{\text{Sh}} [S, U]$ and $x$ is free and independent from $U$ in $[\sigma]_U$, then:

$$\begin{aligned}&\alpha_{\text{Sh}}(\text{mgu}([\sigma]_U, [x/t]_U))\\ &\qquad\qquad \sqsubseteq_{\text{Sh}} [(S \setminus (\mathbf{rel}(S,x) \cup \mathbf{rel}(S,t))) \cup \mathbf{bin}(\mathbf{rel}(S,x), \mathbf{rel}(S,t)), U] \ .\end{aligned}$$

*Proof*
The proof can be found in the Appendix as Prop. B.4 □

Now we analyze the case when $x$ is not guaranteed to be free and independent from $U$ in $[\sigma]_U$. We show that it is possible to consider three distinct cases depending on the set of variables $Y = \{y \in \text{vars}(t) | \text{vars}(\sigma(y)) \subseteq \text{uvars}(x\sigma = t\sigma)\}$, that is the set of variables $y$ such that all the variables in $\text{vars}(\sigma(y))$ appear once in $x\sigma = t\sigma$. Such variables play a special role in the unification process. Generally speaking, we can form new sharing groups by merging sets from $\mathbf{rel}(S, x)$ and $\mathbf{rel}(S, t)$. Obviously, any new sharing group must be formed by choosing at least one element from $\mathbf{rel}(S, x)$ and at least one from $\mathbf{rel}(S, t)$. We show that, if we do not include any variable from $\text{vars}(t) \setminus Y$, then we may avoid to include more than one sharing group from $\mathbf{rel}(S, x)$. Intuitively speaking, variables from $Y$ do not allow to merge different sharing groups from $\mathbf{rel}(S, x)$ since such variables appear only once and thus cannot be bound to different occurrences of $x$.

*Example 6.9*
Let $\sigma = \{x/f(u, v)\}$, $U = \{u, v, x, y, z\}$ and consider the binding $x/f(f(y, z), z)$. We have that $Y = \{y\}$, $\alpha_{\text{Sh}}([\sigma]_U) = [S, U] = [\{\mathtt{ux}, \mathtt{vx}, \mathtt{y}, \mathtt{z}\}, U]$, $\mathbf{rel}(S, x) = \{\mathtt{ux}, \mathtt{vx}\}$ and $\mathbf{rel}(S, t) = \{\mathtt{y}, \mathtt{z}\}$. In the standard definition of abstract unification, `uvxy` would be one of the possible resultant sharing groups. However, since `uvxy` is obtained by joining two sharing groups in $\mathbf{rel}(S, x)$ and it does not contain any variable in $\text{vars}(t) \setminus Y$, it cannot be generated. In fact, the result of the unification is $\eta = \{x/f(f(y, z), z), u/f(y, z), v/z\}$ and $\alpha_{\text{Sh}}([\eta]_U) = [\{\mathtt{uxy}, \mathtt{uvxz}\}, U]$. The variables $u$ and $v$ occur in the same sharing group thanks to the two occurrences of $z$. □

*Proposition 6.10*
Let $[\sigma]_U \in \mathit{ISubst}_\sim$ and $\{x/t\} \in \mathit{ISubst}$ such that $\mathrm{vars}(\{x/t\}) \subseteq U$ and $\sigma$ and $\{x/t\}$ unify. Given $Y \subseteq \mathrm{vars}(t)$ such that, for all $y \in Y$, $\mathrm{vars}(\sigma(y)) \subseteq \mathrm{uvars}(x\sigma = t\sigma)$, if $\alpha_{\mathrm{Sh}}([\sigma]_U) \sqsubseteq_{\mathrm{Sh}} [S, U]$ then

$$\begin{aligned}\alpha_{\mathrm{Sh}}(\mathrm{mgu}([\sigma]_U, [x/t]_U)) \sqsubseteq_{\mathrm{Sh}} [&(S \setminus (\mathbf{rel}(S,t) \cup \mathbf{rel}(S,x)))\\ &\cup \mathbf{bin}(\mathbf{rel}(S,x), \mathbf{rel}(S,Y)^*) \cup \mathbf{bin}(\mathbf{rel}(S,x)^*, \mathbf{rel}(S,Z)^*)\\ &\cup \mathbf{bin}(\mathbf{bin}(\mathbf{rel}(S,x)^*, \mathbf{rel}(S,Z)^*), \mathbf{rel}(S,Y)^*), U]\ ,\end{aligned}$$

where $Z = \mathrm{vars}(t) \setminus Y$.

*Proof*
The proof can be found in the Appendix as Prop. B.6 □

Now, by combining the results from Propositions 6.8 and 6.10 we can show the correctness of $\mathsf{unif}_{\mathrm{Sh}}$.

*Theorem 6.11*
(Correctness of $\mathsf{unif}_{\mathrm{Sh}}$) The unification operator $\mathsf{unif}_{\mathrm{Sh}}$ is correct w.r.t. $\mathsf{unif}_{\mathrm{Ps}}$.

*Proof*
The proof can be found in the Appendix as Theorem B.8 □

### *6.3 Optimality of Forward Unification*

In this section we prove that the abstract unification operator $\mathsf{unif}_{\mathrm{Sh}}$ is optimal w.r.t. the concrete operator $\mathsf{unif}_{\mathrm{Ps}}$, that is to say that, given $[S_1, U_1] \in$ `Sharing` and $\theta \in \mathit{ISubst}$, it holds:

$$\alpha_{\mathrm{Sh}}(\mathsf{unif}_{\mathrm{Ps}}(\gamma_{\mathrm{Sh}}([S_1, U_1]), \theta)) \sqsupseteq_{\mathrm{Sh}} \mathsf{unif}_{\mathrm{Sh}}([S_1, U_1], \theta).$$

Let $\mathsf{unif}_{\mathrm{Sh}}([S_1, U_1], \theta) = [S, U]$ where $U = U_1 \cup \mathrm{vars}(\theta)$. In the rest of this section, we assume fixed $S, S_1, U, U_1, \theta$ as defined above.

For each $X \in S$, we need to exhibit a substitution $\delta$ such that $\alpha_{\mathrm{Sh}}([\delta]_{U_1}) \sqsubseteq_{\mathrm{Sh}} [S_1, U_1]$ and $\alpha_{\mathrm{Sh}}(\mathrm{mgu}([\delta]_{U_1}, [\theta]_U)) \sqsupseteq_{\mathrm{Sh}} [\{X\}, U]$. Any resultant sharing group is obtained by merging together sharing groups from $S_1$ and variables in $\mathrm{vars}(\theta) \setminus U_1$. We show that two sharing groups $B_1$ and $B_2$ may be joined by the abstract unification algorithm only if there are two variables $x_1 \in B_1$, $x_2 \in B_2$ such that $\theta(x_1)$ and $\theta(x_2)$ share some variable. Actually, we need to be careful when $x_1 = x_2$, since we need a variable which occurs at least twice in $\theta(x_1)$. More formally, given $X \in \wp_f(\mathcal{V})$ and $\theta \in \mathit{ISubst}$, we define a relation $\mathcal{R}_{\theta X} \subseteq S_1 \times S_1$ as follows:

$$\begin{aligned}B_1 \mathcal{R}_{\theta X} B_2 \iff \exists x_1 \in B_1\ \exists x_2 \in B_2\ \exists y.\ (y \in \mathrm{vars}(\theta(x_1)) \cap \mathrm{vars}(\theta(x_2)) \cap X) \wedge \\ (x_1 = x_2 \implies y \notin \mathrm{uvars}(\theta(x_1)))\ . \quad (32)\end{aligned}$$

We say that $X$ is $\theta$-connected when there exist $B_1, \ldots, B_n \in S_1$ s.t. $\cup_{1 \leq j \leq n} B_j = X \cap U_1$ and $B_1 \mathcal{R}^*_{\theta X} B_2 \ldots \mathcal{R}^*_{\theta X} B_n$, where $\mathcal{R}^*_{\theta X}$ is the transitive closure of $\mathcal{R}_{\theta X}$.

*Lemma 6.12*
For each $X \in S$, $X$ is $\theta$-connected.

*Proof*
The proof can be found in the Appendix as Lemma C.4 □

Now we will exploit the relation $\mathcal{R}_{\theta X}$ in order to find a substitution $\delta$ such that the concrete unification of $\theta$ with $\delta$ mimics the behavior of the abstract unification of $\theta$ with $[S_1, U_1]$. We define a $\delta$ which has exactly the sharing groups $B_1, \ldots, B_n$ and which is obtained by instantiating $\theta$. The idea is that if $B_1 \mathcal{R}_{\theta X} B_2$ due to $x_1 \in B_1$, $x_2 \in B_2$ and the common variable $y \in \theta(x_1) \cap \theta(x_2)$, then the occurrences of $y$ in $\theta(x_1)$ and $\theta(x_2)$ are replaced by two suitable terms which unify and merge together the two sharing groups $B_1$ and $B_2$.

*Example 6.13*
Let $\theta = \{x/f(u), y/g(u)\}$ and $[S_1, U_1] = [\{\mathtt{xw}, \mathtt{yz}\}, \{w, x, y, z\}]$. Consider $B_1 = \mathtt{xw}$ and $B_2 = \mathtt{yz}$. We choose the variables $x \in B_1$ and $y \in B_2$. Since $u \in \theta(x) \cap \theta(y)$, we can choose the substitution $\delta = \{x/f(w_1), y/g(w_2), w/w_1, z/w_2\}$ obtained from $\theta$ by replacing each occurrence of $u, w, z$ with suitable new terms. It is easy to verify that $\theta$ and $\delta$ unify and that $\alpha_{\mathrm{Sh}}(\mathrm{mgu}([\delta]_{\{w,x,y,z\}}, [\theta]_{\{u,w,x,y,z\}})) \sqsupseteq_{\mathrm{Sh}} [\{\mathtt{uwxyz}\}, \{u, w, x, y, z\}]$. □

*Example 6.14*
Let $\theta = \{x/f(u, u)\}$ and $[S_1, U_1] = [\{\mathtt{xw}, \mathtt{xy}, \mathtt{xz}\}, \{w, x, y, z\}]$. Consider $B_1 = \mathtt{xw}$, $B_2 = \mathtt{xy}$ and $B_3 = \mathtt{xz}$. We choose the variable $x \in B_1 \cap B_2 \cap B_3$. Then $u \notin \mathrm{uvars}(\theta(x))$, and we can choose as $\delta$ the substitution

$$\{x/f(t(w_1, w_1), t(w_2, w_3)), w/w_1, y/w_2, z/w_3\} \ ,$$

obtained from $\theta$ by replacing each occurrence of $u, w, y, z$ with suitable new terms. It is easy to see that $\theta$ and $\delta$ unify and that $\alpha_{\mathrm{Sh}}(\mathrm{mgu}([\delta]_{\{w,x,y,z\}}, [\theta]_{\{u,w,x,y,z\}})) \sqsupseteq_{\mathrm{Sh}} [\{\mathtt{uwxyz}\}, \{u, w, x, y, z\}]$. □

Following this idea we can now prove that mgu and $\mathsf{unif}_{\mathrm{Sh}}$ are optimal.

*Proposition 6.15*
For all $X \in S$ there exists $[\delta]_{U_1} \in \mathit{ISubst}_{\sim}$ such that $\alpha_{\mathrm{Sh}}([\delta]_{U_1} \leq_{\mathrm{Sh}} [S_1, U_1]$ and

$$\alpha_{\mathrm{Sh}}(\mathrm{mgu}([\delta]_{U_1}, [\theta]_U)) \sqsupseteq_{\mathrm{Sh}} [\{X\}, U] \ .$$

*Proof*
The proof can be found in the Appendix as Prop. C.6 □

The optimality result for $\mathsf{unif}_{\mathrm{Sh}}$ w.r.t. $\mathsf{unif}_{\mathrm{Ps}}$ immediately follows from the above proposition.

*Theorem 6.16*
(Optimality of $\mathsf{unif}_{\mathrm{Sh}}$) $\mathsf{unif}_{\mathrm{Sh}}$ is optimal w.r.t. $\mathsf{unif}_{\mathrm{Ps}}$.

Optimality of $\mathsf{unif}_{\mathrm{Sh}}$ also implies the following corollary:

*Corollary 6.17*
The result of $\mathsf{unif}_{\mathrm{Sh}}$ does not depend on the order of the bindings in its second argument.

### *6.4 Summing Up*

We may put together all the results of correctness, optimality and completeness shown so far to prove the main theorem of this section.

*Theorem 6.18*
$\mathbf{U}^f_{\mathrm{Sh}}$ is well defined, correct and optimal w.r.t. $\mathbf{U}^f_{\mathrm{Ps}}$.

*Proof*
The proof can be found in the Appendix as Theorem C.7 □

Generally speaking, in order to obtain optimality, it is always a better choice to abstract a concrete operator "as a whole", instead of abstracting each component and then composing the abstract operators. According to this rule, we could think that a better approximation may be reached by abstracting $\mathbf{U}^f_{\mathrm{Ps}}$ as a whole. However, since abstract projection/renaming is complete and $\gamma$-complete, this does not happen, as shown by the previous theorem. Studying the direct abstraction of this composition would still be useful to find a direct implementation which is more efficient than computing $\mathsf{unif}_{\mathrm{Sh}}$ and projecting later, but we do not consider this problem here.

Since $\mathbf{U}^f_{\mathrm{Sh}}$ generates less sharing groups then $\mathbf{U'}^f_{\mathrm{Sh}}$ and since checking whether a variable is in $U$ is easy, we can expect an improvement in the efficiency of the analysis by replacing $\mathbf{U'}^f_{\mathrm{Sh}}$ with $\mathbf{U}^f_{\mathrm{Sh}}$ in the computation of the entry substitution. If computing $Y$ and $Z$ at each step of $\mathbf{u}^f_{\mathrm{Sh}}$ seems difficult, it is always possible to precompute these values before the actual analysis begins, since they depend on the syntax of the program only. Moreover, in the definition of $\mathbf{u}^f_{\mathrm{Sh}}$, when $x \in U$ we know that $\mathbf{rel}(S, x) = \{\{x\}\}$, since $\theta$ is an idempotent substitution and $x \notin U_1$.

A further optimization is obtained by replacing $\mathbf{rel}(S, Y)$ with the set of all the sharing groups whose variables are all contained in $Y$. Clearly, this is a subset of $\mathbf{rel}(S, Y)$ and it is immediate to check that the result of $\mathbf{u}^f_{\mathrm{Sh}}$ does not change. In fact, all the sharing groups in $\mathbf{bin}(\mathbf{rel}(S, x), \mathbf{rel}(S, Y)^*)$ which are not generated anymore, may be found in $\mathbf{bin}(\mathbf{rel}(S, x)^*, \mathbf{rel}(S, Z)^*)$.

We said before that this operator introduces new optimizations which, to the best of our knowledge, are not used even in more complex domains for sharing analysis which include linearity and freeness information. We give here one example which shows their effects.

*Example 6.19*
Let us consider the following unification.

$$\mathbf{U}^f_{\mathrm{Sh}}([\{\mathtt{xw}, \mathtt{xz}, \mathtt{yw}, \mathtt{yz}\}, \{x, y, w, z\}], p(x, y, w, z), p(f(u, h), f(u, k), s, t)) \ .$$

By applying the optimizations suggested from the unification algorithm in presence of linearity and freeness information in (Hans and Winkler 1992), we may start from the abstract object $S = \{\mathtt{xw}, \mathtt{xz}, \mathtt{yw}, \mathtt{yz}, \mathtt{u}, \mathtt{h}, \mathtt{k}, \mathtt{s}, \mathtt{t}\}$ and process the bindings one at a time, keeping in mind that $u, h, k, s, t$ are initially free. This means that in the binding $x/f(u, h)$, the term $f(u, h)$ is linear, and therefore we can avoid to compute the star union in $\mathbf{rel}(S, x)$, thus obtaining:

$$\{\mathtt{k}, \mathtt{s}, \mathtt{t}, \mathtt{yw}, \mathtt{yz}\} \cup \mathbf{bin}(\{\mathtt{xw}, \mathtt{xz}\}, \{\mathtt{u}, \mathtt{h}, \mathtt{uh}\}) = \\ \{\mathtt{k}, \mathtt{s}, \mathtt{t}, \mathtt{yw}, \mathtt{yz}, \mathtt{xwu}, \mathtt{xwh}, \mathtt{xzu}, \mathtt{xzh}, \mathtt{xwuh}, \mathtt{xzuh}\} \ .$$

However, after this unification, the variable $u$ can be bound to a non-linear term. Therefore, when we consider the next binding $y/f(u, k)$, according to (Hans and Winkler 1992), we are forced to compute all the star unions, obtaining

$$\{\mathtt{s}, \mathtt{t}\} \cup \mathbf{bin}(\{\mathtt{yw}, \mathtt{yz}\}^*, (\{\mathtt{k}\} \cup \mathbf{bin}(\{\mathtt{xw}, \mathtt{xz}\}, \{\mathtt{u}, \mathtt{uh}\}))^*) \cup \{\mathtt{xwh}, \mathtt{xzh}\} \ .$$

Finally, in the bindings $w/s$ and $z/t$ we may omit all the star unions since $t$ and $s$ are free, and we get the final result

$$\mathbf{bin}(\{\mathtt{yws}, \mathtt{yzt}\}^*, (\{\mathtt{k}\} \cup \mathbf{bin}(\{\mathtt{xws}, \mathtt{xzt}\}, \{\mathtt{u}, \mathtt{uh}\}))^*) \cup \{\mathtt{xwsh}, \mathtt{xzth}\} \ .$$

Observe that we obtain the sharing group $\mathtt{ywsztk}$, and thus, after projecting on $\{u, h, k, s, t\}$, we obtain the sharing group $\mathtt{stk}$. However, when we consider the second binding, we know that $k$ is free and independent from $y$, and this is enough to apply a new optimization. In fact, $k$ can share with more than one sharing group related to $y$ only if $k$ shares with $u$. If we compute the abstract unification with our algorithm, we obtain

$$\{\mathtt{ywsk}, \mathtt{yztk}\} \cup \mathbf{bin}(\{\mathtt{yws}, \mathtt{yzt}\}^*, \mathbf{bin}(\{\mathtt{xws}, \mathtt{xzt}\}, \{\mathtt{u}, \mathtt{uh}\})^*) \\ \cup \mathbf{bin}(\mathbf{bin}(\{\mathtt{yws}, \mathtt{yzt}\}^*, \mathbf{bin}(\{\mathtt{xws}, \mathtt{xzt}\}, \{\mathtt{u}, \mathtt{uh}\})^*), \{k\}) \cup \{\mathtt{xwsh}, \mathtt{xzth}\}$$

and when we project on $\{u, h, k, s, t\}$, the sharing group $\mathtt{stk}$ does not appear. In fact, note that any sharing group generated by

$$\mathbf{bin}(\mathbf{bin}(\{\mathtt{yws}, \mathtt{yzt}\}^*, \mathbf{bin}(\{\mathtt{xws}, \mathtt{xzt}\}, \{\mathtt{u}, \mathtt{uh}\})^*), \{k\})$$

contains the variable $u$. The result does not change by permuting the order of the bindings. If we consider the binding $y/f(u, k)$ before $x/f(u, h)$, with the standard operators we get:

$$\mathbf{bin}(\{\mathtt{xws}, \mathtt{xzt}\}^*, (\{\mathtt{h}\} \cup \mathbf{bin}(\{\mathtt{yws}, \mathtt{yzt}\}, \{\mathtt{u}, \mathtt{uk}\}))^*) \cup \{\mathtt{ywsk}, \mathtt{yztk}\}$$

and, when we project on $\{u, h, k, s, t\}$, we obtain the sharing group $\mathtt{sth}$, which does not appear in our result. □

## 7 Matching and Backward Unification

To the best of our knowledge, in all the collecting denotational semantics for logic programs, backward unification is performed by using unification instead of matching. This means that, instead of $\mathbf{U}^b_{\mathrm{Ps}}$, the concrete semantics uses a backward unification operator which unifies two concrete objects in `Psub` with a substitution:

$$\mathbf{U}'^{b}_{\mathrm{Ps}}([\Delta_1, U_1], [\Delta_2, U_2], A_1, A_2) = \\ \pi_{\mathrm{Ps}}(\mathsf{unif}''_{\mathrm{Ps}}(\rho([\Delta_1, U_1]), [\Delta_2, U_2], \mathrm{mgu}(\rho(A_1) = A_2)), U_2 \cup \mathrm{vars}(A_2)) \ , \quad (33)$$

where $\rho$ is a renaming such that $\rho(U_1 \cup \mathrm{vars}(A_1)) \cap (U_2 \cup \mathrm{vars}(A_2)) = \emptyset$ and

$$\mathsf{unif}''_{\mathrm{Ps}}([\Delta_1, U_1], [\Delta_2, U_2], \delta) = \\ [\{\mathrm{mgu}([\theta_1]_{U_1}, [\theta_2]_{U_2}, [\delta]_{\mathrm{vars}(\delta)}) \mid [\theta_1]_{U_1} \in \Delta_1, [\theta_2]_{U_2} \in \Delta_2\}, U_1 \cup U_2] \quad (34)$$

is simply the pointwise extension of mgu over `Psub`. It is worth observing that $\mathsf{unif}''_{\mathrm{Ps}}(\rho([\Delta_1, U_1]), [\Delta_2, U_2], \delta)$ is a very specific kind of unification, since $\rho(U_1)$ and $U_2$ are disjoint. The optimal abstract operator $\mathbf{U}'^{b}_{\mathrm{Sh}}$ w.r.t. $\mathbf{U}'^{b}_{\mathrm{Ps}}$ is very similar to that proposed in (Cortesi and Filé 1999) (see Section 8.2 for further details), and it is given by:

$$\mathbf{U}'^{b}_{\mathrm{Sh}}([S_1, U_1], [S_2, U_2], A_1, A_2) = \\ \pi_{\mathrm{Sh}}(\mathsf{unif}_{\mathrm{Sh}}([\rho(S_1) \cup S_2, \rho(U_1) \cup U_2], \mathrm{mgu}(\rho(A_1) = A_2)), U_2 \cup \mathrm{vars}(A_2)) \ . \quad (35)$$

As said before, this choice results in a loss of precision already at the concrete level, which leads to a loss of precision in the abstract counterpart. When we compute $\mathbf{U}'^{b}_{\mathrm{Ps}}([\Delta_1, U_1], [\Delta_2, U_2], A_1, A_2)$, we essentially unify all pairs $\theta_1$ and $\theta_2$, elements of $\Delta_1$ and $\Delta_2$, with $\delta = \mathrm{mgu}(A_1 = A_2)$ (assuming we do not need renamings). However, it could be possible to consider only the pairs in which $\theta_1$ is an instance of $\mathrm{mgu}(\theta_2, \delta)$ w.r.t. the variables of interest in $U_1 \cap U_2$. If this does not hold, then $\theta_1$ cannot be a success substitution corresponding to the call substitution $\theta_2$, and therefore we are unifying two objects which pertain to different computational paths, with an obvious loss of precision, already at the concrete level. This problem has been pointed out by Marriott et al. (1994, Section 5.5).

We now want to define the optimal abstract operator $\mathbf{U}^{b}_{\mathrm{Sh}}$ corresponding to $\mathbf{U}^{b}_{\mathrm{Ps}}$. This is accomplished by composing the forward unification operator $\mathsf{unif}_{\mathrm{Sh}}$ with a new operator $\mathsf{match}_{\mathrm{Sh}}$, which is the abstract counterpart of $\mathsf{match}_{\mathrm{Ps}}$.

*Definition 7.1*
Given $[S_1, U_1], [S_2, U_2] \in$ `Sharing`, we define

$$\mathsf{match}_{\mathrm{Sh}}([S_1, U_1], [S_2, U_2]) = \\ [S'_1 \cup S'_2 \cup \{X_1 \cup X_2 \mid X_1 \in S''_1, X_2 \in (S''_2)^*, X_1 \cap U_2 = X_2 \cap U_1\}, U_1 \cup U_2]$$

where $S'_1 = \{B \in S_1 \mid B \cap U_2 = \emptyset\}$ and $S''_1 = S_1 \setminus S'_1$, $S'_2 = \{B \in S_2 \mid B \cap U_1 = \emptyset\}$ and $S''_2 = S_2 \setminus S'_2$

The idea is that we may freely combine those sharing groups in $S_2$ that have some variable in common with $U_1$, i.e., $X_2 \in (S''_2)^*$, if the projection of the result on $U_1$ is equal to some sharing group in $S_1$, when projected on $U_2$. This means that new aliasings between variables may arise in the concrete counterpart of $S_2$ (the entry substitution), as long as they do not affect the variables of the exit substitution.

*Definition 7.2*
The abstract backward unification may be defined as

$$\mathbf{U}^b_{\mathrm{Sh}}([S_1,U_1],[S_2,U_2],A_1,A_2) = \pi_{\mathrm{Sh}}(\mathsf{match}_{\mathrm{Sh}}(\rho([S_1,U_1]), \mathsf{unif}_{\mathrm{Sh}}([S_2,U_2],\mathrm{mgu}(\rho(A_1)=A_2))), U_2 \cup \mathrm{vars}(A_2)) \ . \quad (36)$$

where $\rho$ is a renaming such that $\rho(U_1 \cup \mathrm{vars}(A_1)) \cap (U_2 \cup \mathrm{vars}(A_2)) = \emptyset$.

*Example 7.3*
Let $U_1 = \{u,v,w\}$, $U_2 = \{x,y,z\}$, $\Theta_1 = \{[\{v/t(u,w,w)\}]_{U_1}, [\{v/t(u,u,w)\}]_{U_1}\}$. $\Theta_2 = \{[\{y/t(x,z,z)\}]_{U_2}, [\{y/t(x,x,z)\}]_{U_2}\}$ and $\rho = id$. We have

$$\mathbf{U'}^b_{\mathrm{Ps}}([\Theta_1,U_1],[\Theta_2,U_2],p(u,v,w),p(x,y,z)) = \pi_{\mathrm{Ps}}([\Theta,U_1\cup U_2],U_2) \ ,$$

with $[\theta]_{U_1\cup U_2} = [\{y/t(x,x,x), z/x, u/x, v/t(x,x,x), w/x\}]_{U_1\cup U_2} \in \Theta$. Let $[S_1,U_1] = \alpha_{\mathrm{Sh}}([\Theta_1,U_1])$, $[S_2,U_2] = \alpha_{\mathrm{Sh}}([\Theta_2,U_2])$, $S_1 = \{\mathtt{uv},\mathtt{vw}\}$ and $S_2 = \{\mathtt{xy},\mathtt{yz}\}$. We obtain

$$\mathbf{U'}^b_{\mathrm{Sh}}([S_1,U_1],[S_2,U_2],p(u,v,w),p(x,y,z)) = \pi_{\mathrm{Sh}}([S,U_1\cup U_2],U_2) \ ,$$

and $\mathtt{xyzuvw} \in S$. So, it seems that $u, v$ and $w$ may share a common variable. Note that $\theta$ is obtained by unifying $\sigma_2 = \{y/t(x,z,z)\}$ with $\sigma_1 = \{v/t(u,u,w)\}$ but $\sigma_1(v) = t(u,u,w)$ is not an instance of $\mathrm{mgu}(\sigma_2, \mathrm{mgu}(p(x,y,z) = p(u,v,w)))(v) = t(x,z,z)$. Therefore, $\sigma_1$ and $\sigma_2$ do pertain to different computational paths. Using the backward unification with matching, we obtain

$$\begin{aligned}&\mathbf{U}^b_{\mathrm{Ps}}([\Theta_1,U_1],[\Theta_2,U_2],p(u,v,w),p(x,y,z)) = \\ &\quad \pi_{\mathrm{Ps}}([\{[y/t(x,z,z),u/x,v/t(x,z,z),w/z],[y/t(x,x,z),u/x,v/t(x,x,z),w/z]\}, \\ &\qquad \{x,y,z,u,v,w\}],\{u,v,w\}) \ ,\end{aligned}$$

which does not contain $\theta$. In the abstract domain, we have:

$$\mathbf{U}^b_{\mathrm{Sh}}([S_1,U_1],[S_2,U_2],p(u,v,w),p(x,y,z)) = \pi_{\mathrm{Sh}}([\{\mathtt{xyuv},\mathtt{yzvw}\},U_1\cup U_2],U_2) \ .$$

After the unification we know that $x$ and $z$ are independent. Note that the abstract matching operators defined in (King and Longley 1995; Hans and Winkler 1992), cannot establishthis property. The algorithm in (Muthukumar and Hermenegildo 1992) computes the same result of ours in this particular example, but since their matching is partially performed by first projecting the sharing information on the term positions of the calling atom and of the clause head, this does not hold in general. For example, their algorithm states that $x$ and $z$ may possibly share when the unification is performed between the calling atom $p(t(x,y,z))$ and the head $p(t(u,v,w))$, where $t$ is a function symbol, $p$ a unary predicate and the call substitution is the same as before. □

### *7.1 Correctness and Optimality*

We can prove that $\mathbf{U}^b_{\mathrm{Sh}}$ is actually the best correct abstraction of the backward concrete unification $\mathbf{U}^b_{\mathrm{Ps}}$. To prove correctness we only need to show that $\mathsf{match}_{\mathrm{Sh}}$ is correct w.r.t. $\mathsf{match}_{\mathrm{Ps}}$. Correctness of $\mathbf{U}^b_{\mathrm{Sh}}$ will follow from the fact that $\mathbf{U}^b_{\mathrm{Sh}}$ is a composition of correct abstract operators.

*Theorem 7.4*
(CORRECTNESS OF $\mathsf{match}_{\mathrm{Sh}}$) $\mathsf{match}_{\mathrm{Sh}}$ is correct w.r.t. $\mathsf{match}_{\mathrm{Ps}}$.

*Proof*
The proof can be found in the Appendix as Theorem D.1. □

However, the composition of optimal operators may fail to be optimal. Therefore, optimality of $\mathsf{match}_{\mathrm{Sh}}$ does not guarantee optimality of $\mathbf{U}^b_{\mathrm{Sh}}$. In order to prove the optimality result, we need to establish two additional properties on the abstract operators $\mathsf{match}_{\mathrm{Sh}}$ and $\mathsf{unif}_{\mathrm{Sh}}$. The idea is that both these operators are used in a very specific way in the backward unification.

*Proposition 7.5*

1. $\mathsf{match}_{\mathrm{Sh}}$ is optimal w.r.t. $\mathsf{match}_{\mathrm{Ps}}$;
2. when $\mathsf{match}_{\mathrm{Ps}}$ is restricted to the case when the second argument contains a single substitution, then $\mathsf{match}_{\mathrm{Sh}}$ is complete w.r.t. the second argument, i.e.

$$\mathsf{match}_{\mathrm{Sh}}([S_1, U_1], \alpha_{\mathrm{Sh}}([\{[\sigma_2]\}, U_2])) = \alpha_{\mathrm{Sh}}(\mathsf{match}_{\mathrm{Ps}}(\gamma_{\mathrm{Sh}}([S_1, U_1]), [\{[\sigma_2]\}, U_2]))$$

3. $\mathsf{unif}_{\mathrm{Sh}}$ is optimal in a very strong way: given $[S_1, U_1] \in$ `Sharing` and $\theta \in$ *ISubst*, there exists a substitution $\delta \in$ *ISubst* such that $\alpha_{\mathrm{Sh}}([\delta]_{U_1}) \sqsubseteq_{\mathrm{Sh}} [S_1, U_1]$ and

$$\alpha_{\mathrm{Sh}}(\mathsf{unif}_{\mathrm{Ps}}([\{[\delta]\}, U_1], \theta)) = \mathsf{unif}_{\mathrm{Sh}}([S_1, U_1], \theta) \ .$$

*Proof*
Proofs of these properties can be found in the Appendix as Theorems D.2, D.3 and D.4. □

On the last point, note that the standard definition of optimality for $\mathsf{unif}_{\mathrm{Sh}}$ only assures the existence of a set of substitutions $\Delta$ such that $\alpha_{\mathrm{Sh}}([\Delta, U_1]) \sqsubseteq_{\mathrm{Sh}} [S_1, U_1]$ and $\alpha_{\mathrm{Sh}}(\mathsf{unif}_{\mathrm{Ps}}([\Delta, U_1], \theta)) = \mathsf{unif}_{\mathrm{Sh}}([S_1, U_1], \theta)$. However, we show that any set $\Delta$ can be reduced to a singleton. This allows us to find a single substitution to be used for proving the optimality result for all the resultant sharing groups. Finally, using Theorem 7.4 and Prop. 7.5 we may prove the expected result.

*Theorem 7.6*
$\mathbf{U}^b_{\mathrm{Sh}}$ is correct and optimal w.r.t. $\mathbf{U}^b_{\mathrm{Ps}}$.

*Proof*
The proof can be found in the Appendix as Theorem D.5. □

To the best of our knowledge, this is the first abstract matching operator which is optimal for the corresponding concrete operator. We now give an example of a program where the use of $\mathbf{U}^f_{\mathrm{Sh}}$ and $\mathbf{U}^b_{\mathrm{Sh}}$ gives better results than the standard operators $\mathbf{U'}^f_{\mathrm{Sh}}$ and $\mathbf{U'}^b_{\mathrm{Sh}}$.

*Example 7.7*
We keep on Examples 4.4, 6.2 and 6.5 and consider the trivial program with just one clause $\mathtt{p(u,v,w)}$ and the goal $p(x,y,z)$ with $\{\mathtt{xy},\mathtt{yz}\}$. Using our abstract operators, we obtain the entry substitution $\{\mathtt{uv},\mathtt{vw}\}$ and the success substitution $\{\mathtt{xy},\mathtt{yz}\}$ (see Ex. 6.5 and 7.3), thus proving that $x$ and $z$ are independent.

We now compute the abstract semantics of the goal $p(x,y,z)$ with $\{\mathtt{xy},\mathtt{yz}\}$. From Example 4.4, we have that the abstract semantics of $P$ is

$$\lambda A.\lambda\chi.\mathbf{U}^b_{\mathrm{Sh}}(\mathbf{U}^f_{\mathrm{Sh}}(\chi,A,p(u,v,w)),\chi,p(u,v,w),A) \enspace .$$

Thus, in order to compute the semantics of the goal $p(x,y,z)$ with $\{\mathtt{xy},\mathtt{yz}\}$, we need to compute

$$\mathbf{U}^b_{\mathrm{Sh}}(\mathbf{U}^f_{\mathrm{Sh}}([\{\mathtt{xy},\mathtt{yz}\},\{x,y,x\}],p(x,y,z),p(u,v,w)), \\ [\{\mathtt{xy},\mathtt{yz}\},\{x,y,x\}],p(u,v,w),p(x,y,z)) \enspace .$$

From Example 6.5, we know that

$$\mathbf{U}^f_{\mathrm{Sh}}([\{\mathtt{xy},\mathtt{yz}\},\{x,y,x\}],p(x,y,z),p(u,v,w)) = [\{\mathtt{uv},\mathtt{vw}\},\{u,v,w\}] \enspace ,$$

from which we obtain (see Example 7.3):

$$\mathbf{U}^b_{\mathrm{Sh}}([\{\mathtt{uv},\mathtt{vw}\},\{u,v,w\}],[\{\mathtt{xy},\mathtt{yz}\},\{x,y,x\}],p(u,v,w),p(x,y,z)) = \\ [\{\mathtt{xy},\mathtt{yz}\},\{x,y,z\}] \enspace ,$$

which shows that $x$ and $y$ are independent.

If we replace either $\mathbf{U}^b_{\mathrm{Sh}}$ or $\mathbf{U}^f_{\mathrm{Sh}}$ with $\mathbf{U'}^f_{\mathrm{Sh}}$ or $\mathbf{U'}^b_{\mathrm{Sh}}$, then the success substitution will contain the sharing group $\mathtt{xyz}$. In fact, as shown in Ex. 6.2, the entry substitution in the latter case would be $[\{\mathtt{uv},\mathtt{vw},\mathtt{uvw}\},\{u,v,w\}]$. If we compute the success substitution we obtain:

$$\mathbf{U'}^b_{\mathrm{Sh}}([\{\mathtt{uv},\mathtt{vw},\mathtt{uvw}\},\{u,v,w\}],[\{\mathtt{xy},\mathtt{yz}\},\{x,y,z\}],p(u,v,w),p(x,y,z)),\{x,y,z\}) \\ = [\{\mathtt{xy},\mathtt{yz},\mathtt{xyz}\},\{x,y,z\}] \enspace ,$$

which contains the sharing group $\mathtt{xyz}$. □

### *7.2 Programs in Head Normal Form*

It is worth noting that the improvement in the previous example is obtained with a program in *head normal form*. Usually, when programs are in head normal form, the forward and backward unification may be replaced by renamings, which are complete and do not cause any loss in precision. However, there is the need of an unification operator for the explicit constraints which appear in the body of the clauses. In general, the analyses we obtain in our framework are more precise than those which can be obtained by using the standard domain `Sharing` by translating the same program to the head normal form.

*Example 7.8*

Consider again Ex. 7.7 and the program $\mathtt{p(u, f(s), w)} \leftarrow$ which is not in head normal form. Using our abstract operators, we obtain the success substitution $\{\mathtt{xy}, \mathtt{yz}\}$, as in Ex. 7.7. If we normalize the program, we obtain the clause $\mathtt{p(u, v, w)} \leftarrow \mathtt{v} = \mathtt{f(s)}$. The entry substitution obtained from $\{\mathtt{xy}, \mathtt{yz}\}$ by simply renaming the variables $x, y, z$ to $u, v, w$ and introducing the new variable $s$ is $\{\mathtt{uv}, \mathtt{vw}, \mathtt{s}\}$. By using the standard operator for unification, when applying the binding $v/f(s)$ we obtain $\{\mathtt{uvs}, \mathtt{vws}, \mathtt{uvws}\}$, and thus the success substitution will contain the sharing group $\mathtt{xyz}$, resulting in a loss of precision. □

It is possible to use our forward abstract unification in a normalized program by enlarging the set of variables of interest only when new variables are effectively met, instead of adding all the variables which appear in the body of a clause once for all when the entry substitution is computed. In the example above, the variable $s$ can be introduced when unifying the abstract object $\{\mathtt{uv}, \mathtt{vw}\}$ with $v/f(s)$. Since $\mathsf{unif}_{\mathrm{Sh}}([\{\mathtt{uv}, \mathtt{vw}\}, \{u, v, w\}], \{v/f(s)\}) = [\{\mathtt{uvs}, \mathtt{vws}\}, \{u, v, w, s\}]$, we still obtain as success substitution $\{\mathtt{xy}, \mathtt{yz}\}$, thus proving that $x$ and $z$ are independent.

In the general case, translating a program in head normal form will negatively affect the precision of the analysis. To achieve the same precision in both cases, we need to add structural information to the abstract domain (Le Charlier and Van Hentenryck 1994).

## 8 Related Works

### *8.1 Relationship with ESubst*

The domain *ESubst* proposed by Jacobs and Langen (1992) uses a non standard definition of substitution. We may prove that *ESubst* is isomorphic to $\mathit{ISubst}_\sim$. This formalizes the intuition, which has never been proved before, that working with *ESubst* is essentially like working with substitutions. Similar proofs may be developed for ex-equations (Marriott et al. 1994) and existential Herbrand constraints (Levi and Spoto 2003).

We now briefly recall the definition of the domain *ESubst*. For the sake of clarity, in the following, we call E-substitution the nonstandard substitution defined in (Jacobs and Langen 1992). An E-substitution $\sigma$ is a mapping from a finite set of variables $\mathrm{dom}(\sigma) \subseteq \mathcal{V}$ to $\mathsf{Terms}$. This approach differs from the standard definition of substitutions, which are mappings from $\mathcal{V}$ to $\mathsf{Terms}$ that are almost everywhere the identity. The preorder on E-substitutions is defined as follows:

$$\sigma \leq_E \theta \iff \mathrm{dom}(\theta) \subseteq \mathrm{dom}(\sigma) \wedge \big(\forall t \in \mathsf{Terms}.\ \mathrm{vars}(t) \subseteq \mathrm{dom}(\theta) \Rightarrow \exists \delta \text{ an E-substitution s.t. } \sigma t = \delta(\theta(t))\big) \ , \quad (37)$$

where the application of an E-substitution to a term is defined as usual.

Let $\sim_E$ be the equivalence relation on E-substitutions induced by $\leq_E$. The domain *ESubst* is defined as the set of equivalence classes of E-substitutions w.r.t. $\sim_E$, that is $\mathit{ESubst} = \{[\sigma]_{\sim_E} \mid \sigma \text{ is an E-substitution}\}$. The next theorem shows that *ESubst* is isomorphic to $\mathit{Subst}_\sim$ which, as shown in Prop. 3.7, is isomorphic to $\mathit{ISubst}_\sim$.

*Theorem 8.1*
*ESubst* and $Subst_{\sim}$ are isomorphic posets.

*Proof*
To each E-substitution $\theta$ we may associate a substitution $\theta'$ such that $\theta'(x) = \theta(x)$ if $x \in \mathrm{dom}(\theta)$ and $\theta'(x) = x$ otherwise. Note that, for each term $t$, $\theta(t) = \theta'(t)$: an E-substitution and the corresponding standard substitution behave in the same way on terms.

We may prove that, if $\theta_1 \leq_E \theta_2$, then $\theta'_1 \preceq_{\mathrm{dom}(\theta_2)} \theta'_2$. By definition, if $\theta_1 \leq_E \theta_2$ then $\mathrm{dom}(\theta_2) \subseteq \mathrm{dom}(\theta_1)$ and $\forall t \in \mathsf{Terms}$ with $\mathrm{vars}(t) \subseteq \mathrm{dom}(\theta_2)$, there exists an E-substitution $\delta$ such that $\theta_1(t) = \delta(\theta_2(t))$. Let $\mathrm{dom}(\theta_2) = \{x_1, \ldots, x_n\}$ and consider a term $t$ such that $\mathrm{vars}(t) = \{x_1, \ldots, x_n\}$ (note that $t$ exists iff there is at least a term symbol of arity strictly greater than 1). By definition, there exists an E-substitution $\delta$ such that $\theta_1(t) = \delta(\theta_2(t))$, that is, for any $v \in \mathrm{dom}(\theta_2)$ it holds $\theta_1(v) = \delta(\theta_2(v))$. This means that $\theta'_1(v) = \delta'(\theta'_2(v))$ and therefore $\theta'_1 \preceq_{\mathrm{dom}(\theta_2)} \theta'_2$.

On the converse, for each $\theta \in \mathit{Subst}$ and $U \in \wp_f(\mathcal{V})$, we associate a corresponding E-substitution $\theta^{*U}$ such that $\mathrm{dom}(\theta^{*U}) = U$ and $\theta^{*U}(v) = \theta(v)$ for each $v \in U$. As for the previous case, we have that if $\theta_1 \preceq_U \theta_2$, then $\theta_1^{*U} \leq_E \theta_2^{*U}$. First of all, note that $\mathrm{dom}(\theta_1^{*U}) = U = \mathrm{dom}(\theta_2^{*U})$. Moreover, by definition of $\preceq_U$, there is $\delta \in \mathit{Subst}$ such that $\theta_1(v) = \delta(\theta_2(v))$ for each $v \in U$. Now, given a term $t$ such that $\mathrm{vars}(t) \subseteq U$, we may check that $\theta_1^{*U}(t) = \delta^{*\mathrm{vars}(\theta_2(U))}(\theta_2^{*U}(t))$ and this proves $\theta_1^{*U} \leq_E \theta_2^{*U}$.

Now, we may lift these operations to equivalence classes to obtain the function $\iota : \mathit{ESubst} \rightarrow \mathit{Subst}_{\sim}$ such that

$$\iota([\theta]_{\sim_E}) = [\theta']_{\mathrm{dom}(\theta)} \quad .$$

The map $\iota$ is well defined: if $\theta_1 \sim_E \theta_2$ then $\mathrm{dom}(\theta_1) = \mathrm{dom}(\theta_2)$ and, by the above property, $\theta'_1 \sim_{\mathrm{dom}(\theta_2)} \theta'_2$. Moreover, there is an inverse $\iota^{-1}$ given by

$$\iota^{-1}([\theta]_U) = [\theta^{*U}]_{\sim_E} \quad .$$

It is easy to check that $\iota^{-1}$ is well defined: if $\theta_1 \preceq_U \theta_2$, then $\theta_1^{*U} \leq_E \theta_2^{*U}$.

It is immediate to check, given the properties above, that $\iota$ and $\iota^{-1}$ are one the inverse of the other. Moreover, they are both monotonic. If $[\theta_1]_E \leq_E [\theta_2]_E$ then $\mathrm{dom}(\theta_2) \subseteq \mathrm{dom}(\theta_1)$ and $\theta'_1 \preceq_{\mathrm{dom}(\theta_2)} \theta'_2$, i.e., $\iota([\theta_1]_{\sim_E}) = [\theta'_1]_{\mathrm{dom}(\theta_1)} \preceq [\theta'_2]_{\mathrm{dom}(\theta_2)} = \iota([\theta_2]_{\sim_E})$. On the converse, if $[\theta_1]_U \preceq [\theta_2]_V$ then $[\theta_1]_V \preceq [\theta_2]_V$ and therefore $\iota^{-1}([\theta_1]_V) \leq_E \iota^{-1}([\theta_2]_V)$. We only need to prove that $\iota^{-1}([\theta_1]_U) \leq_E \iota^{-1}([\theta_1])_V$. This follows from that fact that, given a term $t$ with $\mathrm{vars}(t) \subseteq V$, $\theta_1^{*U}(t) = \theta_1^{*V}(t)$.
□

It is worth noting that the most general unifier as defined in (Jacobs and Langen 1992) corresponds to mgu in $\mathit{ISubst}_{\sim}$. In formulas, given term $t_1$ and $t_2$, we have that

$$\iota([\mathrm{mgu}(t_1, t_2)]_{\sim_E}) = [\mathrm{mgu}(\{t_1 = t_2\})]_{\mathrm{vars}(t_1 = t_2)} \quad , \tag{38}$$

where mgu on the left is the operator in Definition 1 of (Jacobs and Langen 1992) and $\iota : \mathit{ESubst} \rightarrow \mathit{ISubst}_{\sim}$ is the isomorphism defined in the proof of Theorem 8.1.

To the best of our knowledge, this is the first proof of the relationship between the mgu in a domain of existential substitutions and the standard mgu for substitutions. Moreover, it is worth noting that by adding a bottom element to $\mathit{ISubst}_\sim$ and *ESubst*, they turn out to be isomorphic complete lattices.

### *8.2 A Case Study*

In Section 3 we said that, in order to define a good collecting semantics for correct answer substitutions, there are several possible directions. We may work with a domain of existentially quantified substitutions like $\mathit{ISubst}_\sim$, or we may work with standard substitutions, being careful to keep enough representatives for each equivalence class. We have already discussed the benefits of using equivalence classes. In order to show the kind of problems which arise from the use of domains of substitutions, without any equivalence relation, we want to show a small flaw of the semantic framework defined in (Cortesi and Filé 1999) for the analysis of sharing, and widely used in several other works on program analysis such as (Bagnara et al. 2002; Hill et al. 2004).

The framework is based upon the domain $\mathtt{Rsub} = (\wp(\mathit{Subst}) \times \wp_f(\mathcal{V})) \cup \{\top_{\mathrm{Rs}}, \bot_{\mathrm{Rs}}\}$ which is a complete lattice, partially ordered as follows: $\top_{\mathrm{Rs}}$ is the top element, $\bot_{\mathrm{Rs}}$ is the bottom element and $[\Theta_1, U_1] \sqsubseteq_{\mathrm{Rs}} [\Theta_2, U_2]$ if and only if $U_1 = U_2$ and $\Theta_1 \subseteq \Theta_2$. An object $[\Theta, U]$ is a set of substitution $\Theta$ where the set of variables of interest $U$ is explicitly provided.

The main operation in Rsub is the concrete unification $\mathbf{U}_{\mathrm{Rs}} : \mathtt{Rsub} \times \mathtt{Rsub} \times \mathit{ISubst} \rightarrow \mathtt{Rsub}$ such that:

$$\begin{aligned}
\mathbf{U}_{\mathrm{Rs}}(\bot_{\mathrm{Rs}}, \xi, \delta) =& \mathbf{U}_{\mathrm{Rs}}(\xi, \bot_{\mathrm{Rs}}, \delta) = \bot_{\mathrm{Rs}} \\
\mathbf{U}_{\mathrm{Rs}}(\xi, \top_{\mathrm{Rs}}, \delta) =& \mathbf{U}_{\mathrm{Rs}}(\top_{\mathrm{Rs}}, \xi, \delta) = \top_{\mathrm{Rs}} \qquad \text{if } \xi \neq \bot_{\mathrm{Rs}} \\
\mathbf{U}_{\mathrm{Rs}}([\Theta_1, U_1], [\Theta_2, U_2], \delta) =& [\{\mathrm{mgu}(\sigma_1, \sigma_2, \delta) \mid \sigma_1 \in \Theta_1, \sigma_2 \in \Theta_2, \\
& \mathrm{vars}(\sigma_1) \cap \mathrm{vars}(\sigma_2) = \emptyset\}, U_1 \cup U_2] \ .
\end{aligned} \tag{39}$$

Although it is well defined for all the values of the domain, $\mathbf{U}_{\mathrm{Rs}}([\Theta_1, U_1], [\Theta_2, U_2], \delta)$ may be restricted to those values such that $U_1 \cap U_2 = \emptyset$ and $\mathrm{vars}(\delta) \subseteq U_1 \cup U_2$, since this is the only way $\mathbf{U}_{\mathrm{Rs}}$ is used in the semantics defined in (Cortesi and Filé 1999).

The abstract domain is the same Sharing we use in our paper, with abstraction map $\alpha_{\mathrm{Sh}} : \mathtt{Rsub} \rightarrow \mathtt{Sharing}$ and unification $\mathbf{U}_{\mathrm{Sh}} : \mathtt{Sharing} \times \mathtt{Sharing} \times \mathit{ISubst} \rightarrow \mathtt{Sharing}$ defined by:

$$\alpha_{\mathrm{Sh}}([\Theta, U]) = \bigsqcup\nolimits_{\mathrm{Sh}} \{\alpha_{\mathrm{Sh}}([\sigma]_U) \mid \sigma \in \Theta\} \ , \tag{40}$$

$$\mathbf{U}_{\mathrm{Sh}}([\Theta_1, U_1], [\Theta_2, U_2], \delta) = \mathsf{unif}_{\mathrm{Sh}}([\Theta_1 \cup \Theta_2, U_1 \cup U_2], \delta) \tag{41}$$

The domain of $\mathbf{U}_{\mathrm{Sh}}$ is restricted to the case $U_1 \cap U_2 = \emptyset$ and $\mathrm{vars}(\delta) \subseteq U_1 \cup U_2$.

By looking at the paper, we think that, in the idea of the authors, $[\Theta, U] \in \mathtt{Rsub}$ should have been treated as $[\{[\sigma]_U \mid \sigma \in \Theta\}, U] \in \mathtt{Psub}$ is in our framework. However, the condition $\mathrm{vars}(\sigma_1) \cap \mathrm{vars}(\sigma_2) = \emptyset$, introduced in $\mathbf{U}_{\mathrm{Rs}}$ in order to avoid variable clashes between the two chosen substitutions, is not enough for this purpose. Actually, $\mathbf{U}_{\mathrm{Rs}}$ only checks that $\sigma_1$ and $\sigma_2$ do not have variables in common, without considering their sets of variables of reference $U_1$ and $U_2$. This unification can lead to counterintuitive results.

*Example 8.2*
Consider the following concrete unification:

$$\mathbf{U}_{\mathrm{Rs}}([\{\{x/y\}\},\{x\}],[\{\epsilon\},\{y\}],\epsilon) = [\{\{x/y\}\},\{x,y\}] \ . \tag{42}$$

Being $\mathrm{vars}(\epsilon)=\emptyset$, the concrete unification operator allows us to unify $\{x/y\}$ with $\epsilon$ without renaming the variable $y$, which is not a variable of interest in the first element but it is treated as if it was. This also causes the incorrectness of $\mathbf{U}_{\mathrm{Sh}}$. If we consider Eq. (42) and compute the result on the abstract side by using the abstract unification operator $\mathbf{U}_{\mathrm{Sh}}$, we have:

$$\begin{aligned} & \mathbf{U}_{\mathrm{Sh}}(\ \alpha_{\mathrm{Sh}}([\{\{x/y\}\},\{x\}]),\ \ \alpha_{\mathrm{Sh}}([\{\epsilon\},\{y\}]),\ \ \epsilon) \\ = \ & \mathbf{U}_{\mathrm{Sh}}(\qquad [\{\mathtt{x}\},\{x\}], \qquad\qquad [\{\mathtt{y}\},\{y\}], \quad \epsilon) = [\{\mathtt{x},\mathtt{y}\},\{x,y\}] \ . \end{aligned}$$

This is not a correct approximation of the concrete result, since:

$$\alpha_{\mathrm{Sh}}([\{\{x/y\}\},\{x,y\}]) = [\{\mathtt{xy}\},\{x,y\}] \not\sqsubseteq_{\mathrm{Sh}} [\{\mathtt{x},\mathtt{y}\},\{x,y\}] \ . \quad \square$$

This counterexample proves that the abstract unification operator $\mathbf{U}_{\mathrm{Sh}}$ is not correct w.r.t. the concrete one $\mathbf{U}_{\mathrm{Rs}}$, invalidating the Theorem 6.3 in (Cortesi and Filé 1999). The problem can be solved by introducing a stronger check on variable clashes, namely by replacing the condition $\mathrm{vars}(\sigma_1)\cap\mathrm{vars}(\sigma_2)=\emptyset$ with $(\mathrm{vars}(\sigma_1)\cup U_1)\cap(\mathrm{vars}(\sigma_2)\cup U_2)=\emptyset$ in the definition of $\mathbf{U}_{\mathrm{Rs}}$, thus obtaining the following operator:

$$\begin{aligned} \mathbf{U}^*_{\mathrm{Rs}}([\Theta_1,U_1],[\Theta_2,U_2],\delta) = [\{ & \mathrm{mgu}(\sigma_1,\sigma_2,\delta) \mid \sigma_1\in\Theta_1, \sigma_2\in\Theta_2, \\ & (\mathrm{vars}(\sigma_1)\cup U_1)\cap(\mathrm{vars}(\sigma_2)\cup U_2)=\emptyset\}, U_1\cup U_2] \ . \end{aligned} \tag{43}$$

By using $\mathbf{U}^*_{\mathrm{Rs}}$ instead of $\mathbf{U}_{\mathrm{Rs}}$, the proof of Theorem 6.3 in (Cortesi and Filé 1999) becomes valid.

*Theorem 8.3*
$\mathbf{U}_{\mathrm{Sh}}$ is correct w.r.t. $\mathbf{U}^*_{\mathrm{Rs}}$.

*Proof*
If we look at the proof of Theorem 6.3 in (Cortesi and Filé 1999), it appears that the problem is in the base case of the inductive argument, when $i=0$. Here, it is stated that given $[A_1,U_1]$ and $[A_2,U_2]$ in `Sharing` with $U_1\cap U_2=\emptyset$, $\sigma_i\in\gamma_{\mathrm{Sh}}([A_i,U_i])$ for $i\in\{1,2\}$ with $\mathrm{vars}(\sigma_1)\cap\mathrm{vars}(\sigma_2)=\emptyset$, then it holds that $[\{\rho_0\},U_0]\sqsubseteq_{\mathrm{Rs}}\gamma_{\mathrm{Sh}}([R_0,U_0])$ where $\rho_0=\sigma_1\uplus\sigma_2$, $U_0=U_1\cup U_2$ and $R_0=A_1\cup A_2$. However, the substitutions $\sigma_1=\{x/y\}\in\gamma_{\mathrm{Sh}}([\{\mathtt{x}\},\{x\}])$ and $\sigma_2=\epsilon\in\gamma_{\mathrm{Sh}}([\{\mathtt{y}\},\{y\}])$ of the previous example make the statement false. On the contrary, when $\mathbf{U}^*_{\mathrm{Rs}}$ is used instead of $\mathbf{U}_{\mathrm{Rs}}$, then $\sigma_1$ and $\sigma_2$ are required to satisfy the condition $(\mathrm{vars}(\sigma_1)\cup U_1)\cap(\mathrm{vars}(\sigma_2)\cup U_2)=\emptyset$. From this, it truly follows that $[\{\rho_0\},U_0]=[\{\sigma_1\uplus\sigma_2\},U_0]\sqsubseteq_{\mathrm{Rs}}\gamma_{\mathrm{Sh}}([R_0,U_0])$. The inductive case for $i>0$ is identical to that in (Cortesi and Filé 1999), since for any $A,B\in$ `Rsub` and $\delta\in \mathit{ISubst}$ it holds that $\mathbf{U}^*_{\mathrm{Rs}}(A,B,\delta)\sqsubseteq_{\mathrm{Rs}}\mathbf{U}_{\mathrm{Rs}}(A,B,\delta)$. $\square$

Observer that, in order to define a real semantics for logic programs, a renaming operation should be introduced in the framework of Cortesi and Filé (1999). This can be done along the line of Cortesi et al. (1994). Due to the kind of renamings involved, by replacing $\mathbf{U}_{\mathrm{Rs}}$ with $\mathbf{U}^{*}_{\mathrm{Rs}}$, the semantics in (Cortesi et al. 1994) does not change. Therefore this flaw does not affect the real analysis of logic programs.

## *8.3 Other Related Works*

### *8.3.1 Backward Unification*

The idea of using a refined operator for computing answer substitutions is not new, and may be traced back to the frameworks in (Bruynooghe 1991; Le Charlier and Van Hentenryck 1994). The abstract domains considered in these papers contain structural information, freeness, groundness and pair-sharing, but no set-sharing information. Working within these frameworks, Hans and Winkler (1992) and King and Longley (1995) propose correct abstract operators w.r.t. matching for the domain `SFL`. Muthukumar and Hermenegildo (1991; 1992) use a refined algorithm for backward unification in `Sharing`, although it is not presented in algebraic form. However, to the best of our knowledge, this is the first paper which formally introduces matching from the point of view of a collecting denotational semantics, deriving the abstract operator from the concrete one, and proving correctness and optimality. Moreover, this is the first paper which presents optimal abstract matching for a domain for set-sharing analysis (see Example 7.3).

### *8.3.2 Forward/Backward Unification and* `PSD`

Although the usual goal of sharing analyses is to discover the pairs of variables which may possibly share, `Sharing` is a domain that keeps track of set-sharing information. Bagnara et al. (2002) propose a new domain, called `PSD`, which is the complete shell (Giacobazzi et al. 2000) of pair sharing w.r.t. `Sharing`. They recognize that, in an abstract object $[S, U]$, some sharing groups in $S$ may be *redundant* as far as pair sharing is concerned. Although our forward unification is more precise than the standard unification, it could be the case that they have the same precision in `PSD`. This would mean that $\mathbf{U}^{f}_{\mathrm{Sh}}([S_1, U_1], A_1, A_2)$ and $\mathbf{U'}^{f}_{\mathrm{Sh}}([S_1, U_1], A_1, A_2)$ only differ for redundant sharing groups. However, this is not the case, and Examples 6.2, 6.3 and 6.19 show improvements which are still significant in `PSD`. The same holds for backward unification in Example 7.3. It is not clear whether `PSD` is still complete w.r.t. pair-sharing when our specialized operators are used.

### *8.3.3 Domains with Freeness and Linearity*

Although the use of freeness and linearity information has been pursued in several papers, e.g., (Muthukumar and Hermenegildo 1991; Hans and Winkler 1992), optimal operators for these domains have never been developed. All the abstract unification operators for `SFL`, e.g., (Muthukumar and Hermenegildo 1992; Hans and Winkler 1992; Hill et al. 2004), when unifying with a binding $\{x/t\}$ where neither $x$ nor $t$ are linear, does compute all the star unions. On the contrary, in $\mathbf{u}^f_{\mathrm{Sh}}$ we apply an optimization which is able to avoid some sharing groups (see e.g., Example 6.19). This optimization could be integrated in a domain which explicitly contains freeness and linearity information.

Actually, Hill et al. (2004) include some optimizations for the standard abstract unification of `SFL` which are similar to ours, in the case of a binding $\{x/t\}$ with $x$ linear. In addition, in (Hill et al. 2004; Howe and King 2003) the authors propose to remove the check for independence between $x$ and $t$. We think it should be possible to devise an optimal abstract unification for an enhanced domain including linearity information, by combining these improvements with our results.

A first optimality result is shown in (Amato and Scozzari 2003), which is based on a preliminary version of the framework we present here. The authors consider two domains for set-sharing and linearity (without freeness), namely the standard reduced product of `Sharing` and linearity, and the domain proposed by King (1994). The paper presents the abstract operators for forward unification, which turn out to be optimal in the case of a single-binding substitution. These are the only operators in the literature which are strictly more precise than our optimized forward unification operator for `Sharing`.

#### *8.3.4 Another Optimality Proof*

Codish et al. (2000) provide an alternative approach to the analysis of sharing by using set logic programs and ACI1 unification. They define abstract operators which are proved to be correct and optimal, and examine the relationship between set substitutions and `Sharing`, proving that they are essentially isomorphic. However, they do not extend this correspondence to the abstract operators, so that a proof of optimality of $\mathbf{U}^f_{\mathrm{Sh}}$ w.r.t. $\mathbf{U}^f_{\mathrm{Ps}}$ starting from their results should be feasible but it is not immediate. Moreover, since they provide a goal-independent analysis, they do not have different operators for forward and backward unification.

## 9 Conclusions

We think that there are three major contributions in this paper.

- We integrate the framework of Cortesi et al. (1996) with several different proposals appeared in the literature for goal-dependent analysis of logic programs. We give formal proofs of the correctness of the resulting analysis and of optimality of the abstract operators. The aim is to clarify the relationships between these proposals and to provide a clear guidance for the development of static analysis for logic programs.
- We introduce a new concrete domain of equivalence classes of substitutions which address the problem of variable clashes by taking into account sets of variables of interest. This problem has been considered by many authors but, in our opinion, none of them fully developed a corresponding theory of substitutions, in the style of Palamidessi (1990).

- Our definition of abstract forward unification sheds new light on the role of freeness and linearity information, suggesting new optimizations which can also be used in more powerful domains such as `SFL`.

Although sharing analysis with more complex domains, including freeness and linearity information, will likely be more precise than the analysis performed with `Sharing` in our optimized framework, we think that this article may be a guideline for developing new analysis for logic programs. The main ideas contained in this paper are not tied to the abstract domain in use. The framework we propose may be instantiated with more precise abstract domains to further improve the result of the abstract analysis. Moreover, the algorithm for the abstract forward unification can be easily slotted into other analysis frameworks based on different concrete semantics, including goal-independent ones.

To the best of our knowledge, this is the first work which optimizes the abstract forward unification for sharing analysis using freeness and linearity information implicitly, i.e., without using a domain which contains such information.

This is also the first work where an abstract backward unification operator using matching is proved to be optimal. We have shown that, to the best of our knowledge, all the abstract backward unification operators proposed so far for `Sharing` or more powerful domains (Hans and Winkler 1992; King and Longley 1995; Muthukumar and Hermenegildo 1992) were not optimal.

As a future work, we think that our results could be easily generalized for designing optimal unification operators for more complex domains possibly including linearity, freeness and structural information. Preliminary results have appeared in (Amato and Scozzari 2003). Moreover, the problem of efficiently implementing the refined backward unification could be addressed.

## A Correctness of the Goal-Dependent Collecting Semantics

In this appendix we provide a tedious proof that the collecting semantics we define is correct w.r.t. computed answers. We begin by formally introducing a notation for SLD-derivations, following (Lloyd 1987; Apt 1990). Given a goal $G = g_1 \dots g_k$ and a clause $cl = H \leftarrow B$ such that $\mathrm{vars}(G) \cap \mathrm{vars}(cl) = \emptyset$, we write

$$G \xrightarrow[\sigma]{cl} (g_1 \dots g_{i-1} B g_{i+1} \dots g_k)\sigma \tag{A1}$$

when $\sigma = \mathrm{mgu}(g_i, H)$. Given a goal $G$ and a program $P$, an *SLD-derivation* of $G$ in $P$ is given by a sequence of clauses $cl_1, \dots, cl_n$ and idempotent substitutions $\sigma_1, \dots, \sigma_n$, such that

$$G \xrightarrow[\sigma_1]{cl_1} G_1 \xrightarrow[\sigma_2]{cl_2} \cdots \xrightarrow[\sigma_n]{cl_n} G_n \ , \tag{A2}$$

where each $cl_i$ is the renaming of a clause in $P$ apart from $G, cl_1, \dots, cl_{i-1}$. The goal $G_n$ is called the *end-goal*, $n$ is the length of the derivation and $(\sigma_n \circ \sigma_{n-1} \circ \dots \circ \sigma_2 \circ \sigma_1)_{|\mathrm{vars}(G)}$ is the (partial) computed answer. An SLD-refutation is an SLD-derivation with the empty end-goal (denoted by $\Box$). A *leftmost* SLD-derivation is an SLD-derivation where we always rewrite the leftmost atom in the goal (i.e., such that $i = 1$ at every step in (A1)).

We write $G \xrightarrow[\sigma]{*} G'$ to denote an SLD-derivation with end-goal $G'$ and partial computed answer $\sigma$. We also write $G \xrightarrow[\sigma]{\leq i} G'$ to denote an SLD-derivation with end-goal $G'$, partial computed answer $\sigma$ and whose length is less or equal then $i$. A substitution $\sigma$ is a *computed answer* for $G$ in $P$ if there is an SLD-refutation $G \xrightarrow[\sigma]{*} \Box$.

In this appendix we will prove the relationship between the set of computed answers for $P$ and its collecting semantics $\mathcal{P}[\![P]\!]$.

### A.1 Relevant Denotations

We have defined a denotation as a continuous map in $\mathsf{Atoms} \to \mathtt{Psub} \xrightarrow{c} \mathtt{Psub}$. We now want to characterize the denotations which may arise as the results of our collecting semantics.

*Definition A.1*
A denotation $d \in \mathcal{D}en$ is said to be *relevant* when

- $d$ is strict, i.e., $dA\bot_{\mathrm{Ps}} = \bot_{\mathrm{Ps}}$ ;
- $dA[\Delta, V]$ is either $\bot_{\mathrm{Ps}}$ or $[\Delta', V \cup \mathrm{vars}(A)]$ for some $\Delta'$.

Note that the least denotation $\lambda A.\lambda[\Delta, V].\bot_{\mathrm{Ps}}$ is relevant. A relevant denotation is well-behaved, in the sense that either it does not say anything, or gives information for all and only the variables which occur in the atom $A$ and the entry substitution $[\Delta, V]$.

*Proposition A.2*
If $d$ is relevant, then

1. $\mathcal{B}[\![B]\!]d\bot_{\mathrm{Ps}} = \bot_{\mathrm{Ps}}$;
2. $\mathcal{B}[\![B]\!]d[\Delta, V]$ is either $\bot_{\mathrm{Ps}}$ or $[\Delta', V \cup \mathrm{vars}(B)]$ for some $\Delta'$;
3. $\mathcal{C}[\![H \leftarrow B]\!]d$ is relevant;
4. $\mathcal{P}[\![P]\!]$ is relevant.

*Proof*
The first two points easily follow by induction on the structure of the body $B$. For the third point, consider the definition of $\mathcal{C}$. Note that

$$\mathbf{U}^{f}_{\mathrm{Ps}}(x, A, H) = \pi_{\mathrm{Ps}}(\mathsf{unif}_{\mathrm{Ps}}(\rho(x), \mathrm{mgu}(\rho(A) = H)), \mathrm{vars}(H)) \enspace .$$

Since $\mathrm{vars}(\rho(A))$ is disjoint from $H$ by definition of $\rho$, and since we consider relevant mgus, then either $\mathrm{vars}(\mathrm{mgu}(\rho(A) = H)) = \mathrm{vars}(\rho(A)) \cup \mathrm{vars}(H)$ or $\mathrm{mgu}(\rho(A) = H) = \bot$. In the latter case, $\mathcal{C}[\![H \leftarrow B]\!]dA = \bot_{\mathrm{Ps}}$, otherwise $\mathbf{U}^{f}_{\mathrm{Ps}}(x, A, H) = [\Delta', \mathrm{vars}(H)]$ for some $\Delta'$. By the previous point, we have that $\mathcal{B}[\![B]\!]d(\mathbf{U}^{f}_{\mathrm{Ps}}(x, A, H))$ is either $\bot_{\mathrm{Ps}}$ or $[\Delta'', \mathrm{vars}(H) \cup \mathrm{vars}(B)]$ for some $\Delta''$. In the first case, $\mathcal{C}[\![H \leftarrow$

$B]\!]dA = \bot_{\mathrm{Ps}}$, otherwise, assuming $x = [\Theta, V]$, we have

$$\mathcal{C}[\![H \leftarrow B]\!]dAx = \mathbf{U}^b_{\mathrm{Ps}}([\Delta'', \mathrm{vars}(H) \cup \mathrm{vars}(B)], x, H, A) = \\ \pi_{\mathrm{Ps}}\Big(\mathsf{match}_{\mathrm{Ps}}(\rho([\Delta'', \mathrm{vars}(H) \cup \mathrm{vars}(B)]), \\ \mathsf{unif}_{\mathrm{Ps}}([\Theta, V], \mathrm{mgu}(\rho(H) = A))), V \cup \mathrm{vars}(A)\Big) \ .$$

For the same reason explained above, and since we can ignore the case in which $\rho(H)$ and $A$ do not unify, we have that $\mathsf{unif}_{\mathrm{Ps}}([\Theta, V], \mathrm{mgu}(\rho(H) = A)) = [\Theta', V \cup \mathrm{vars}(A)]$ and therefore

$$\pi_{\mathrm{Ps}}(\mathsf{match}_{\mathrm{Ps}}(\rho([\Delta'', \mathrm{vars}(H) \cup \mathrm{vars}(B)]), [\Theta', V \cup \mathrm{vars}(A)]), V \cup \mathrm{vars}(A)) = \\ [\Theta'', V \cup \mathrm{vars}(A)] \ ,$$

which is what we wanted to prove.

The forth point follows by the fact that, given the proof of the third point, $\mathcal{C}[\![cl]\!]d$ is relevant for each clause *cl*, and that least upper bound of relevant denotations are easily seen to be relevant. □

### *A.2 Unused variables*

*Definition A.3*
Given $[\phi]_V \in \mathit{ISubst}_\sim$ and $x \in \mathcal{V}$, we say that $x$ is *unused* in $[\phi]_V$ when $[\phi]_V = \mathrm{mgu}\big(\pi_{V \setminus \{x\}}([\phi]_V), [\epsilon]_{\{x\}}\big)$.

First of all, note that this definition does not depend on the choice of representatives. If a variable $x$ is unused in $[\phi]_V$, it means that $[\phi]_V$ does not constraint in any way its value. In other words, $x$ is free and independent from all the other variables in $V$. This is made clear by the following characterization:

*Proposition A.4*
The variable $x \in V$ is unused in $[\phi]_V$ iff it is free and independent in $[\phi]_V$.

*Proof*
If $x$ is free and independent in $[\phi]_V$, we may assume without loss of generality that $x \notin \mathrm{vars}(\phi)$. Let $V' = V \setminus \{x\}$. We have that

$$\mathrm{mgu}(\pi_{V'}([\phi]_V), [\epsilon]_{\{x\}}) = \mathrm{mgu}([\phi]_{V'}, [\epsilon]_{\{x\}}) = [\phi_{|V'}]_V = [\phi]_V \ ,$$

which proves that $x$ is unused. On the other hand, assume $\phi$ is a canonical representative and $\mathrm{mgu}([\phi]_{V'}, [\epsilon]_{\{x\}}) = [\phi]_V$. Then $\phi_{|V'} \sim_V \phi$. It is obvious that $x$ is free and independent in $[\phi_{|V'}]_V = [\phi]_V$, since $x \notin \mathrm{dom}(\phi_{|V'})$ and $x \notin \mathrm{rng}(\phi)$. □

### *A.3 ISubst$_\sim$ and composition*

The operations described in Section 3.2 are those required to provide a collecting semantics for logic programs over the domain $\mathit{ISubst}_\sim$. Note that we do not define any notion of composition, although it plays a central role with the standard substitutions. Actually, composition cannot be defined in our framework since, given any element of $\mathit{ISubst}_\sim$, variables not of interest are considered up to renaming only, and therefore cannot be named. Nonetheless, in order to prove the equivalence between the standard semantics based on SLD-resolution and our collecting semantics, we will need to relate the composition of substitutions with unification in $\mathit{ISubst}_\sim$.

*Lemma A.5*
(Composition Lemma) Let $\sigma_1, \sigma_2, \sigma_3 \in \mathit{Subst}$, $U, V \in \wp_f(\mathcal{V})$. Then it holds that:

$$\mathrm{mgu}([\sigma_3 \circ \sigma_2]_U, [\sigma_2 \circ \sigma_1]_V) = [\sigma_3 \circ \sigma_2 \circ \sigma_1]_{U \cup V}$$

provided that:

- $\mathrm{dom}(\sigma_1) \cap U = \emptyset$;
- if $y \in \sigma_2(\sigma_1(V)) \setminus \sigma_2(\sigma_1(U \cap V))$ then $y \notin \mathrm{dom}(\sigma_3) \cup \sigma_3(\sigma_2(U))$.

*Proof*
Let $\theta \in [\sigma_3 \circ \sigma_2]_U$, $\eta \in [\sigma_2 \circ \sigma_1]_V$ be canonical representatives such that $(\mathrm{vars}(\theta) \cup U) \cap (\mathrm{vars}(\eta) \cup V) \subseteq U \cap V$. By definition, there exist $\rho, \rho' \in \mathit{Ren}$ such that $\theta = (\rho' \circ \sigma_3 \circ \sigma_2)_{|U}$ and $\eta = (\rho \circ \sigma_2 \circ \sigma_1)_{|V}$.

Then $\mathrm{mgu}([\sigma_3 \circ \sigma_2]_U, [\sigma_2 \circ \sigma_1]_V) = [\mathrm{mgu}(\theta, \eta)]_{U \cup V}$. It holds that $\mathrm{mgu}(\theta, \eta) = \mathrm{mgu}(\eta(\mathrm{Eq}(\theta))) \circ \eta$. It follows that $\eta(\mathrm{Eq}(\theta)) = \{\eta(x) = \eta(\theta(x)) \mid x \in U\} = \{\eta(x) = \theta(x) \mid x \in U\}$ since $\theta$ is a canonical representative. If $x \in U \cap V$, then $\eta(x) = \theta(x)$ becomes $\rho \circ \sigma_2 \circ \sigma_1(x) = \rho' \circ \sigma_3 \circ \sigma_2(x)$ which is $\rho \circ \sigma_2(x) = \rho' \circ \sigma_3 \circ \sigma_2(x)$ since $\mathrm{dom}(\sigma_1) \cap U = \emptyset$ by hypothesis. Thus $\{\eta(x) = \theta(x) \mid x \in U \cap V\}$ and $\{\rho(y) = \rho' \circ \sigma_3(y) \mid y \in \sigma_2(U \cap V)\}$ have the same set of solutions. If $x \notin V$ then $\{\eta(x) = \theta(x) \mid x \in U \setminus V\} = \{x = \theta(x) \mid x \in U \setminus V\}$.

Now $\delta = \{\rho(y)/\rho' \circ \sigma_3(y) \mid y \in \sigma_2(U \cap V)\} \cup \{x/\theta(x) \mid x \in U \setminus V\}$ is an idempotent substitution. Actually, all the $\rho(y)$'s are distinct variables and different from $U \setminus V$ therefore $\delta$ is a substitution. Moreover, $\mathrm{dom}(\delta) \subseteq \mathrm{vars}(\eta(V)) \cup (U \setminus V)$ is disjoint from $\mathrm{rng}(\delta) = \mathrm{vars}(\theta(U))$.

Let $\rho''$ be the substitution

$$\rho''(x) = \begin{cases} \rho'(x) & \text{if } x \in \sigma_3\sigma_2(U) \\ \rho(x) & \text{if } x \in \sigma_2(\sigma_1(V)) \setminus \sigma_2(\sigma_1(U \cap V)) \\ x & \text{otherwise} \end{cases}$$

Note that, thanks to the second hypothesis of the lemma, we are sure that the first and second case in the definition of $\rho''$ may not occur together. We want to prove that $\delta(\eta(x)) = \rho''(\sigma_3(\sigma_2(\sigma_1(x))))$ for each $x \in U \cup V$. Since $\rho''$ restricted to $\mathrm{vars}(\sigma_3(\sigma_2(\sigma_1(U \cup V))))$ is an injective map from variables to variables, by Lemma 3.4 this implies $\delta \circ \eta \sim_{U \cup V} \sigma_3 \circ \sigma_2 \circ \sigma_1$, which is the statement of the theorem.

Thus if $x \in U \setminus V$ then $\eta(x) = x$ and $\delta(\eta(x)) = \theta(x) = \rho'(\sigma_3(\sigma_2(x))) = \rho''(\sigma_3(\sigma_2(x))) = \rho''(\sigma_3(\sigma_2(\sigma_1(x))))$ since $\mathrm{dom}(\sigma_1) \cap U = \emptyset$ and by definition of $\rho''$.

If $x \in U \cap V$ then $\delta(\eta(x)) = \delta(\rho(\sigma_2(x)))$ since $\mathrm{dom}(\sigma_1) \cap U = \emptyset$ and thus $\delta(\eta(x)) = \rho'(\sigma_3(\sigma_2(x)))$, which is equal to $\rho''(\sigma_3(\sigma_2(\sigma_1(x))))$ since $\mathrm{dom}(\sigma_1) \cap U = \emptyset$ and by definition of $\rho''$.

If $x \in V \setminus U$ then $\delta(\eta(x)) = \delta(\rho(\sigma_2\sigma_1(x)))$. Let $y \in \mathrm{vars}(\sigma_2(\sigma_1(x)))$. If we assume that $y \in \mathrm{vars}(\sigma_2(U \cap V))$, then $\delta(\rho(y)) = \rho'(\sigma_3(y)) = \rho''(\sigma_3(y))$ by definition of $\delta$ and $\rho''$. If $y \notin \mathrm{vars}(\sigma_2(U \cap V))$ then $\delta(\rho(y)) = \rho(y) = \rho''(y) = \rho''(\sigma_3(y))$ by definition of $\rho''$ and the second condition in the theorem. In both cases we obtain $\delta(\rho(y)) = \rho''(\sigma_3((y)))$ for each $y \in \mathrm{vars}(\sigma_2(\sigma_1(x)))$. Therefore, for each $x \in U \cap V$, $\delta(\eta(x)) = \delta(\rho(\sigma_2(\sigma_1(x)))) = \rho''(\sigma_3(\sigma_2(\sigma_1(x))))$ and this concludes the proof. □

### *A.4 Proof of Correctness*

Let $D_P$ be defined as $\lambda d. \bigsqcup_{\mathrm{Ps}}\{\mathcal{C}[\![cl]\!]d \mid cl \in P\}$ and let $D_P^i$ be the $i$-th iteration of $D_P$ with $D_P^0 = \lambda A.\lambda x.\bot_{\mathrm{Ps}}$. Note that $D_P^\omega = \mathcal{P}[\![P]\!]$ and $D_P^i$ is relevant for each $i$.

*Lemma A.6*
(Correctness Lemma) Let $i \in \mathbb{N}$, $[\phi]_V \in \mathit{ISubst}_\sim$, $G \in \mathsf{Bodies}$ and $P \in \mathsf{Progs}$. If $[\phi]_{V \cup G} = \mathrm{mgu}([\phi]_V, [\epsilon]_G)$ and $G\phi \xrightarrow[\sigma]{*} \square$ is a leftmost SLD-refutation, with at most $i$ steps, where all clauses are renamed apart from $V$, $G$, $\phi$ and the program $P$, then $\mathcal{B}[\![G]\!]D_P^i[\{[\phi]\}, V] \sqsupseteq_{\mathrm{Ps}} [\{[\sigma \circ \phi]\}, V \cup \mathrm{vars}(G)]$.

*Remark A.7*
The condition $[\phi]_{V \cup G} = \mathrm{mgu}([\phi]_V, [\epsilon]_G)$ is used to check that the chosen representative $\phi$ does not bind any variable in $\mathrm{vars}(G) \setminus V$. All the variables in $\mathrm{vars}(G) \setminus V$ are forced to be unused, according to Definition A.3.

*Remark A.8*
The theorem probably holds under weaker conditions on the variables of the SLD-resolution. However, proving the result in this case would be more difficult. Since the obtained generalization is not very interesting, we valued that it was not worth the effort.

*Proof*
The proof is by double induction on $i$ and on the structure of the goal $G$. Assume fixed $\Phi = \{[\phi]_V\}$ such that $[\phi]_{V \cup G} = \mathrm{mgu}([\phi]_V, [\epsilon]_G)$.

We start with the case $i = 0$. The only SLD-refutation of length 0 is the SLD-derivation for the empty goal $\square$, whose computed answer substitution is $\epsilon$. In the collecting semantics, we have $\mathcal{B}[\![\square]\!]D_P^i[\{[\phi]\}, V] = [\{[\phi]\}, V] = [\{[\epsilon \circ \phi]\}, V]$ which is the required result.

If $i > 0$, assume the lemma holds for all $j < i$ and we prove it for $i$, by induction on the structure of goals. The case for the empty goal has been already examined, so we assume $G = A, G'$ where $A$ is an atom. To ease the exposition, we first consider the atomic case where $G' = \square$ and then we analyze the general one.

*Atomic goal.* Given the not-empty SLD-derivation $G\phi \xrightarrow[\sigma]{*} \square$, we may decompose it as:

$$G\phi \xrightarrow[\sigma_1]{\rho(cl)} (C_1 \dots C_n)\rho\sigma_1 \xrightarrow[\sigma_2]{*} \square$$

where $cl = H \leftarrow C_1 \dots C_n$ is a program clause, $\sigma_1 = \mathrm{mgu}(G\phi, H\rho)$ and $\rho$ is a renaming of $cl$ apart from $G$, $V$, $\phi$ and the program $P$. Note that this implies the standard renaming condition for SLD-resolutions, i.e., that $\rho(cl)$ is renamed apart from $G\phi$. Since $G$ is atomic, then

$$\mathcal{B}[\![G]\!]D_P^i[\Phi, V] = D_P^i G[\Phi, V] \sqsupseteq_{\mathrm{Ps}} \mathcal{C}[\![H \leftarrow C_1 \dots C_n]\!]D_P^{i-1}G[\Phi, V] \ ,$$

which, in turn, is equal to $\mathbf{U}^b_{\mathrm{Ps}}(\mathcal{B}[\![C_1 \dots C_n]\!]D_P^{i-1}(\mathbf{U}^f_{\mathrm{Ps}}([\Phi, V], G, H)), [\Phi, V], H, G)$. We know that

$$\mathbf{U}^f_{\mathrm{Ps}}([\{[\phi]\}, V], G, H) = \pi_{\mathrm{Ps}}(\mathrm{mgu}(\rho'([\phi]_V), [\mathrm{mgu}(\rho'(G) = H)]_{\rho'(G) \cup H}), \mathrm{vars}(H))$$

where $\rho'$ is any renaming such that $\rho'(\mathrm{vars}(G) \cup V) \cap \mathrm{vars}(H) = \emptyset$. We can choose as $\rho'$ the renaming $\rho^{-1}$ since $\rho(\mathrm{vars}(cl)) \cap \mathrm{vars}(G) = \emptyset$ and $\rho(\mathrm{vars}(cl)) \cap V = \emptyset$ implies that $\rho^{-1}(\mathrm{vars}(G) \cup V) \cap \mathrm{vars}(H) = \emptyset$. In turn, this implies that

$$\begin{aligned}
& \mathrm{mgu}(\rho'([\phi]_V), [\mathrm{mgu}(\rho'(G) = H)]_{\rho'(G) \cup H}) \\
= \ & \rho^{-1}(\mathrm{mgu}([\phi]_V, [\mathrm{mgu}(G = \rho(H))]_{G \cup \rho(H)}) \\
= \ & \rho^{-1}(\mathrm{mgu}([\phi]_V, [\mathrm{mgu}(G = \rho(H))]_{G \cup \rho(H)}, [\epsilon]_G)) \\
= \ & \rho^{-1}(\mathrm{mgu}([\phi]_{V \cup G}, [\mathrm{mgu}(G = \rho(H))]_{G \cup \rho(H)}) \\
= \ & \rho^{-1}([\mathrm{mgu}(\phi, \mathrm{mgu}(G = \rho(H)))]_{V \cup G \cup \rho(H)}) \ .
\end{aligned}$$

The last pass is only valid when $(V \cup \mathrm{vars}(G) \cup \mathrm{vars}(\phi)) \cap (\mathrm{vars}(G) \cup \mathrm{vars}(\rho(H)) \subseteq (V \cup \mathrm{vars}(G)) \cap (\mathrm{vars}(G) \cup \mathrm{vars}(\rho(H))) = \mathrm{vars}(G)$. This is the case since $\mathrm{vars}(\phi) \cap \rho(\mathrm{vars}(cl)) = \emptyset$, thanks to our choice of $\rho$.

By standard properties of substitutions, we obtain:

$$\begin{aligned}
& \rho^{-1}([\mathrm{mgu}(\phi, \mathrm{mgu}(G = \rho(H)))]_{V \cup G \cup \rho(H)}) \\
= \ & \rho^{-1}([\mathrm{mgu}(G\phi = (\rho(H))\phi) \circ \phi]_{V \cup G \cup \rho(H)}) \\
= \ & \rho^{-1}([\mathrm{mgu}(G\phi = \rho(H)) \circ \phi]_{V \cup G \cup \rho(H)}) \\
= \ & \rho^{-1}([\sigma_1 \circ \phi]_{V \cup G \cup \rho(H)}) \ ,
\end{aligned}$$

since $\mathrm{vars}(\phi) \cap \mathrm{vars}(\rho(H)) = \emptyset$. For the same reason, $\sigma_1 \circ \phi \sim_{\mathrm{vars}(\rho(H))} \sigma_1$. It follows that

$$\rho^{-1}(\sigma_1 \circ \phi) \sim_{\mathrm{vars}(H)} \rho^{-1}(\sigma_1) = \rho^{-1} \circ \sigma_1 \circ \rho \sim_{\mathrm{vars}(H)} \sigma_1 \circ \rho \ .$$

Therefore $\mathbf{U}^f_{\mathrm{Ps}}([\{[\phi]\}, V], G, H) = [\{[\sigma_1 \circ \rho]\}, \mathrm{vars}(H)]$ and

$$\begin{aligned}
& \mathbf{U}^b_{\mathrm{Ps}}(\mathcal{B}[\![C_1 \dots C_n]\!]D_P^{i-1}(\mathbf{U}^f_{\mathrm{Ps}}([\Phi, V], G, H)), [\Phi, V], H, G) \sqsupseteq_{\mathrm{Ps}} \\
& \qquad \mathbf{U}^b_{\mathrm{Ps}}(\mathcal{B}[\![C_1 \dots C_n]\!]D_P^{i-1}[\{[\sigma_1 \circ \rho]\}, \mathrm{vars}(H)], [\Phi, V], H, G) \ .
\end{aligned}$$

Note that the SLD resolution $(C_1 \dots C_n)\rho\sigma_1 \xrightarrow[\sigma_2]{*} \Box$ can be seen as $(C_1 \dots C_n)(\sigma_1 \circ \rho) \xrightarrow[\sigma_2]{*} \Box$. In order to apply the inductive hypothesis on the latter derivation, we need to verify that $[\sigma_1 \circ \rho]_{\mathrm{vars}(cl)} = \mathrm{mgu}([\sigma_1 \circ \rho]_{\mathrm{vars}(H)}, [\epsilon]_{\mathrm{vars}(C_1 \dots C_n)})$. By definition $\sigma_1 \circ \rho = \mathrm{mgu}(G\phi, H\rho) \circ \rho$. Moreover, since $\rho(\mathrm{vars}(cl)) \cap \mathrm{vars}(G\phi) = \emptyset$ and $\rho(\mathrm{vars}(cl)) \cap \mathrm{vars}(H\rho) = \mathrm{vars}(H\rho)$, it follows that for all $v \in \rho(\mathrm{vars}(cl) \setminus \mathrm{vars}(H))$, $v \notin \mathrm{vars}(\sigma_1)$. Hence, for each $v \in \mathrm{vars}(cl) \setminus \mathrm{vars}(H)$, $\sigma_1(\rho(v)) = \rho(v)$. Moreover, if $\rho(v)$ occurs in $(\sigma_1 \circ \rho)(x)$ for some $x$, then $\rho(v)$ occurs in $\rho(x)$ and this is only possible if $x = v$. By Prop. A.4, this proves that $\text{mgu}([\sigma_1 \circ \rho]_{\text{vars}(H)}, [\epsilon]_{\text{vars}(C_1 \ldots C_n)}) = [\sigma_1 \circ \rho]_{\text{vars}(cl)}$. Thus, by inductive hypothesis, we have that:

$$\mathbf{U}^b_{\text{Ps}}(\mathcal{B}[\![C_1 \ldots C_n]\!] D_P^{i-1}[\{[\sigma_1 \circ \rho]\}, \text{vars}(H)], [\Phi, V], H, G) \sqsupseteq_{\text{Ps}} \mathbf{U}^b_{\text{Ps}}([\{[\sigma_2 \circ \sigma_1 \circ \rho]\}, \text{vars}(cl)], [\Phi, V], H, G) \enspace .$$

We know that $\mathsf{unif}_{\text{Ps}}([\{[\phi]\}, V], \text{mgu}(\rho(H) = G)) = [\{[\sigma_1 \circ \phi]\}, V \cup \text{vars}(G) \cup \text{vars}(\rho(H))]$. Therefore, choosing $\rho$ as the renaming for $\mathbf{U}^b_{\text{Ps}}$, we obtain

$$\begin{aligned} & \mathsf{match}_{\text{Ps}}(\rho([\{[\sigma_2 \circ \sigma_1 \circ \rho]\}, \text{vars}(cl)]), [\{[\sigma_1 \circ \phi]\}, V \cup \text{vars}(G) \cup \text{vars}(\rho(H))]) \\ = \; & \mathsf{match}_{\text{Ps}}([\{[\rho \circ \sigma_2 \circ \sigma_1]\}, \text{vars}(\rho(cl))]), [\{[\sigma_1 \circ \phi]\}, V \cup \text{vars}(G) \cup \text{vars}(\rho(H))]) \\ = \; & \mathsf{match}_{\text{Ps}}([\{[\sigma_2 \circ \sigma_1]\}, \text{vars}(\rho(cl))]), [\{[\sigma_1 \circ \phi]\}, V \cup \text{vars}(G) \cup \text{vars}(\rho(H))]) \enspace . \end{aligned}$$

Since $\text{vars}(\rho(cl)) \cap (V \cup \text{vars}(G) \cup \text{vars}(\rho(H))) = \text{vars}(\rho(H))$ and $\sigma_2 \circ \sigma_1 \preceq_{\text{vars}(\rho(H))} \sigma_1 \circ \phi$ (being $\text{vars}(\phi) \cap \text{vars}(\rho(H)) = \emptyset$), it holds:

$$\mathsf{match}_{\text{Ps}}([\{[\sigma_2 \circ \sigma_1]\}, \text{vars}(\rho(cl))]), [\{[\sigma_1 \circ \phi]\}, V \cup \text{vars}(G) \cup \text{vars}(\rho(H))]) = [\text{mgu}([\sigma_2 \circ \sigma_1]_{\rho(cl)}, [\sigma_1 \circ \phi]_{V \cup G \cup \rho(H)}), V \cup \text{vars}(G) \cup \text{vars}(\rho(H))]$$

We would like to apply the Composition Lemma (Lemma A.5) to this unification. We need to check that:

- $\text{dom}(\phi) \cap \rho(cl) = \emptyset$;
- $y \in \sigma_1\phi(V \cup \text{vars}(G) \cup \rho(H)) \setminus \sigma_1\phi(\rho(H))$ then $y \notin \text{dom}(\sigma_2) \cup \sigma_2\sigma_1(\rho(cl))$.

The first property trivially follows by the hypothesis that $\rho$ renames $cl$ apart from $\phi$. For the second condition, note that, since $\sigma_1 = \text{mgu}(G\phi, H\rho)$, if $y \in \sigma_1(\phi(G))$ then $y \in \sigma_1(\rho(H)) = \sigma_1(\phi(\rho(H)))$. Therefore $y \in \sigma_1(\phi(V \cup \text{vars}(G))) \setminus \sigma_1(\phi(\rho(H)))$ iff $y \in \sigma_1(\phi(V \setminus G)) = \phi(V \setminus G)$. However, since such a variable does not appear in the initial goal of the SLD-resolution $G\phi$ and since the resolution is renamed apart from $\phi$, it happens that it does not appear in $\text{vars}(\sigma_2)$, and thus in $\text{dom}(\sigma_2)$. We now show that $y \notin \sigma_2(\sigma_1(\rho(cl)))$. By hypothesis, $y \notin \sigma_1(\phi(\rho(cl)))$, and since $\rho(cl)$ is renamed apart from $\phi$, it follows that $y \notin (\sigma_1(\rho(cl)))$. Moreover, as we have seen before, $y \notin \text{vars}(\sigma_2)$, hence $y \notin \text{vars}(\sigma_2(\sigma_1(\rho(cl))))$.

It turns out that we may apply the Composition Lemma (Lemma A.5) and we obtain

$$[\text{mgu}([\sigma_2 \circ \sigma_1]_{\rho(cl)}, [\sigma_1 \circ \phi]_{V \cup G \cup \rho(H)}), V \cup \text{vars}(G) \cup \text{vars}(\rho(H))] = [\{\sigma_2 \circ \sigma_1 \circ \phi\}, \rho(cl) \cup V \cup G] \enspace .$$

By projecting on $G \cup V$ we obtain

$$\mathcal{B}[\![G]\!] D_P^i[\Phi, V] \sqsupseteq_{\text{Ps}} [\{\sigma_2 \circ \sigma_1 \circ \phi]\}, V \cup \text{vars}(G)] \enspace ,$$

which concludes the proof of the atomic case.

*Non-atomic goal.* In this case, decompose the (leftmost) SLD-resolution for $G = A, G'$ in the following way:

$$A\phi, G'\phi \xrightarrow[\sigma_1]{*} G'\phi\sigma_1 \xrightarrow[\sigma_2]{*} \Box \enspace , \tag{A3}$$

where both the sub-derivations have length strictly less than $i$. Note that, since the complete derivation is renamed apart from $V, G, \phi$ and the program $P$, the same holds for the first sub-derivation. Moreover, since $[\phi]_{V\cup G} = \mathrm{mgu}([\phi]_V, [\epsilon]_G)$, each $v \in A$ is free and independent in $[\phi]_{V\cup G}$, i.e., $[\phi]_{V\cup A} = \mathrm{mgu}([\phi]_V, [\epsilon]_A)$. Therefore, we may apply what proved in the atomic case above, obtaining

$$D^i_P A[\Phi, V] \sqsupseteq_{\mathrm{Ps}} [\{\sigma_1 \circ \phi\}, V \cup \mathrm{vars}(A)] \ .$$

The second sub-derivation in (A3) is renamed apart from

- $V$ since the complete derivation is renamed apart from $V$;
- $A$ and $G'$ since the complete derivation is renamed apart from $G$;
- $\sigma_1 \circ \phi$ since the complete derivation is renamed apart from $\phi$ and the second part is renamed apart from $\sigma_1$;
- $P$, since the complete derivation is renamed apart from $P$.

Moreover, assume $x \in \mathrm{vars}(G') \setminus \mathrm{vars}(V \cup A)$ and $x \neq y \in \mathrm{vars}(V \cup G)$. By hypothesis, $[\phi]_{V\cup G} = \mathrm{mgu}([\phi]_V, [\epsilon]_G)$, which implies that $\phi(x) \in \mathcal{V}$ and $\phi(x) \notin \mathrm{vars}(\phi(y))$. Since $\mathrm{vars}(\sigma_1) = W \cup X$ where $W$ is a fresh set of variables disjoint from $V \cup G$ and $\phi$ and $X \subseteq \mathrm{vars}(A\phi)$, it happens that $\phi(x) \notin \mathrm{vars}(\sigma_1)$. Therefore $\sigma_1(\phi(x)) = \phi(x)$ and $\phi(x) \notin \mathrm{vars}(\sigma_1(\phi(y)))$. This implies that $[\sigma_1 \circ \phi]_{V\cup G} = \mathrm{mgu}([\sigma_1 \circ \phi]_{V\cup A}, [\epsilon]_{G'})$ by Prop. A.4. This means that we may apply the inductive hypothesis on the second sub-derivation, obtaining:

$$\mathcal{B}[\![G']\!] D^i_P[\{\sigma_1 \circ \phi\}, V \cup \mathrm{vars}(A)] \sqsupseteq_{\mathrm{Ps}} [\{\sigma_2 \circ \sigma_1 \circ \phi\}, V \cup \mathrm{vars}(G)] \ .$$

Since $\mathcal{B}[\![A, G']\!] D^i_P[\Phi, V] = \mathcal{B}[\![G']\!] D^i_P(D^i_P A[\Phi, V])$ by the above disequalities and monotonicity of $\mathcal{B}$, we obtain

$$\mathcal{B}[\![A, G']\!] D^i_P[\Phi, V] \sqsupseteq_{\mathrm{Ps}} [\{\sigma_2 \circ \sigma_1 \circ \phi\}, V \cup \mathrm{vars}(G)] \ .$$

which concludes the proof. □

Now we may use standard properties of SLD-resolution together with Lemma A.6 to prove the required correctness theorem.

*Theorem A.9*
(Semantic Correctness) Given a program $P$ and an goal $G$, if $\theta$ is a computed answer for the goal $G$, then

$$\mathcal{B}[\![G]\!](\mathcal{P}[\![P]\!])G[\{\epsilon\}, \mathrm{vars}(G)] \sqsupseteq_{\mathrm{Ps}} [\{[\theta]\}, \mathrm{vars}(G)] \ .$$

*Proof*
If $\theta$ is a computed answer for a goal $G$, and $\rho$ is a renaming, then $\theta' = (\rho \circ \theta)_{|\mathrm{vars}(G)}$ is a computed answer too (Apt 1990) and $\theta \sim_{\mathrm{vars}(G)} \theta'$. Consider any such $\theta'$ with the property that $\mathrm{vars}(\theta') \cap \mathrm{vars}(P) = \emptyset$ and let $G \xrightarrow[\theta']{*} \Box$ be a leftmost SLD-resolution for $\theta'$. Since there exists a leftmost SLD-resolution $G \xrightarrow[\theta']{*} \Box$ which is renamed apart from $P$, then, by Lemma A.6, the thesis follows. □

## B Correctness of Forward Unification

*Lemma B.1*
Given $\delta, \sigma \in \mathit{Subst}$, $v \in \mathcal{V}$, it is the case that $\mathit{occ}(\delta \circ \sigma, v) = \mathit{occ}(\sigma, \mathit{occ}(\delta, v))$.

*Proof*
By definition, $x \in \mathit{occ}(\delta \circ \sigma, v)$ iff $v \in \delta(\sigma(x))$, i.e., there exists $w \in \mathcal{V}$ such that $w \in \sigma(x)$ and $v \in \delta(w)$. In other words, $x \in \mathit{occ}(\delta \circ \sigma, v)$ iff there exists $w \in \mathcal{V}$ s.t. $w \in \mathit{occ}(\delta, v)$ and $x \in \mathit{occ}(\sigma, w)$ iff $x \in \mathit{occ}(\sigma, \mathit{occ}(\delta, v))$. □

*Proposition B.2*
Let $t \in \mathsf{Terms}$, $\sigma \in \mathit{Subst}$ and $U \in \wp_f(\mathcal{V})$ such that $\mathrm{vars}(t) \subseteq U$. Let $\alpha_{\mathrm{Sh}}([\sigma]_U) \sqsubseteq_{\mathrm{Sh}} [S, U]$. Then the following property holds:

$$\forall v \in \mathcal{V}. v \in \mathrm{vars}(t\sigma) \iff \mathit{occ}(\sigma, v) \cap U \in \mathbf{rel}(S, t) \ .$$

*Proof*
Note that $v \in \mathrm{vars}(t\sigma)$ iff $\exists u \in t$ such that $v \in \sigma(u)$. In turn, this holds iff $\exists u \in t$ s.t. $u \in \mathit{occ}(\sigma, v)$ iff $\mathit{occ}(\sigma, v) \cap \mathrm{vars}(t) \neq \emptyset$ iff $(\mathit{occ}(\sigma, v) \cap U) \cap \mathrm{vars}(t) \neq \emptyset$. Note that $X = \mathit{occ}(\sigma, v) \cap U \in S$ and therefore $X \cap \mathrm{vars}(t) \neq \emptyset$ iff $X \in \mathbf{rel}(S, t)$ by definition of **rel**. □

*Proposition B.3*
Let $[\sigma]_U \in \mathit{ISubst}_\sim$, $\{x/t\} \in \mathit{ISubst}$ such that $\mathrm{vars}(\{x/t\}) \subseteq U$ and $\sigma$ and $\{x/t\}$ unify. If $\alpha_{\mathrm{Sh}}([\sigma]_U]) \sqsubseteq_{\mathrm{Sh}} [S, U]$ and $\delta = \mathrm{mgu}(x\sigma = t\sigma)$, we obtain:

$$\begin{aligned}\alpha_{\mathrm{Sh}}(\mathrm{mgu}([\sigma]_U, [x/t]_U)) \sqsubseteq_{\mathrm{Sh}} &[(S \setminus (\mathbf{rel}(S, x) \cup \mathbf{rel}(S, t))) \\ &\cup \{\mathit{occ}(\sigma, \mathit{occ}(\delta, v)) \cap U \mid v \in \mathrm{vars}(x\sigma = t\sigma)\}, U] \ .\end{aligned}$$

*Proof*
Since $\mathrm{vars}(\{x/t\}) \subseteq U$, we have $\mathrm{mgu}([\sigma]_U, [x/t]_U) = [\mathrm{mgu}(\sigma, \{x/t\})]_U$. Then, by definition of $\delta$, it holds that $\mathrm{mgu}(\sigma, x = t) = \mathrm{mgu}(\mathrm{Eq}(\sigma) \cup x\sigma = t\sigma) = \mathrm{mgu}(x\sigma = t\sigma) \circ \sigma = \delta \circ \sigma$ (Palamidessi 1990, Prop. 6.1). Therefore, we only need to show that:

$$\begin{aligned}\alpha_{\mathrm{Sh}}([\delta \circ \sigma]_U) \sqsubseteq_{\mathrm{Sh}} &[(S \setminus (\mathbf{rel}(S, x) \cup \mathbf{rel}(S, t))) \\ &\cup \{\mathit{occ}(\sigma, \mathit{occ}(\delta, v)) \cap U \mid v \in \mathrm{vars}(x\sigma = t\sigma)\}, U] \ .\end{aligned} \tag{B1}$$

By definition of $\alpha_{\mathrm{Sh}}$, we have to show that, for all $v \in \mathcal{V}$, $\mathit{occ}(\delta \circ \sigma, v) \cap U \in (S \setminus (\mathbf{rel}(S, x) \cup \mathbf{rel}(S, t))) \cup \{\mathit{occ}(\sigma, \mathit{occ}(\delta, v)) \cap U \mid v \in \mathrm{vars}(x\sigma = t\sigma)\}$. Let $v \in \mathcal{V}$. We have the following cases:

- $v \in \mathrm{vars}(x\sigma = t\sigma)$: by Lemma B.1, $\{\mathit{occ}(\delta \circ \sigma, v) \cap U \mid v \in \mathrm{vars}(x\sigma = t\sigma)\} = \{\mathit{occ}(\sigma, \mathit{occ}(\delta, v)) \cap U \mid v \in \mathrm{vars}(x\sigma = t\sigma)\}$.
- $v \notin \mathrm{vars}(x\sigma = t\sigma)$: thus $v \notin \mathrm{vars}(\delta)$ and $\mathit{occ}(\delta \circ \sigma, v) = \mathit{occ}(\sigma, v)$. We know that $\mathit{occ}(\sigma, v) \cap U \in S$, by definition of $S$. Moreover, we show that $\mathit{occ}(\sigma, v) \cap U \notin \mathbf{rel}(S, x) \cup \mathbf{rel}(S, t)$. Since $v \notin \mathrm{vars}(x\sigma = t\sigma)$, we can apply Prop. B.2 twice to the terms $x$ and $t$, and obtain $\mathit{occ}(\sigma, v) \cap U \notin \mathbf{rel}(S, x) \cup \mathbf{rel}(S, t)$.

By collecting the results of the two cases, Equation (B1) is proved. □

*Proposition B.4*
Let $[\sigma]_U \in \mathit{ISubst}_\sim$, $\{x/t\} \in \mathit{ISubst}$ such that $\mathrm{vars}(\{x/t\}) \subseteq U$ and $\sigma$ and $\{x/t\}$ unify. If $\alpha_{\mathrm{Sh}}([\sigma]_U) \sqsubseteq_{\mathrm{Sh}} [S,U]$ and $x$ is free and independent from $U$ in $[\sigma]_U$, then:

$$\begin{aligned}&\alpha_{\mathrm{Sh}}(\mathrm{mgu}([\sigma]_U,[x/t]_U))\\ &\qquad\sqsubseteq_{\mathrm{Sh}} [(S \setminus (\mathbf{rel}(S,x) \cup \mathbf{rel}(S,t))) \cup \mathbf{bin}(\mathbf{rel}(S,x), \mathbf{rel}(S,t)), U] \ .\end{aligned}$$

*Proof*
First of all note that, without loss of generality, we may assume $x \notin \mathrm{vars}(\sigma)$. Then, by Prop.B.3, we have that:

$$\begin{aligned}\alpha_{\mathrm{Sh}}(\mathrm{mgu}([\sigma]_U,[x/t]_U)) \sqsubseteq_{\mathrm{Sh}}&[(S \setminus (\mathbf{rel}(S,x) \cup \mathbf{rel}(S,t)))\\ &\cup \{\mathit{occ}(\sigma, \mathit{occ}(\delta,v)) \cap U \mid \in \mathrm{vars}(x\sigma = t\sigma)\}, U] \ ,\end{aligned}$$

where $\delta = \mathrm{mgu}(x\sigma = t\sigma)$. Since $x \notin \mathrm{vars}(\sigma)$, we have that $x\sigma = t\sigma$ is equal to $x = t\sigma$. Moreover, $x \notin \mathrm{vars}(t\sigma)$ since $x \notin \mathrm{vars}(t)$ and $x \notin \mathrm{vars}(\sigma)$ by hypothesis. Thus $\delta = \mathrm{mgu}(x = t\sigma) = \{x/t\sigma\}$. It follows that $\mathrm{vars}(x\sigma = t\sigma) = \{x\} \cup \mathrm{vars}(t\sigma)$. Therefore, the following equalities hold:

$$\begin{aligned}&\{\mathit{occ}(\sigma,\mathit{occ}(\delta,v)) \cap U \mid v \in \mathrm{vars}(x\sigma = t\sigma)\}\\ =\ &\{\mathit{occ}(\sigma,\mathit{occ}(\delta,v)) \cap U \mid v \in \{x\} \cup \mathrm{vars}(t\sigma)\}\\ =\ &\{\mathit{occ}(\sigma,\mathit{occ}(\delta,v)) \cap U \mid v \in \mathrm{vars}(t\sigma)\} && [\text{since } x \in \mathrm{dom}(\delta),\ \mathit{occ}(\delta,x) = \emptyset]\\ =\ &\{\mathit{occ}(\sigma,\{x,v\}) \cap U \mid v \in \mathrm{vars}(t\sigma)\} && [\text{since } \delta = \{x/t\sigma\}]\\ =\ &\{(\mathit{occ}(\sigma,x) \cup \mathit{occ}(\sigma,v)) \cap U \mid v \in \mathrm{vars}(t\sigma)\}\\ =\ &\{(\{x\} \cup \mathit{occ}(\sigma,v)) \cap U \mid v \in \mathrm{vars}(t\sigma)\} && [\text{since } x \notin \mathrm{vars}(\sigma)]\end{aligned}$$

Moreover, for each $v \in \mathrm{vars}(t\sigma)$, by Prop. B.2 it holds that $\mathit{occ}(\sigma,v) \cap U \in \mathbf{rel}(S,t)$. Therefore, $\{(\{x\} \cup \mathit{occ}(\sigma,v)) \cap U \mid v \in \mathrm{vars}(t\sigma)\} \subseteq \mathbf{bin}(\{x\}, \mathbf{rel}(S,t))$. Since $x \notin \mathrm{vars}(\sigma)$ and $x \in U$, it follows that $\mathit{occ}(\sigma,x) = \{x\}$ and thus $\{x\} \in \mathbf{rel}(S,x)$ being $\alpha_{\mathrm{Sh}}([\sigma]_U) \sqsubseteq_{\mathrm{Sh}} [S,U]$. As a consequence $\mathbf{bin}(\{x\}, \mathbf{rel}(S,t)) \subseteq \mathbf{bin}(\mathbf{rel}(S,x), \mathbf{rel}(S,t))$ from which it follows that $\alpha_{\mathrm{Sh}}([\mathrm{mgu}(\mathrm{Eq}(\sigma) \cup x = t)]_U) \sqsubseteq_{\mathrm{Sh}} [(S \setminus (\mathbf{rel}(S,x) \cup \mathbf{rel}(S,t))) \cup \mathbf{bin}(\mathbf{rel}(S,x), \mathbf{rel}(S,t)), U]$. □

*Proposition B.5*
Given $s,t \in \mathsf{Terms}$ and $W, Y \in \wp_f(\mathcal{V})$ such that $s$ and $t$ unify, $\mathrm{vars}(s = t) \subseteq W$ and $Y \subseteq \mathrm{uvars}(s = t)$, then $\delta = \mathrm{mgu}(s = t)$ enjoys the following properties:

1. $\forall v \in \mathrm{vars}(s).\ \mathit{occ}(\delta,v) \cap \mathrm{vars}(s) \neq \emptyset \Rightarrow \mathit{occ}(\delta,v) \cap \mathrm{vars}(t) \neq \emptyset$ ,
2. $\forall v \in \mathrm{vars}(s).\ \mathit{occ}(\delta,v) \cap \mathrm{vars}(s) \supseteq \{x_1,x_2\} \wedge x_1 \neq x_2 \Rightarrow \mathit{occ}(\delta,v) \cap Z \neq \emptyset$ .

where $Z = \mathrm{vars}(t) \setminus Y$.

*Proof*
We prove the two points separately.

1. If $\mathit{occ}(\delta,v) \cap \mathrm{vars}(s) \neq \emptyset$ then $v \notin \mathrm{dom}(\delta)$ and therefore $v \in \delta(s)$. Since $\delta$ is an unifier for $s$ and $t$, it should be $v \in \delta(t)$, and therefore there exists $y \in t$ such that $y \in \mathit{occ}(\delta,v)$.

2. First of all, note that, given two terms $s$ and $t$ in a given signature $\Sigma$, the result of $\mathrm{mgu}(s = t)$ does not change if we enlarge $\Sigma$ with a new constant symbol. Therefore, assume without loss of generality that there is a constant symbol $a$ in the signature. The proof proceeds by contradiction.
   Assume that there exist $x_1, x_2 \in \mathrm{vars}(s)$, $v \in W$ such that $x_1, x_2 \in \mathit{occ}(\delta, v)$ and $\mathit{occ}(\delta, v) \cap Z = \emptyset$. Let $\sigma = \{x = a \mid x \in W\}$ and consider the substitution $\delta' = \{z/(\delta(z))\sigma \mid z \in Z\}$. Note that this is an idempotent substitution since it is ground. Now consider $\delta'' = \mathrm{mgu}(\mathrm{Eq}(\delta) \cup \mathrm{Eq}(\delta'))$, which clearly exists and, by definition of $\delta'$, is $\delta'' = \{x/a \mid x \in \mathrm{vars}(\delta(Z))\} \circ \delta$. Therefore, $\mathit{occ}(\delta'', v) = \mathit{occ}(\delta, v)$ because $v \notin \mathrm{vars}(\delta(Z))$ being $\mathit{occ}(\delta, v) \cap Z = \emptyset$.
   Moreover, $\delta'' = \mathrm{mgu}(\mathrm{Eq}(\delta) \cup \mathrm{Eq}(\delta')) = \mathrm{mgu}(\{s = t\} \cup \mathrm{Eq}(\delta')) = \mathrm{mgu}(s\delta' = t\delta') \circ \delta' = \delta' \uplus \mathrm{mgu}(s\delta' = t\delta')$. By definition of $\delta'$, it holds that $\mathrm{vars}(t\delta') \cap Z = \emptyset$, and thus $\mathrm{vars}(t\delta') \subseteq Y$. From the definition of $Y$ it follows that $\mathrm{vars}(t\delta') \subseteq \mathrm{uvars}(s = t)$, and thus $\mathrm{vars}(t\delta') \subseteq \mathrm{uvars}(s\delta' = t\delta')$, since $\mathrm{rng}(\delta') = \emptyset$. Therefore the term $t\delta'$ is linear and independent from $s\delta'$ and $\mathit{occ}(\mathrm{mgu}(s\delta' = t\delta'), v) = \mathit{occ}(\mathrm{mgu}(s\delta' = t\delta') \uplus \delta', v) = \mathit{occ}(\delta, v)$.
   If we apply the result for linear and independent terms, e.g., (King 2000, Prop. 3.1), we obtain an absurd, since it is not possible that both $x_1$ and $x_2$ are elements of $\mathit{occ}(\mathrm{mgu}(s\delta' = t\delta'), v)$.

This concludes the proof. □

*Proposition B.6*
Let $[\sigma]_U \in \mathit{ISubst}_{\sim}$, $\{x/t\} \in \mathit{ISubst}$ such that $\mathrm{vars}(\{x/t\}) \subseteq U$ and $\sigma$ and $\{x/t\}$ unify. Given $Y \subseteq \mathrm{vars}(t)$ such that, for all $y \in Y$, $\mathrm{vars}(\sigma(y)) \subseteq \mathrm{uvars}(x\sigma = t\sigma)$, if $\alpha_{\mathrm{Sh}}([\sigma]_U) \sqsubseteq_{\mathrm{Sh}} [S, U]$ then

$$\begin{aligned}\alpha_{\mathrm{Sh}}(\mathrm{mgu}([\sigma]_U, [x/t]_U)) \sqsubseteq_{\mathrm{Sh}} [&(S \setminus (\mathbf{rel}(S,t) \cup \mathbf{rel}(S,x)))\\ &\cup \mathbf{bin}(\mathbf{rel}(S,x), \mathbf{rel}(S,Y)^*) \cup \mathbf{bin}(\mathbf{rel}(S,x)^*, \mathbf{rel}(S,Z)^*)\\ &\cup \mathbf{bin}(\mathbf{bin}(\mathbf{rel}(S,x)^*, \mathbf{rel}(S,Z)^*), \mathbf{rel}(S,Y)^*), U]\ ,\end{aligned}$$

where $Z = \mathrm{vars}(t) \setminus Y$.

*Proof*
By Prop. B.3, we have that

$$\begin{aligned}\alpha_{\mathrm{Sh}}(\mathrm{mgu}([\sigma]_U, [x/t]_U)) \sqsubseteq_{\mathrm{Sh}} [&(S \setminus (\mathbf{rel}(S,x) \cup \mathbf{rel}(S,t)))\\ &\cup \{\mathit{occ}(\sigma, \mathit{occ}(\delta, v)) \cap U \mid v \in \mathrm{vars}(x\sigma = t\sigma)\}, U]\ ,\end{aligned}$$

where $\delta = \mathrm{mgu}(x\sigma = t\sigma)$. We show that

$$\begin{aligned}\{\mathit{occ}(\sigma, \mathit{occ}(\delta, v)) \cap U \mid v \in \mathrm{vars}(x\sigma = t\sigma)\}\\ \subseteq \mathbf{bin}(\mathbf{rel}(S,x), \mathbf{rel}(S,Y)^*) \cup \mathbf{bin}(\mathbf{rel}(S,x)^*, \mathbf{rel}(S,Z)^*)\\ \cup \mathbf{bin}(\mathbf{bin}(\mathbf{rel}(S,x)^*, \mathbf{rel}(S,Z)^*), \mathbf{rel}(S,Y)^*) \cup \{\emptyset\}\ ,\end{aligned}$$

from which the thesis follows. The following equalities hold, for all $v \in \mathrm{vars}(x\sigma =$

$t\sigma$).

$$\begin{aligned} & occ(\sigma, occ(\delta, v)) \cap U \\ = & \bigcup\{occ(\sigma, w) \cap U \mid w \in occ(\delta, v)\} \\ = & \bigcup\{occ(\sigma, w) \cap U \mid w \in occ(\delta, v) \cap \mathrm{vars}(x\sigma)\} \\ & \cup \bigcup\{occ(\sigma, w) \cap U \mid w \in occ(\delta, v) \cap \mathrm{vars}(t\sigma)\} \\ & \qquad [\![\text{by partitioning the variables in } occ(\delta, v) \subseteq \mathrm{vars}(\delta) \cup \{v\}]\!] \end{aligned}$$

By applying Prop. B.5 (1) to the equation $x\sigma = t\sigma$ we get $occ(\delta, v) \cap \mathrm{vars}(x\sigma) \neq \emptyset$ iff $occ(\delta, v) \cap \mathrm{vars}(t\sigma) \neq \emptyset$. Since the case $occ(\delta, v) = \emptyset$ is trivial, it only remain to consider the case $occ(\delta, v) \neq \emptyset$ which implies $occ(\delta, v) \cap \mathrm{vars}(t\sigma) \neq \emptyset \neq occ(\delta, v) \cap \mathrm{vars}(x\sigma)$. In the following, let $A = \bigcup\{occ(\sigma, w) \cap U \mid w \in occ(\delta, v) \cap \mathrm{vars}(x\sigma)\}$ and $B = \bigcup\{occ(\sigma, w) \cap U \mid w \in occ(\delta, v) \cap \mathrm{vars}(t\sigma)\}$. Note that, by Prop. B.2, $occ(\sigma, w) \cap U \in \mathbf{rel}(S, \{x\})$ if $w \in \mathrm{vars}(x\sigma)$ and $x \in U$, which implies $A \in \mathbf{rel}(S, \{x\})^*$. For the same reason, $B \in \mathbf{rel}(S, \mathrm{vars}(t))^*$, i.e.,

$$occ(\sigma, occ(\delta, v)) \cap U \in \mathbf{bin}(\mathbf{rel}(S, \{x\})^*, \mathbf{rel}(S, \mathrm{vars}(t))^*) \ ,$$

which is the standard result for abstract unification without considering freeness or linearity. We can do better if we proceed by cases on $occ(\delta, v) \cap \mathrm{vars}(t\sigma)$.

$\boldsymbol{occ(\delta, v) \cap \mathrm{vars}(t\sigma) \subseteq \mathrm{vars}(\sigma(Y))}$ Let $Z' = \mathrm{vars}(t\sigma) \setminus \mathrm{vars}(\sigma(Y))$ it follows that $occ(\delta, v) \cap Z' = \emptyset$. Therefore, by Prop. B.5(2) applied to the terms $x\sigma$ and $t\sigma$, we have that $\nexists x_1, x_2 \in \mathrm{vars}(x\sigma)$ such that $x_1, x_2 \in occ(\delta, v)$. Since $occ(\delta, v) \cap \mathrm{vars}(x\sigma) \neq \emptyset$, it follows that there exists $x' \in \mathrm{vars}(x\sigma)$ such that $occ(\delta, v) \cap \mathrm{vars}(x\sigma) = \{x'\}$. This implies that $A \in \mathbf{rel}(S, \{x\})$. Moreover, by Prop. B.2 applied to the set of variables $Y$, $B \in \mathbf{rel}(S, Y)^*$ and this proves

$$occ(\sigma, occ(\delta, v)) \cap U \in \mathbf{bin}(\mathbf{rel}(S, \{x\}), \mathbf{rel}(S, Y)^*) \ .$$

**otherwise** We are in the case that $occ(\delta, v) \cap \mathrm{vars}(t\sigma) \nsubseteq \mathrm{vars}(\sigma(Y))$, i.e., $occ(\delta, v) \cap \mathrm{vars}(\sigma(Z)) \neq \emptyset$. Therefore, there exists $w \in occ(\delta, v) \cap \mathrm{vars}(\sigma(Z))$ and using Prop. B.2 we have that $occ(\sigma, w) \cap U \in \mathbf{rel}(S, Z)$. This implies that $B \in \{B_1 \cup \ldots B_n \cup C_1 \cup \ldots C_p \mid B_i \in \mathbf{rel}(S, Y), n \geq 0, C_i \in \mathbf{rel}(S, Z), p \geq 1\} = \mathbf{rel}(S, Z)^* \cup \mathbf{bin}(\mathbf{rel}(S, Y)^*, \mathbf{rel}(S, Z)^*)$. As a final result we have that:

$$\begin{aligned} occ(\sigma, occ(\delta, v)) \cap U \in\ & \mathbf{bin}(\mathbf{rel}(S, \{x\})^*, \mathbf{rel}(S, Z)^* \cup \mathbf{bin}(\mathbf{rel}(S, Y)^*, \mathbf{rel}(S, Z)^*)) \\ =\ & \mathbf{bin}(\mathbf{rel}(S, \{x\})^*, \mathbf{rel}(S, Z)^*) \cup \\ & \mathbf{bin}(\mathbf{bin}(\mathbf{rel}(S, \{x\})^*, \mathbf{rel}(S, Z)^*), \mathbf{rel}(S, Y)^*) \ , \end{aligned}$$

which proves the theorem. □

*Lemma B.7*
Let $[\sigma]_V \in \mathit{ISubst}_\sim$, $\theta \in \mathit{ISubst}$ such that $\mathrm{vars}(\theta) \subseteq V$ and $\sigma$ and $\theta$ unify. Assume given $U \subseteq V$ such that, for each $x \in U$,

1. $x$ is free in $[\sigma]_V$;
2. $x$ is independent from $\mathrm{vars}(\theta)$ in $[\sigma]_V$;
3. if $x \in \mathrm{dom}(\theta)$ then $x$ is independent in $[\sigma]_V$.

If $\alpha_{\mathrm{Sh}}([\sigma]_V) \sqsubseteq_{\mathrm{Sh}} [S, V]$ then $\alpha_{\mathrm{Sh}}(\mathrm{mgu}([\sigma]_V, [\theta]_V)) \sqsubseteq_{\mathrm{Sh}} [\mathbf{u}^f_{\mathrm{Sh}}(S, U, \theta), V]$.

*Proof*

The proof is by induction on $|\mathrm{dom}(\theta)|$. Assume $|\mathrm{dom}(\theta)| = 0$, then $\theta = \epsilon$ and $\alpha_{\mathrm{Sh}}(\mathrm{mgu}([\sigma]_V, [\epsilon]_V)) = \alpha_{\mathrm{Sh}}([\sigma]_V) \sqsubseteq_{\mathrm{Sh}} [S, V] = [\mathbf{u}^f_{\mathrm{Sh}}(S, U, \epsilon), V]$.

Now assume that it holds for $|\mathrm{dom}(\theta)| \leq n$ and we show it holds for $|\mathrm{dom}(\theta)| = n + 1$, too. Let $\theta$ be $\theta' \uplus \{x/t\}$. We distinguish two cases: either $x \in U$ or $x \notin U$.

1. ($x \in U$) By definition of $\mathbf{u}^f_{\mathrm{Sh}}$ we have that

$$\begin{aligned}&\mathbf{u}^f_{\mathrm{Sh}}(S, U, \{x/t\} \uplus \theta') \\ &= \mathbf{u}^f_{\mathrm{Sh}}((S \setminus (\mathbf{rel}(S, x) \cup \mathbf{rel}(S, t))) \cup \mathbf{bin}(\mathbf{rel}(S, x), \mathbf{rel}(S, t)), U \setminus \{x\}, \theta') \enspace .\end{aligned}$$

   Since $x \in U \cap \mathrm{dom}(\theta)$, by hypothesis $x$ is free and independent in $[\sigma]_V$. Thus we can apply Prop. B.4, from which we obtain that:

$$\begin{aligned}&\alpha_{\mathrm{Sh}}(\mathrm{mgu}([\sigma]_V, [x/t]_V)) \\ &\sqsubseteq_{\mathrm{Sh}} [S \setminus (\mathbf{rel}(S, x) \cup \mathbf{rel}(S, t)) \cup \mathbf{bin}(\mathbf{rel}(S, x), \mathbf{rel}(S, t)), V] \enspace .\end{aligned}$$

   Let $[\sigma']_V = \mathrm{mgu}([\sigma]_V, [x/t]_V)$ and $U' = U \setminus \{x\}$. We may assume without loss of generality that $\mathrm{vars}(\sigma) \cap U = \emptyset$ and we obtain $\sigma' = \mathrm{mgu}(\mathrm{Eq}(\sigma) \cup \{x = t\}) = \sigma \uplus \{x/t\sigma\}$. Given $u \in U'$, we have $\sigma'(u) = \sigma(u) = u \in \mathcal{V}$, hence $u$ is free in $[\sigma]_V$. If $u \neq v \in \mathrm{vars}(\theta')$, then $v \neq x$ and therefore $u \notin \sigma'(v) = \sigma(v)$. Thus $u$ is independent from $\mathrm{vars}(\theta')$ in $[\sigma']_V$. Moreover, if $u \in \mathrm{dom}(\theta')$, then $u \neq x$, $u \notin t$ and $u \notin \mathit{vars}(\sigma)$, and therefore $u \notin \mathrm{vars}(\sigma') \subseteq \mathrm{vars}(\sigma) \cup \mathrm{vars}(x = t)$. This means that $u$ is independent in $[\sigma']_U$. Therefore, by inductive hypothesis,

$$\begin{aligned}\alpha_{\mathrm{Sh}}(\mathrm{mgu}([\sigma]_V, [\theta]_V)) &= \alpha_{\mathrm{Sh}}(\mathrm{mgu}([\sigma']_V, [\theta']_V)) \\ &\sqsubseteq_{\mathrm{Sh}} [\mathbf{u}^f_{\mathrm{Sh}}(S', U', \theta'), V] = [\mathbf{u}^f_{\mathrm{Sh}}(S, U, \theta), V] \enspace ,\end{aligned}$$

   which concludes this part of the proof.

2. ($x \notin U$) By definition of $\mathbf{u}^f_{\mathrm{Sh}}$ we have that:

$$\begin{aligned}\mathbf{u}^f_{\mathrm{Sh}}(S, U, \{x/t\} \uplus \theta) = \mathbf{u}^f_{\mathrm{Sh}}(&(S \setminus (\mathbf{rel}(S, x) \cup \mathbf{rel}(S, t))) \\ &\cup \mathbf{bin}(\mathbf{rel}(S, x), \mathbf{rel}(S, Y)^*) \cup \mathbf{bin}(\mathbf{rel}(S, x)^*, \mathbf{rel}(S, Z)^*) \\ &\cup \mathbf{bin}(\mathbf{bin}(\mathbf{rel}(S, x)^*, \mathbf{rel}(S, Y)^*), \mathbf{rel}(S, Z)^*)), U \setminus \mathrm{vars}(\{x/t\}), \delta) \enspace ,\end{aligned}$$

   where $Y = \mathrm{uvars}(t) \cap U$ and $Z = \mathrm{vars}(t) \setminus Y$. Since $Y \subseteq U$, then for all $u \in Y$ and for all $v \in \mathrm{vars}(x = t)$ with $v \neq u$, it is the case that $v$ and $u$ do not share variables, i.e., $v \neq u \Rightarrow \sigma(u) \notin \sigma(v)$. Therefore $\sigma(u) \in \mathrm{uvars}(x\sigma = t\sigma)$. Then we can apply Prop. B.6 to obtain

$$\begin{aligned}\alpha_{\mathrm{Sh}}([\sigma]_V, [x/t]_V]) \sqsubseteq_{\mathrm{Sh}} &(S \setminus (\mathbf{rel}(S, t) \cup \mathbf{rel}(S, x))) \\ &\cup \mathbf{bin}(\mathbf{rel}(S, x), \mathbf{rel}(S, Y)^*) \cup \mathbf{bin}(\mathbf{rel}(S, x)^*, \mathbf{rel}(S, Z)^*) \\ &\cup \mathbf{bin}(\mathbf{bin}(\mathbf{rel}(S, x)^*, \mathbf{rel}(S, Z)^*), \mathbf{rel}(S, Y)^*), V] \enspace .\end{aligned}$$

   Again, assume $\mathrm{vars}(\sigma) \cap U = \emptyset$, $\sigma' = \mathrm{mgu}(\mathrm{Eq}(\sigma) \cup \{x = t\}) = \mathrm{mgu}(x\sigma = t\sigma) \circ \sigma$ and $U' = U \setminus \mathrm{vars}(\{x/t\})$. Given $u \in U'$, $u \notin \mathrm{vars}(x = t)$ and since $u$ by hypothesis does not share with any variable in $x = t$, we have $u \notin \mathrm{vars}(\{x\sigma/t\sigma\})$. As a result $\sigma'(u) = \sigma(u) = u \in \mathcal{V}$. Moreover, for each variable $v$, $u \in \sigma'(v)$ iff $u \in \sigma(v)$. Therefore, if $v \in \mathrm{vars}(\theta')$ and $v \neq u$, $v$ and $u$ are independent in $[\sigma']_V$. Finally, if $u \in \mathrm{dom}(\theta')$, then $u \notin \mathrm{vars}(\sigma)$ which implies $u \notin \mathrm{vars}(\sigma')$. By inductive hypothesis we have

$$\alpha_{\mathrm{Sh}}(\mathrm{mgu}([\sigma]_V, [\theta]_V)) = \alpha_{\mathrm{Sh}}(\mathrm{mgu}([\sigma']_V, [\theta']_V)) \sqsubseteq_{\mathrm{Sh}} [\mathbf{u}^f_{\mathrm{Sh}}(S', U', \theta'), V] = [\mathbf{u}^f_{\mathrm{Sh}}(S, U, \theta), V] \ ,$$

which proves the lemma. □

*Theorem B.8*
(Correctness of $\mathsf{unif}_{\mathrm{Sh}}$) The unification operator $\mathsf{unif}_{\mathrm{Sh}}$ is correct w.r.t. $\mathsf{unif}_{\mathrm{Ps}}$.

*Proof*
Given $[\Delta, V] = \mathsf{unif}_{\mathrm{Ps}}([\Delta_1, V_1], \delta)$, we know that, if $[\theta]_V \in \Delta$, then

$$[\theta]_V = \mathrm{mgu}([\theta_1]_{V_1}, [\delta]_{\mathrm{vars}(\delta)}) = \mathrm{mgu}([\theta_1]_{V_1}, [\epsilon]_{V_1 \cup \mathrm{vars}(\delta)}, [\delta]_{\mathrm{vars}(\delta)}) \ .$$

Note that, if $\alpha_{\mathrm{Sh}}([\theta_1]_{V_1}) \sqsubseteq_{\mathrm{Sh}} [S, V_1]$, then

$$\alpha_{\mathrm{Sh}}(\mathrm{mgu}([\theta_1]_{V_1}, [\epsilon]_{V_1 \cup \mathrm{vars}(\delta)})) \sqsubseteq_{\mathrm{Sh}} [S \cup \{\{x\} \mid x \in \mathrm{vars}(\delta) \setminus V_1\}, V_1 \cup \mathrm{vars}(\delta)]$$

and each variable in $\mathrm{vars}(\delta) \setminus V_1$ is free and independent in $\mathrm{mgu}([\theta_1]_{V_1}, [\epsilon]_{V_1 \cup \mathrm{vars}(\delta)})$. Therefore, by applying Lemma B.7, we obtain

$$\alpha_{\mathrm{Sh}}([\theta]_V) \sqsubseteq_{\mathrm{Sh}} \mathsf{unif}_{\mathrm{Sh}}([S, V_1], \delta) \ .$$

The theorem follows by the pointwise extension of $\alpha_{\mathrm{Sh}}$ to elements of `Psub`. □

## C Optimality of Forward Unification

We first introduce some notations. Given $[S_1, U_1] \in$ `Sharing` and $\theta \in \mathit{ISubst}$, let $\mathsf{unif}_{\mathrm{Sh}}([S_1, U_1], \theta) = [S, U_1 \cup \mathrm{vars}(\theta)]$ and $X \in S$. To ease notation, let us define $U_2 = \mathrm{vars}(\theta) \setminus U_1$, $S_2 = \{\{x\} \mid x \in U_2\}$, $U = U_1 \cup U_2$, $X_1 = X \cap U_1$ and $X_2 = X \cap U_2$.

We begin by checking some properties of the unification algorithm in $\mathbf{u}^f_{\mathrm{Sh}}$. To simplify the notation, in the rest of this section we will use a slightly modified version of the operator $\mathbf{u}^f_{\mathrm{Sh}}$ which uses the rule $\mathbf{u}^f_{\mathrm{Sh}}(T, V, \epsilon) = (T, V)$ (instead of the original rule $\mathbf{u}^f_{\mathrm{Sh}}(T, V, \epsilon) = T$). The only consequence of this modification is that the new operator returns a pair whose first argument is the same as in the original operator and whose second argument is a set of variables guaranteed to be free after the unification.

*Remark C.1*
Given $(T', V') = \mathbf{u}^f_{\mathrm{Sh}}(T, V, \theta)$ the following properties are easily checked from the definition:

1. $V' \subseteq V$;
2. if $x \in V' \cap \mathrm{rng}(\theta)$ and $x \in \theta(v)$, then $v \in V$.
3. $\mathbf{u}^f_{\mathrm{Sh}}(T, V, \theta \uplus \theta') = \mathbf{u}^f_{\mathrm{Sh}}(T', V', \theta')$

Let $[H, U] = \alpha_{\mathrm{Sh}}([\theta]_U)$. We want to prove that each $X \in S$ is obtained as union of a number of sharing groups in $H$. However, these sharing groups cannot be joined freely but only according to some conditions.

*Lemma C.2*
For each $X \in S$, either $X \in H$ or there are $B_1, \dots, B_k \in H$ s.t. $\cup_{i \leq k} B_i = X$ and for each $i \leq k$, $B_i \cap U_1 \neq \emptyset$.

*Proof*
The proof proceeds by induction on the number of bindings $n$ in $\theta$. If $n = 0$, then $\theta = \epsilon$, $S = S_1 \cup S_2$ and $H = \{\{x\} \mid x \in U_1 \cup U_2\}$. If $X \in S_2$ then $X = \{x\}$ for some $x \in U_2$, i.e., $X \in H$. Otherwise, if $X \in S_1$, then $X = \bigcup\{\{x\} \mid x \in X\}$. Since $x \in \text{vars}(S_1)$ entails $x \in U_1$, we may take as $B_i$'s the singletons $\{x\}$ for each $x \in X$ and we have the required result.

If $n \neq 0$ then $\theta = \theta' \uplus \{x/t\}$ and $\mathbf{u}^f_{\text{Sh}}(S_1 \cup S_2, U_2, \theta) = \mathbf{u}^f_{\text{Sh}}(T, V, \{x/t\})$ where $(T, V) = \mathbf{u}^f_{\text{Sh}}(S_1 \cup S_2, U_2, \theta')$. Let $[H', U] = \alpha_{\text{Sh}}([\theta']_U)$. We distinguish the cases $x \in V$ and $x \notin V$.

Assume $x \in V$. If $X \in T \setminus (\mathbf{rel}(T, t) \cup \mathbf{rel}(T, x))$ then $X \cap \text{vars}(\{x/t\}) = \emptyset$. By inductive hypothesis, $X = B_1 \cup \ldots \cup B_h$ where each $B_j \in H'$. Since $B_j \cap \text{vars}(\{x/t\}) = \emptyset$, we have $B_j \in H$ and therefore the property is satisfied. Otherwise, $X = A_1 \cup A_2$ where $A_1 \in \mathbf{rel}(T, x)$ and $A_2 \in \mathbf{rel}(T, t)$. Note that since $x \notin \text{vars}(\theta')$, then $\mathbf{rel}(H', x) = \{\{x\}\}$. Since $\{x\} \cap U_1 = \emptyset$, it is not possible to join $\{x\}$ with any other sharing group in $H'$, and therefore $\mathbf{rel}(T, x) = \{\{x\}\}$ and $A_1 = \{x\}$. Now assume, without loss of generality, $A_2 \in \mathbf{rel}(T, y)$, with $y \in \text{vars}(t)$. By inductive hypothesis $A_2 = C_1 \cup \ldots \cup C_h$ with each $C_j \in H'$. First of all, note that, for each $j$, either $C_j \cap \text{vars}(\{x/t\}) = \emptyset$ which entails $C_j \in H$, or $C_j = \mathit{occ}(\theta', w)$ for some $w \in \text{vars}(t)$, which entails $\{x\} \cup C_j = \mathit{occ}(\theta, w) \in H$. Therefore, it is possible to take $k = h$ and $B_j$ equals either to $C_j$ or $C_j \cup \{x\}$ so that $B_j \in H$. Since there is at least one index $l$ such that $y \in C_l$, then $C_l = \mathit{occ}(\theta', y)$ and $x \in B_l$. Therefore $\cup_j B_j = X$. Moreover, either $h = 1$ or $h > 1$ and $C_j \cap U_1 \neq \emptyset$ for each $j \leq h$.

Now assume $x \notin V$. If $X \in T \setminus (\mathbf{rel}(T, t) \cup \mathbf{rel}(T, x))$ then $X \cap \text{vars}(\{x/t\}) = \emptyset$ and everything is as for the case $x \in V$. Otherwise, the three cases in the definition of $\mathbf{u}^f_{\text{Sh}}$ may be subsumed saying that $X = A_1 \cup A_2$ where $A_1 \in \mathbf{rel}(S, x)^*$ and $A_2 \in \mathbf{rel}(S, t)^*$. Assume, by inductive hypothesis, that $A_1 = C^1_1 \cup \ldots \cup C^1_h$ where each $C^1_j \in H'$ and $A_2 = C^2_1 \cup \ldots \cup C^2_l$ where each $C^2_j \in H'$. Since $x \notin \text{vars}(\theta')$ then $\mathbf{rel}(H', x) = \{\{x\}\}$. Therefore there exists $C^1_j$ such that $C^1_j = \{x\}$. We assume without loss of generality that $C^1_1 = \{x\}$. As for the case with $x \in V$, we may define $B^2_j$ equals to either $C^2_j$ or $C^2_j \cup \{x\}$ so that $B^2_j \in H$. The same holds for all the elements of the kind $C^1_j$ for $j > 1$. Moreover, there is at least one $j$ such that $C^2_j = \mathit{occ}(\theta', y)$ for some $y \in \text{vars}(t)$, i.e., such that $x \in B^2_j$. Then, we have a collection of elements $B^1_j$ and $B^2_j$ such that each $B^1_j, B^2_j \in H$ and whose union gives $X$. We only need to prove that $B^1_j \cap U_1 \neq \emptyset$ and $B^2_j \cap U_1 \neq \emptyset$ for each $j$. Note that if $C^2_j \cap U_1 \neq \emptyset$, then $B^2_j \cap U_1 \neq \emptyset$. Assume $C^2_j \cap U_1 = \emptyset$. By inductive hypothesis, this happens if $C^2_j \in \mathbf{rel}(S, t)$ (otherwise $C^2_j$ is obtained by joining more than one element in $H'$, and therefore it must contains some variable in $U_1$) . Thus, there exists $y \in \text{vars}(t)$ such that $y \in C^2_j$, and therefore $B^2_j = C^2_j \cup \{x\}$ and $B^2_j \cap U_1 \neq \emptyset$. In the same way, if $C^1_j \cap U_1 \neq \emptyset$ the same holds for $B^1_j$. Note that, given $C^1_j$, by inductive hypothesis either $C^1_j \notin \mathbf{rel}(S, x)$ and therefore $C^1_j \cap U_1 \neq \emptyset$, or $C^1_j \in \mathbf{rel}(S, x)$, and therefore $x \in C^1_j$ which entails again $C^1_j \cap U_1 \neq \emptyset$. □

*Corollary C.3*
$X = \{x \mid \text{vars}(\theta(x)) \cap X \neq \emptyset\}$.

*Proof*
By Lemma C.2 we know $X = B_1 \cup \dots \cup B_N$ with $B_i \in H$. If $x \in X$ then $x \in B_i$ for some $i \leq N$. Assume $B_i = occ(\theta, w)$. Then $w \in B_i \subseteq X$ and $w \in \text{vars}(\theta(x)) \cap X$. In the opposite direction, assume $z \in \text{vars}(\theta(x)) \cap X$. Since there is only one sharing group $B$ in $H$ such that $z \in B$, namely $B = occ(\theta, z)$, it must be the case that $B = B_j$ for some $j \in \{1, \dots, N\}$ and therefore $x \in B_j \subseteq X$. □

*Lemma C.4*
For each $X \in S$, $X$ is $\theta$-connected.

*Proof*
First note that, if $X$ is $\theta$-connected and $Y \subseteq U_2$, then given $\theta' = \theta \uplus \theta''$, it holds that $X \cup Y$ is $\theta'$-connected.

The proof is by induction on the number of bindings in $\theta$. If $\theta = \epsilon$ there is nothing to prove since $X \in S_1 \cup S_2$, and thus $X_1 \in S_1$.

Let $\theta = \theta' \uplus \{x/t\}$, $[H', U] = \alpha_{\text{Sh}}([\theta']_U)$, and $(S, V') = \mathbf{u}^f_{\text{Sh}}(T, V, \{x/t\})$ where $\mathbf{u}^f_{\text{Sh}}(S_1 \cup S_2, U_2, \theta') = (T, V)$.

We distinguish two cases according to the fact that $x \in V$ or not. Consider the case $x \in V$, which implies $x \in U_2$. By hypothesis $x \notin \text{vars}(\theta')$ therefore, by Lemma C.2, $\mathbf{rel}(T, x) = \{\{x\}\}$. Therefore $S$ is obtained by joining to each $Q \in \mathbf{rel}(T, t)$ the new sharing group $\{x\}$ and removing $\{x\}$ from $T$. It happens that each $Q \in S$ is $\theta$-connected since: 1) either $Q \in T$; 2) or $Q = Q' \cup \{x\}$ for $Q' \in T$ and $x \in U_2$. In the first case, $Q$ is $\theta'$-connected by inductive hypothesis, hence it is also $\theta$ connected and the thesis follows. In the latter case, $Q'$ is $\theta'$-connected, and thus $Q' \cup \{x\}$ is $\theta$-connected since $x \in U_2$.

The other case is when $x \notin V$. If we take $Q \in S$ and assume $Q \in T \setminus (\mathbf{rel}(T, x) \cup \mathbf{rel}(T, t))$, then it is $\theta'$-connected by inductive hypothesis, and thus it is $\theta$-connected. Otherwise, take $Q = Q_1 \cup Q_2$ with $Q_1 \in \mathbf{rel}(T, x)$ and $Q_2 \in \mathbf{rel}(T, Y)^*$ where $Y = \text{uvars}(t) \cap V$. Given $y \in Y$, since $y \in V$, then for each binding $x'/t'$ in $\theta'$, if $y \in \text{vars}(t')$ then $x' \in U_2$ (see Remark C.1). Therefore $\mathbf{rel}(H, y) = \{K\}$ with $K \subseteq U_2$, and by Lemma C.2, the same holds for $\mathbf{rel}(T, y)$. This means $Q_2 \subseteq U_2$. Thus $Q \cap U_1 = Q_1 \cap U_1$. Since $Q_1$ is $\theta'$-connected by inductive hypothesis, it follows that $Q_1$ is $\theta$-connected.

Now, take $Q_1 \in \mathbf{rel}(T, x)^*$ and $Q_2 \in \mathbf{rel}(T, Z)^*$, where $Z = \text{vars}(t) \setminus Y$. Thus $Q_1 = A_1 \cup \ldots \cup A_k$ with $A_i \in \mathbf{rel}(T, x)$. By inductive hypothesis, $A_i$ is $\theta'$-connected, and therefore it is $\theta$-connected. It follows that for each $i \leq k$ there exist $B^i_1, \dots, B^i_{k_i} \in S_1$ such that $\cup_{j \leq k_i} B^i_j = A_i \cap U_1$ and $B^i_{j_1} \mathcal{R}^*_{\theta A_i} B^i_{j_2}$ for $j_1, j_2 \leq k_i$. The same holds for $Q_2 = C_1 \cup \ldots \cup C_h$ with $C_i \in \mathbf{rel}(T, Z)$: for any $C_i \cap U_1 \neq \emptyset$ we have that $C_i \cap U_1 = \cup_{j \leq h_i} D^i_j$ with $D^i_{j_1} \mathcal{R}^*_{\theta C_i} D^i_{j_2}$ for all $j_1, j_2 \leq h_i$.

We need to show that given any $B^i_m, D^j_n$ then $B^i_m \mathcal{R}^*_{\theta Q} D^j_n$. Actually, it is enough to show that for each $i \leq k, j \leq h$ such that $C_j \cap U_1 \neq \emptyset$, there are $m, n$ s.t. $B^i_m \mathcal{R}_{\theta Q} D^j_n$.

Since $x \in A_i$ and $x \in U_1$, without loss of generality we may assume that $x \in B^i_1$.

In the other hand, although $\text{vars}(t) \cap C_j \neq \emptyset$, we cannot infer that there exists any $D^j_n$ s.t. $\text{vars}(t) \cap D^j_n \neq \emptyset$ since it may well happen that $\text{vars}(t) \cap C_j \subseteq U_2$ although $U_1 \cap C_j \neq \emptyset$.

Assume $C_j \in \mathbf{rel}(T, z)$ for some $z \in Z \cap U_1$. Then, we may assume without loss of generality that $z \in D^j_1$, and $B^i_1 \mathcal{R}_{\theta Q} D^j_1$ follows from the definition of $\mathcal{R}_{\theta Q}$, being $z \in Q$. Otherwise, $C_j \in \mathbf{rel}(T, z)$ for some $z \in Z \cap U_2$. By applying Lemma C.2, we have $C_j = E_1 \cup \cdots \cup E_p$ with $E_i \in H'$ and $E_i \cap U_1 \neq \emptyset$ (this holds even if $p = 1$ since $C_j \cap U_1 \neq \emptyset$). Since $\mathbf{rel}(H', z) = \{occ(\theta', z)\}$, then $occ(\theta', z) \cap U_1 \neq \emptyset$, i.e., there exists $z' \in U_1$ such that $z \in \text{vars}(\theta'(z'))$. Then $z' \in C_j$ and we may assume, without loss of generality, that $z' \in D^j_1$. Again, we have $B^i_1 \mathcal{R}_{\theta Q} D^j_1$ by definition of $\mathcal{R}_{\theta Q}$.

Observe that, if $Q_2 \cap U_1 \neq \emptyset$, by symmetry and transitivity, this alone proves that $B^i_m \mathcal{R}^*_{\theta X} B^{i'}_{m'}$ and $D^j_n \mathcal{R}^*_{\theta Q} D^{j'}_{n'}$ for each $i, m, i', m'$ and $j, n, j', n'$. Otherwise, there is no $D^j_n$ and we need to prove in other ways that $B^i_m \mathcal{R}^*_{\theta Q} B^{i'}_{m'}$. Since $Q_2 \cap U_1 = \emptyset$, then $C_i \subseteq U_2$ for each $i$. This means $C_i = occ(\theta', y)$ for some $y \in U_2$ and since $C_i \subseteq U_2$ it follows immediately that $y \in V$. Then, since $y \in Z$, it must be the case that $y \notin \text{uvars}(t)$ and therefore $B^i_1 \mathcal{R}_{\theta Q} B^{i'}_1$ by definition of $\mathcal{R}_{\theta Q}$.

It remains the case $Q = Q_1 \cup Q_2 \cup Q_3$ with $Q_1 \in \mathbf{rel}(T, x)^*$, $Q_2 \in \mathbf{rel}(T, Y)^*$ and $Q_3 \in \mathbf{rel}(T, Z)^*$. However, this is a trivial corollary of the previous two cases, since we know that $Q_1 \cup Q_3$ is $\theta$-connected and $Q_2 \subseteq U_2$. □

Fixed $X \in S$, our aim is to provide a substitution $\delta$ with $\alpha_{\text{Sh}}([\delta]_{U_1}) \sqsubseteq [S_1, U_1]$ and $\alpha_{\text{Sh}}(\text{mgu}([\delta]_{U_1}, [\theta]_U)) \sqsupseteq [\{X\}, U]$. By Lemma C.4, $X_1 = B_1 \cup \ldots \cup B_n$ with $B_i \in S_1$ and $B_i \mathcal{R}^*_{\theta X} B_j$ for each $i, j \leq n$ (where $X_1 = X \cap U_1$). We let $K_1 = \{B_1, \ldots, B_n\}$. We now want to define a substitution $\delta$ such that $\alpha_{\text{Sh}}([\delta]_{U_1}) = [K_1, U_1]$. For each sharing group $B \in K_1$, let us consider a fresh variable $w_B$. Let $W = \{w_B \mid B \in K_1\}$. For each variable $x$, let $B_x = \{B^1_x, \ldots, B^k_x\}$ be the set $\mathbf{rel}(K_1, x)$. Let $N$ be the maximum cardinality of all the $B_x$ for $x \in X_1$ i.e., $N = \max_{x \in X_1} |B_x|$. For each $x \in X_1$, we define two terms:

$$s_x = t(\underbrace{c(w_{B^1_x}, w_{B^1_x}), c(w_{B^2_x}, w_{B^2_x}), \ldots, c(w_{B^k_x}, w_{B^k_x})}_{k = |B_x| \text{ times}}, \underbrace{c(w_{B^1_x}, w_{B^1_x}), \ldots, c(w_{B^1_x}, w_{B^1_x})}_{N - |B_x| \text{ times}})$$

$$s'_x = t(\underbrace{c(w_{B^1_x}, w_{B^2_x}), c(w_{B^2_x}, w_{B^3_x}), \ldots, c(w_{B^k_x}, w_{B^1_x})}_{k = |B_x| \text{ times}}, \underbrace{c(w_{B^1_x}, w_{B^1_x}), \ldots, c(w_{B^1_x}, w_{B^1_x})}_{N - |B_x| \text{ times}})$$

Note that if $N = 0$ then $X_1 = \emptyset$ and $s_x$, $s'_x$ are undefined for any variable $x$.

We introduce the following notation: given a term $t$ we distinguish different occurrences of the same variable by calling $(y, n)$ the $n$-th occurrence of a variable $y$ in $t$, where the order is lexicographic. For instance, a term $f(x, g(y, y, x))$ can be seen as the term $f((x, 1), g((y, 1), (y, 2), (x, 2)))$. For each $y \in \text{vars}(\theta(U_1)) \cap X$, we choose a variable $x_y \in U_1$ such that $y \in \theta(x_y)$. Let $a$ be a constant. We are now ready to define the substitution $\delta$ in the following way: for each variable $x \in U_1$, $\delta(x)$ is the same as $\theta(x)$ with the difference that each occurrence $(y, i)$ of a variable $y \in \theta(x)$ is replaced by $t_{x,y,i}$ defined as

- $t_{x,y,i} = a$ if $y \notin X$, else

- $t_{x,y,i} = s_x$ if $x = x_y$ and $i = 1$;
- $t_{x,y,i} = s'_x$ otherwise.

Note that, by Corollary C.3, if $x \in X_1$, then $\theta(x)$ is not ground. Therefore, by construction, $\mathrm{dom}(\delta) = U_1$ and $\mathrm{rng}(\delta) = W$. It is easy to check that $\alpha_{\mathrm{Sh}}([\delta]_{U_1}) = [K_1, U_1]$ since given a variable $w_B$, it appears in $\delta(x)$ iff $x \in B$ and therefore $occ(\delta, w_B) \cap U_1 = B$. For all the other variables $occ(\delta, v) = \emptyset$ if $v \in U_1$ and $occ(\delta, v) = \{v\} \not\subseteq U_1$ otherwise. Let us compute the value of $\mathrm{mgu}([\delta]_{U_1}, [\theta]_U)$.

*Lemma C.5*

$$\mathrm{mgu}(\delta, \theta) = \mathrm{mgu}\{w_1 = w_2 \mid w_1, w_2 \in W\} \circ \rho \circ \theta$$

where $\rho = \{v/s_{x_v} \mid v \in \mathrm{vars}(\theta(U_1)) \cap X\} \cup \{v/a \mid v \in \mathrm{vars}(\theta(U_1)) \setminus X\}$.

*Proof*

Since $t_{x_v,v,1} = s_{x_v}$, by using the properties of equation sets it follows that:

$$\begin{aligned} \mathrm{mgu}(\delta, \theta) &= \mathrm{mgu}(\{v = t_{x,v,i} \mid x \in U_1, (v, i) \text{ is an occurrence of } v \text{ in } \theta(x)\}) \circ \theta \\ &= \mathrm{mgu}(E) \circ \rho \circ \theta \ . \end{aligned}$$

where $E = \{t_{x_v,v,1} = t_{x',v,j} \mid x' \in U_1, (v, j) \text{ is an occurrence of } v \text{ in } \theta(x')\}$. Let us define a relation between variables:

$$v\mathcal{R}'u \iff \exists y \in \mathrm{vars}(\theta(v)) \cap X.\ u = x_y \wedge (u = v \Rightarrow y \notin \mathrm{uvars}(\theta(v)))\} \ .$$

Note that $\mathcal{R}'$ is not a symmetric relationship. Moreover, it depends from $\theta$ and $X$, just as $\mathcal{R}_{\theta X}$. However, since in this proof $\theta$ and $X$ are fixed, we decided to omit the indexes in order to simplify notation. By exploiting the above definition, we can rewrite $\mathrm{mgu}(E)$ as follows:

$$\mathrm{mgu}(E) = \mathrm{mgu}(\{s'_v = s_u \mid v, u \in X_1, v\mathcal{R}'u\}) \ . \tag{C1}$$

The above characterization shows that $\mathrm{Eq}(\delta) \cup \mathrm{Eq}(\theta)$ is solvable, since $s_u$ and $s'_v$ are terms which unify by construction. Moreover, note that

$$\mathrm{mgu}\{s_u = s'_v\} = \mathrm{mgu}\{w_B = w_{B'} \mid B \in B_u \wedge B' \in B_v\} \ .$$

We want to prove that $\mathrm{mgu}\{s'_v = s_u \mid v, u \in X_1, v\mathcal{R}'u\} = \mathrm{mgu}\{w_1 = w_2 \mid w_1, w_2 \in W\}$. It is obvious that $\mathrm{mgu}\{s'_v = s_u \mid v, u \in X_1, v\mathcal{R}'u\} = \mathrm{mgu}\{w_B = w_{B'} \mid v, u \in X_1.\ B \in B_v, B' \in B_u, v\mathcal{R}'u\} = \mathrm{mgu}\{w_B = w_{B'} \mid B\hat{\mathcal{R}}B'\}$ where $\hat{\mathcal{R}}$ is the relation on $K_1 \times K_1$ given by

$$B\hat{\mathcal{R}}B' \iff \exists x, y \in X_1.\ B \in B_x \wedge B' \in B_y \wedge x\mathcal{R}'y \ .$$

Since equality is transitive and reflexive, we know that

$$\mathrm{mgu}\{w_B = w_{B'} \mid B\hat{\mathcal{R}}B'\} = \mathrm{mgu}\{w_B = w_{B'} \mid B\hat{\mathcal{R}}^*B'\} \ ,$$

where $\hat{\mathcal{R}}^*$ is the symmetric and transitive closure of $\hat{\mathcal{R}}$. We now prove that $\hat{\mathcal{R}} \subseteq \mathcal{R}_{\theta X} \subseteq \hat{\mathcal{R}}^*$, from which the thesis follows by Lemma C.4.

If $B\hat{\mathcal{R}}B'$ there are $x, y \in X_1$ s.t. $B \in B_x \wedge B' \in B_y \wedge x\mathcal{R}'y$. However $B \in B_x$ iff

$x \in B \in S_1$ and $B' \in B_y$ iff $y \in B' \in S_1$. Now, assume $z \in \text{vars}(\theta(x)) \cap X$ and $y = x_z$. Then $z \in \text{vars}(\theta(x)) \cap \text{vars}(\theta(y)) \cap X$ and this proves that $B\mathcal{R}_{\theta X}B'$. On the other side, assume $B\mathcal{R}_{\theta X}B'$, i.e., there are $x \in B$, $y \in B'$, $z \in \text{vars}(\theta(x)) \cap \text{vars}(\theta(y)) \cap X$ s.t. $x = y \implies z \notin \text{uvars}(\theta(x))$. Since $x \in B$ and $y \in B'$, then $B \in B_x$ and $B' \in B_y$. Since $z \in \text{vars}(\theta(U_1)) \cap X$ then $x_z$ is defined and $B_{x_z} \neq \emptyset$. Assume that $x = y = x_z$. Then $z \notin \text{uvars}(\theta(x))$ and thus $x\mathcal{R}'y$ and $B\hat{\mathcal{R}}B'$. Otherwise, we may assume without loss of generality that $x \neq x_z$. If $y = x_z$ then $x\mathcal{R}'y$ and thus $B\hat{\mathcal{R}}B'$. If $y \neq x_z$ we can choose any $B'' \in B_{x_z}$. We know that $x\mathcal{R}'x_z$, $y\mathcal{R}'x_z$ and thus it holds that $B\hat{\mathcal{R}}B''$ and $B'\hat{\mathcal{R}}B''$, from which $B\hat{\mathcal{R}}^*B'$ follows. The case $y \neq x_z$ is symmetric. □

*Proposition C.6*

$$\alpha_{\text{Sh}}(\text{mgu}([\delta]_{U_1}, [\theta]_U)) \sqsupseteq_{\text{Sh}} [\{X\}, U] \ .$$

*Proof*

First of all, note that $\text{mgu}([\delta]_{U_1}, [\theta]_U) = [\text{mgu}(\delta, \theta)]_U$ since $\text{vars}(\theta) \subseteq U$. We proceed with two different proofs when $W = \emptyset$ and $W \neq \emptyset$. If $W \neq \emptyset$ then, according to Lemma C.5, we can choose $\bar{w} \in W$ and define the substitution $\sigma = \{w'/\bar{w} \mid \bar{w} \neq w' \in W\} = \text{mgu}(E)$. It only remains to prove that $\mathit{occ}(\sigma \circ \rho \circ \theta, \bar{w}) \cap U = X$.

It follows easily that $\mathit{occ}(\sigma \circ \rho \circ \theta, \bar{w}) = \mathit{occ}(\rho \circ \theta, W) = \mathit{occ}(\theta, \text{vars}(\theta(U_1)) \cap X) \cup W) = \mathit{occ}(\theta, \text{vars}(\theta(U_1)) \cap X) \cup W$. Since $U \cap W = \emptyset$ it follows that $\mathit{occ}(\sigma \circ \rho \circ \theta, \bar{w}) \cap U = \mathit{occ}(\theta, \text{vars}(\theta(U_1)) \cap X)$.

By definition, $\mathit{occ}(\theta, \text{vars}(\theta(U_1)) \cap X) = \{y \mid \text{vars}(\theta(y)) \cap \text{vars}(\theta(U_1)) \cap X \neq \emptyset\}$. Thus, for any of such $y$, we have that $\text{vars}(\theta(y)) \cap X \neq \emptyset$ and thus, by Corollary C.3, $y \in X$. It follows that $\mathit{occ}(\theta, \text{vars}(\theta(U_1)) \cap X) \subseteq X$. For the opposite direction, by Lemma C.2 there exist $B_1, \ldots, B_k \in H$ such that $\cup B_i = X$ and $B_i \cap U_1 \neq \emptyset$ for each $i$. Since $B_i \in H$, then there exists $v$ s.t. $B_i = \mathit{occ}(\theta, v)$. Moreover, $v \in X$ since $v \in B_i$ by definition of *occ* and $\theta(v) = v$. Since $B_i \cap U_1 \neq \emptyset$ it follows that there exists $y \in B_i \cap U_1$ such that $v \in \theta(y) \subseteq \theta(U_1)$ and thus $B_i \subseteq \mathit{occ}(\theta, \text{vars}(\theta(U_1)) \cap X)$. Thus $X \subseteq \mathit{occ}(\theta, \text{vars}(\theta(U_1)) \cap X)$.

When $W = \emptyset$, $\text{mgu}(E) = \epsilon$ and $X = X_2$. In this case, by Lemma C.2, $X_2 = \mathit{occ}(\theta, x)$ for some $x \in U_2$. Since $X_2 \cap U_1 = \emptyset$, then $x \notin \text{vars}(\theta(U_1))$, i.e., $x \notin \text{dom}(\rho)$ and therefore $\mathit{occ}(\rho \circ \theta, x) = \mathit{occ}(\theta, x) = X_2$. □

Note that, in this proof, we worked with a signature endowed with a constant $a$ and term symbols $c$ and $t$ of arity two and $N$ respectively. Actually, it is evident that the proof may be easily rewritten for the case when the signature has a constant and a symbol of arity at least two. Given $s$ of arity $n$, we may replace in $\delta$ a term $t(t_1, \ldots, t_N)$ with $c(t_1, c(t_2, c(\ldots, t_N)))$. Then, we replace $c(t_1, t_2)$ with $s(t_1, t_2, a, a, \ldots, a)$ where $a$ is repeated $n - 2$ times.

*Theorem C.7*
$\mathbf{U}^f_{\text{Sh}}$ is well defined, correct and optimal w.r.t. $\mathbf{U}^f_{\text{Ps}}$.

*Proof*
By Equation (31), we need to prove that:

$$\pi_{\mathrm{Sh}}(\mathsf{unif}_{\mathrm{Sh}}(\rho([S_1,U_1]),\mathrm{mgu}(\rho(A_1)=A_2)),\mathrm{vars}(A_2)) = \\ \alpha_{\mathrm{Sh}}(\pi_{\mathrm{Ps}}(\mathsf{unif}_{\mathrm{Ps}}(\rho(\gamma_{\mathrm{Ps}}([S_1,U_1])),\mathrm{mgu}(\rho(A_1)=A_2)),\mathrm{vars}(A_2))) \enspace .$$

By Theorems 5.3 and 5.4, we know that $\pi_{\mathrm{Sh}}$ is correct and complete and that abstract renaming is correct and $\gamma$-complete. Moreover, by Theorem 6.16, abstract unification $\mathsf{unif}_{\mathrm{Sh}}$ is optimal. We have the following equalities.

$$\begin{aligned}
& \alpha_{\mathrm{Sh}}(\pi_{\mathrm{Ps}}(\mathsf{unif}_{\mathrm{Ps}}(\rho(\gamma_{\mathrm{Ps}}([S_1,U_1])),\mathrm{mgu}(\rho(A_1)=A_2)),\mathrm{vars}(A_2))) \\
= \; & \pi_{\mathrm{Sh}}(\alpha_{\mathrm{Sh}}(\mathsf{unif}_{\mathrm{Ps}}(\rho(\gamma_{\mathrm{Ps}}([S_1,U_1])),\mathrm{mgu}(\rho(A_1)=A_2)),\mathrm{vars}(A_2))) && \text{[by Th. 5.3]} \\
= \; & \pi_{\mathrm{Sh}}(\alpha_{\mathrm{Sh}}(\mathsf{unif}_{\mathrm{Ps}}(\gamma_{\mathrm{Ps}}(\rho([S_1,U_1])),\mathrm{mgu}(\rho(A_1)=A_2))),\mathrm{vars}(A_2)) && \text{[by Th. 5.4]} \\
= \; & \pi_{\mathrm{Sh}}(\mathsf{unif}_{\mathrm{Sh}}(\rho([S_1,U_1]),\mathrm{mgu}(\rho(A_1)=A_2)),\mathrm{vars}(A_2)) \enspace . && \text{[by Th. 6.16]}
\end{aligned}$$

Thus $\mathbf{U}^f_{\mathrm{Sh}}$ is correct and optimal w.r.t. $\mathbf{U}^f_{\mathrm{Ps}}$. The fact that it is well defined (i.e., it does not depend on the choice of the renaming $\rho$) is a direct consequence of optimality. □

## D Matching

*Theorem D.1*
(Correctness of $\mathsf{match}_{\mathrm{Sh}}$) $\mathsf{match}_{\mathrm{Sh}}$ is correct w.r.t. $\mathsf{match}_{\mathrm{Ps}}$.

*Proof*
Consider $[\Theta_i,U_i] \sqsubseteq_{\mathrm{Ps}} \gamma_{\mathrm{Sh}}([S_i,U_i])$ for $i \in \{1,2\}$ and $[\sigma]_{U_1\cup U_2} \in \mathsf{match}_{\mathrm{Ps}}([\Theta_1,U_1], [\Theta_2,U_2])$. We need to prove that

$$\alpha_{\mathrm{Sh}}([\sigma]_{U_1\cup U_2}) \in \mathsf{match}_{\mathrm{Sh}}([S_1,U_1],[S_2,U_2]) \enspace .$$

Assume $[\sigma] = \mathrm{mgu}([\sigma_1],[\sigma_2])$ with $[\sigma_1] \in \Theta_1$ and $[\sigma_2] \in \Theta_2$. Let $\sigma_1$ and $\sigma_2$ be two canonical representatives for $[\sigma_1]$ and $[\sigma_2]$ such that $\mathrm{vars}(\sigma_1)\cap\mathrm{vars}(\sigma_2) = U_1\cap U_2$. If $\sigma_1 \preceq_{U_1\cap U_2} \sigma_2$, there exists $\delta \in \mathit{Subst}$ such that $\sigma_1(x) = \delta(\sigma_2(x))$ for each $x \in U_1\cap U_2$. We may assume, without loss of generality, that $\mathrm{dom}(\delta) = \mathrm{vars}(\sigma_2(U_1\cap U_2))$. Now, the following equalities hold.

$$\begin{aligned}
\sigma = & \mathrm{mgu}(\mathrm{Eq}(\sigma_2),\mathrm{Eq}(\sigma_1)) \\
= & \mathrm{mgu}(\{\sigma_2(x) = \sigma_2(\sigma_1(x)) \mid x \in U_1\}) \circ \sigma_2 \\
= & \mathrm{mgu}(\{x = \sigma_1(x) \mid x \in U_1 \setminus U_2\} \cup \{\sigma_1(x) = \sigma_2(x) \mid x \in U_1\cap U_2\}) \circ \sigma_2 \\
& \quad \text{[by partitioning } \mathrm{dom}(\sigma_2)\text{, since } \sigma_2(\sigma_1(x)) = \sigma_1(x) \text{ for } x \in U_1] \\
= & \mathrm{mgu}(\{x = \sigma_1(x) \mid x \in U_1\setminus U_2\}) \circ \delta \circ \sigma_2 \\
& \quad \text{[since } \sigma_1(x) = \delta(\sigma_2(x)) \text{ and } \mathrm{dom}(\delta) = \mathrm{vars}(\sigma_2(U_1\cap U_2))] \\
= & \sigma_{1|U_1\setminus U_2} \circ \delta \circ \sigma_2 \\
= & \sigma_{1|U_1\setminus U_2} \uplus (\delta \circ \sigma_2) \enspace .
\end{aligned} \tag{D1}$$

Now, given a variable $v$, by Lemma B.1, $\mathit{occ}(\sigma,v)\cap(U_1\cup U_2) = (\mathit{occ}(\sigma_{1|U_1\setminus U_2},v)\cap U_1) \cup (\mathit{occ}(\sigma_2,\mathit{occ}(\delta,v))\cap U_2)$. We want to prove that $\mathit{occ}(\sigma,v)\cap(U_1\cup U_2) \in \mathsf{match}_{\mathrm{Sh}}([S_1,U_1],[S_2,U_2])$.

Since $\mathrm{dom}(\sigma) = U_1 \cup U_2$, we may assume that $v \notin U_1 \cup U_2$, otherwise $occ(\sigma, v) \cap (U_1 \cup U_2) = \emptyset$. We recall that $S_1' = \{B \in S_1 \mid B \cap U_2 = \emptyset\}$ and $S_1'' = S_1 \setminus S_1'$, $S_2' = \{B \in S_2 \mid B \cap U_1 = \emptyset\}$ and $S_2'' = S_2 \setminus S_2'$, according to Definition 7.1. We distinguish two cases:

- $v \notin \mathrm{rng}(\delta)$, which implies $v \notin \mathrm{rng}(\sigma_{1|U_2})$. Note that, if $v \in \mathrm{dom}(\delta)$ then $occ(\sigma_2, occ(\delta, v)) = \emptyset \in S_2'$, otherwise $occ(\sigma_2, occ(\delta, v)) = occ(\sigma_2, v) \in S_2'$. So, it always holds that $occ(\sigma_2, occ(\delta, v)) \in S_2'$. We now distinguish some subcases. If $v \in \mathrm{rng}(\sigma_1)$ then $occ(\sigma_{1|U_1 \setminus U_2}, v) = occ(\sigma_1, v)$. Moreover, since $v \in \mathrm{rng}(\sigma_1)$, then $v \notin \mathrm{vars}(\sigma_2)$ and thus $occ(\sigma_2, v) = \{v\}$. We have that $occ(\sigma, v) \cap (U_1 \cup U_2) = occ(\sigma_1, v) \in S_1'$. Otherwise, if $v \in \mathrm{rng}(\sigma_2)$, then $v \notin \mathrm{vars}(\sigma_1)$ and $occ(\sigma_1, v) = \{v\}$. Therefore $occ(\sigma, v) \cap (U_1 \cup U_2) = occ(\sigma_2, occ(\delta, v)) \in S_2'$. Otherwise, if $v \notin \mathrm{rng}(\sigma_1) \cup \mathrm{rng}(\sigma_2)$ then $occ(\sigma, v) \cap (U_1 \cup U_2) = \emptyset$.
- $v \in \mathrm{rng}(\delta)$. We want to prove that $occ(\sigma, v) = X_1 \cup X_2$ where $X_1 = occ(\sigma_1, v)$ and $X_2 = occ(\sigma_2, occ(\delta, v))$ enjoy the following properties: $X_1 \in S_1''$, $X_2 \in S_2''^*$, $X_1 \cap U_2 = X_2 \cap U_1$. First of all, note that $occ(\sigma_{1|U_1 \setminus U_2}, v) \cap U_1 = X_1 \setminus U_2$. Moreover, $occ(\sigma_2, occ(\delta, v)) \cap U_1 = occ(\sigma_{2|U_1}, occ(\delta, v)) \cap U_1$, which in turn is equal to $occ(\delta \circ \sigma_{2|U_1}, v) \cap U_1 = occ(\sigma_{1|U_2}, v) \cap U_1 = occ(\sigma_1, v) \cap U_1 \cap U_2 \supseteq X_1 \cap U_2$. This proves that $occ(\sigma, v) = X_1 \cup X_2$ and $X_1 \cap U_2 = X_2 \cap U_1$.
  While it is obvious that $X_1 \in S_1$ and $X_2 \in S_2^*$, we still need to prove that $X_1 \in S_1''$ and $X_2 \in S_2''^*$. For each $y \in occ(\delta, v)$, by definition of $\delta$ we have that $y \in \sigma_2(U_1 \cap U_2)$ and therefore $occ(\sigma_2, y) \cap U_1 \neq \emptyset$. This proves that $X_2 \in S_2''^*$. Moreover, if $v \in \mathrm{rng}(\delta)$ then $v \in \mathrm{rng}(\sigma_{1|U_2})$ and thus $occ(\sigma_1, v) \in S_1''$. □

*Theorem D.2*
(Weak completeness of $\mathsf{match}_{\mathrm{Sh}}$) The operator $\mathsf{match}_{\mathrm{Sh}}$ is optimal on the first argument and complete on the second one when $\mathsf{match}_{\mathrm{Ps}}$ is restricted to the case when the second argument contains a single substitution. In formulas:

$$\mathsf{match}_{\mathrm{Sh}}([S_1, U_1], \alpha_{\mathrm{Sh}}([\{\sigma_2\}, U_2])) = \alpha_{\mathrm{Sh}}(\mathsf{match}_{\mathrm{Ps}}(\gamma_{\mathrm{Sh}}([S_1, U_1]), [\{[\sigma_2]\}, U_2])) \ .$$

for each $[\{[\sigma_2]\}, U_2] \in \mathtt{Psub}$ and $[S_1, U_1] \in \mathtt{Sharing}$.

*Proof*
Since $\mathsf{match}_{\mathrm{Sh}}$ is correct w.r.t. $\mathsf{match}_{\mathrm{Ps}}$, it follows that:

$$\alpha_{\mathrm{Sh}}(\mathsf{match}_{\mathrm{Ps}}(\gamma_{\mathrm{Sh}}([S_1, U_1]), [\{[\sigma_2]\}, U_2])) \sqsubseteq_{\mathrm{Sh}} \mathsf{match}_{\mathrm{Sh}}([S_1, U_1], \alpha_{\mathrm{Sh}}([\{[\sigma_2]\}, U_2])) \ .$$

So, we only need to prove that:

$$\mathsf{match}_{\mathrm{Sh}}([S_1, U_1], \alpha_{\mathrm{Sh}}([\{[\sigma_2]\}, U_2])) \sqsubseteq_{\mathrm{Sh}} \alpha_{\mathrm{Sh}}(\mathsf{match}_{\mathrm{Ps}}(\gamma_{\mathrm{Sh}}([S_1, U_1]), [\{[\sigma_2]\}, U_2])) \ .$$

Assume, without loss of generality, that $\sigma_2$ is a canonical representative of $[\sigma_2]_{U_2}$ and $\mathrm{rng}(\sigma_2) \cap U_1 = \emptyset$. Take $B \in S$, where $[S, U_1 \cup U_2] = \mathsf{match}_{\mathrm{Sh}}([S_1, U_1], [S_2, U_2])$, with $[S_2, U_2] = \alpha_{\mathrm{Sh}}([\{[\sigma_2]\}, U_2])$. We have three cases.

- If $B \in S_1'$ then $B \in S_1$ and $B \subseteq U_1 \setminus U_2$. Let $\delta = \{x/v \mid x \in B\} \cup \{x/a \mid x \in \mathrm{vars}(\sigma_2(U_1 \setminus B))\}$ and $\sigma_1 = (\delta \circ \sigma_2)_{|U_1}$ where $v$ is a fresh variable. It follows that $\mathrm{dom}(\sigma_1) = U_1$ and $\mathrm{rng}(\sigma_1) = \{v\}$ with $occ(\sigma_1, v) = B$, therefore $[\sigma_1, U_1] \sqsubseteq_{\mathrm{Ps}} \gamma_{\mathrm{Sh}}([S_1, U_1])$. Clearly $\sigma_1 \preceq_{U_1 \cap U_2} \sigma_2$ since $U_1 \cap U_2 \subseteq U_1 \setminus B$. Let $\sigma = \mathrm{mgu}(\sigma_1, \sigma_2)$. Since $B \cap \mathrm{dom}(\sigma_2) = \emptyset$ and $v$ is a fresh variable, it follows that $occ(\sigma, v) = B$, and thus $B \in \alpha_{\mathrm{Sh}}(\mathsf{match}_{\mathrm{Ps}}(\gamma_{\mathrm{Sh}}([S_1, U_1]), [\{[\sigma_2]\}, U_2]))$.

- If $B \in S_2'$, there exists $v \in \mathcal{V}$ such that $occ(\sigma_2, v) \cap U_2 = B$. Let $X = \mathrm{vars}(\sigma_2(U_1)))$ and take $\delta = \{x/a \mid x \in X\}$. Then $\sigma_1 = (\delta \circ \sigma_2)_{|U_1}$ is such that $occ(\sigma_1, v) \cap U_1 = \emptyset$ for each $v \in \mathcal{V}$, therefore $\sigma_1 \in \gamma_{\mathrm{Sh}}([S_1, U_1])$. Moreover $\mathrm{mgu}(\sigma_2, \sigma_1) \in \mathsf{match}_{\mathrm{Ps}}(\gamma_{\mathrm{Sh}}([S_1, U_1]), [\{[\sigma_2]\}, U_2])$. By the proof of Theorem D.1, Equation (D1), we have $\mathrm{mgu}(\sigma_1, \sigma_2) = \delta \circ \sigma_2$. Since $B \cap U_1 = \emptyset$, then $v \notin X = \mathrm{vars}(\delta)$, and therefore $occ(\delta \circ \sigma_2, v) \cap U_2 = occ(\sigma_2, v) \cap U_2 = B$. Hence $B \in \alpha_{\mathrm{Sh}}(\mathsf{match}_{\mathrm{Ps}}(\gamma_{\mathrm{Sh}}([S_1, U_1]), [\{[\sigma_2]\}, U_2]))$.
- We now assume $B = X_1 \cup \bigcup X$ with $X \subseteq S_2''$, $X_1 \in S_1''$, $\bigcup X \cap U_1 = X_1 \cap U_2$. Then, for each $H \in X$, there exists $v_H \in \mathcal{V}$ such that $occ(\sigma_2, v_H) \cap U_2 = H$. Since $H \cap U_1 \neq \emptyset$ for each $H \in X$, then $v_H \in Y = \mathrm{vars}(\sigma_2(U_1))$. Consider the substitution

$$\delta = \{v_H/v \mid H \in X\} \uplus \{w/a \mid w \in Y, \forall H \in X. w \neq v_H\}$$

for a fresh variable $v$ and

$$\sigma_1 = (\delta \circ \sigma_2)_{|U_1} \uplus \{x/v \mid x \in X_1 \setminus U_2\} \ .$$

We want to prove $[\{[\sigma_1]\}, U_1] \in \gamma_{\mathrm{Sh}}([S_1, U_1])$. By definition of $\sigma_1$ we have that $occ(\sigma_1, v) \cap U_1 = (occ(\sigma_2, \{v_H \mid H \in X\}) \cap U_1) \cup X_1 \setminus U_2 = (\bigcup X \cap U_1) \cup X_1 \setminus U_2 = X_1 \in S_1$. Otherwise, for $w \neq v$ we have that either $occ(\sigma_1, w) = \emptyset$ when $w \in U_1$ or $occ(\sigma_1, w) = occ(\sigma_2, w)$ which is disjoint from $U_1$. In both cases, $occ(\sigma_1, w) \cap U_1 = \emptyset \in S_1$. By definition of $\sigma_1$, $[\mathrm{mgu}(\sigma_1, \sigma_2)] \in \mathsf{match}_{\mathrm{Ps}}(\gamma_{\mathrm{Sh}}([S_1, U_1]), [\{[\sigma_2]\}, U_2])$. Moreover, we know from (D1) that

$$\mathrm{mgu}(\sigma_2, \sigma_1) = \delta \circ \sigma_2 \uplus \{x/v \mid x \in X_1 \setminus U_2\} \ .$$

Let $\sigma = \mathrm{mgu}(\sigma_1, \sigma_2)$. Note that $occ(\sigma, v) \cap (U_1 \cup U_2) = X_1 \setminus U_2 \cup occ(\sigma_2, \{v_H \mid H \in X\}) \cap U_2$. By definition of $v_H$, $occ(\sigma_2, v_H) \cap U_2 = H$, hence $occ(\sigma, v) \cap (U_1 \cup U_2) = (X_1 \setminus U_2) \cup \bigcup X = X_1 \cup \bigcup X = B$.

This proves the theorem. □

*Theorem D.3*
(Optimality of $\mathsf{match}_{\mathrm{Sh}}$) $\mathsf{match}_{\mathrm{Sh}}$ is optimal.

*Proof*
Given $[S_1, U_1], [S_2, U_2] \in$ `Sharing`, we have

$$\begin{aligned}
&\alpha_{\mathrm{Sh}}(\mathsf{match}_{\mathrm{Ps}}(\gamma_{\mathrm{Sh}}([S_1, U_1]), \gamma_{\mathrm{Sh}}([S_2, U_2])))\\
&= \alpha_{\mathrm{Sh}}(\sqcup_{\mathrm{Ps}}\{\mathsf{match}_{\mathrm{Ps}}(\gamma_{\mathrm{Sh}}([S_1, U_1]), [\{[\sigma]\}, U_2]) \mid \alpha_{\mathrm{Sh}}([\sigma]_{U_2}) \sqsubseteq_{\mathrm{Sh}} [S_2, U_2]\})\\
&\qquad \text{[since } \mathsf{match}_{\mathrm{Ps}} \text{ is additive]}\\
&= \sqcup_{\mathrm{Sh}} \{\mathsf{match}_{\mathrm{Sh}}([S_1, U_1], [X, U_2]) \mid X = \alpha_{\mathrm{Sh}}([\sigma]_{U_2}) \sqsubseteq_{\mathrm{Sh}} [S_2, U_2]\}\\
&\qquad \text{[by completeness of } \sqcup_{\mathrm{Sh}} \text{ and Theorem D.2]}\\
&= \mathsf{match}_{\mathrm{Sh}}([S_1, U_1], \sqcup_{\mathrm{Sh}}\{[X, U_2] \mid X = \alpha_{\mathrm{Sh}}([\sigma]_{U_2}) \sqsubseteq_{\mathrm{Sh}} [S_2, U_2]\}) \ .\\
&\qquad \text{[since } \mathsf{match}_{\mathrm{Sh}} \text{ is additive]}
\end{aligned}$$

Since $\alpha_{\mathrm{Sh}}$ defines a Galois insertion, it is surjective, and therefore $\sqcup_{\mathrm{Sh}}\{[X, U_2] \mid X = \alpha_{\mathrm{Sh}}([\sigma]_{U_2}) \sqsubseteq_{\mathrm{Sh}} [S_2, U_2]\} = [S_2, U_2]$ and we obtain

$$\alpha_{\mathrm{Sh}}(\mathsf{match}_{\mathrm{Ps}}(\gamma_{\mathrm{Sh}}([S_1, U_1]), \gamma_{\mathrm{Sh}}([S_2, U_2]))) = \mathsf{match}_{\mathrm{Sh}}([S_1, U_1], [S_2, U_2]) \ ,$$

which concludes the proof. □

*Theorem D.4*
(Strong optimality of $\mathsf{unif}_{\mathrm{Sh}}$) Given $[S_1, U_1] \in \mathtt{Sharing}$ and $\theta \in \mathit{ISubst}$, there exists a substitution $\delta \in \mathit{ISubst}$ such that $\alpha_{\mathrm{Sh}}([\delta]_{U_1}) \sqsubseteq_{\mathrm{Sh}} [S_1, U_1]$ and

$$\alpha_{\mathrm{Sh}}(\mathsf{unif}_{\mathrm{Ps}}([\{[\delta]\}, U_1], \theta)) = \mathsf{unif}_{\mathrm{Sh}}([S_1, U_1], \theta) \ .$$

*Proof*
The optimality result proved in Theorem 6.16 shows that there exists $[\Theta_1, U_1] \sqsubseteq_{\mathrm{Ps}} \gamma_{\mathrm{Sh}}([S_1, U_1])$ such that $\alpha_{\mathrm{Sh}}(\mathsf{unif}_{\mathrm{Ps}}([\Theta_1, U_1], \theta)) = \mathsf{unif}_{\mathrm{Sh}}([S_1, U_1], \delta)$. We need a stronger result which proves that $\Theta_1$ can be chosen as a singleton.

Assume $\mathsf{unif}_{\mathrm{Sh}}([S_1, U_1], \theta) = [S, U_1 \cup U_2]$ where $U_2 = \mathrm{vars}(\theta) \setminus U_1$ and $S = \{X^1, \ldots, X^n\}$. Following the construction in Section C, for each $X^i$ let us define $X_1^i$, $X_2^i$, $K^i$, $K_1^i$, $K_2^i$, $W^i$, $s_x^i$, ${s'}_x^i$, $U$ as in the proof of optimality for $\mathsf{unif}_{\mathrm{Sh}}$. We choose $W^i, W^j$ such that $W^i \cap W^j = \emptyset$ if $i \neq j$ and we denote by $w_B^i$ the elements of $W^i$.

For each $y \in \mathrm{vars}(\theta(U_1)) \cap (\cup_{1 \leq i \leq n} X^i)$, we choose a variable $x_y \in U_1$ such that $y \in \theta(x_y)$. Then, we define the substitution $\delta$ in the following way: for each variables $x \in U_1$, $\delta(x)$ is the same as $\theta(x)$, with the exception that each occurrence $(y, j)$ of a variable $y \in \theta(x)$ is replaced by $t_{x,y,j} = t(t_{x,y,j}^1, \ldots, t_{x,y,j}^n)$, where:

- $t_{x,y,j}^i = a$ if $y \notin X^i$,
- $t_{x,y,j}^i = s_x^i$ otherwise, if $x = x_y$ and $j = 1$;
- $t_{x,y,j}^i = {s'}_x^i$ otherwise.

By construction $\mathrm{dom}(\delta) = U_1$ and $\mathrm{rng}(\delta) = \bigcup_{1 \leq i \leq n} W^i$. It is easy to check that $\alpha_{\mathrm{Sh}}([\{\delta\}, U_1]) = [\bigcup_{1 \leq i \leq n} K_1^i, U_1] \sqsubseteq_{\mathrm{Sh}} [S_1, U_1]$. Using the properties of the equation sets we can prove that

$$\begin{aligned} & \mathrm{mgu}(\delta, \theta) \\ = \ & \mathrm{mgu}(\{v = t_{x,v,j} \mid x \in U_1, (v, j) \text{ is an occurrence of } v \text{ in } \theta(x)\}) \circ \theta \\ = \ & \mathrm{mgu}(E) \circ \rho \circ \theta \ , \end{aligned}$$

where

$$\begin{aligned} \rho = \{v/t_{x_v,v,1} \mid v \in \mathrm{vars}(\theta(U_1))\} \ , \\ E = \{t_{x_v,v,1}^i = t_{x',v,j}^i \mid i \in \{1, \ldots, n\}, v \in X^i, x' \in U_1, \\ (v, j) \text{ is an occurrence of } v \text{ in } \theta(x')\} \ . \end{aligned}$$

Now, each $E^i = \{t_{x_v,v,1}^i = t_{x',v,j}^i \mid x' \in U_1, (v, j) \text{ is an occurrence of } v \text{ in } \theta(x'), v \in X^i\}$ is the same equation which appears in (C1) for $X = X^i$. Therefore, for each $i \in \{1, \ldots, n\}$ such that $W^i \neq \emptyset$, we choose a single $w^i \in W^i$ and define $\eta^i$ with $\mathrm{dom}(\eta^i) = W^i \setminus \{w^i\}$ and $\eta^i(w^i_B) = w^i$ for each $w^i_B \in W^i$. If $W^i = \emptyset$, we choose $\eta^i = \epsilon$. We know from the proof of Lemma C.5 that $\eta^i = \mathrm{mgu}(E^i)$, and $\mathrm{mgu}(E) = \eta = \biguplus_{1 \leq i \leq n} \eta^i$ since $\mathrm{vars}(E^i) \cap \mathrm{vars}(E^j) = \emptyset$ for $i \neq j$. Therefore

$$\mathrm{mgu}(\delta, \theta) = \eta \circ \rho \circ \theta \enspace .$$

We now want to prove that $\alpha_{\mathrm{Sh}}([\eta \circ \rho \circ \theta]_{U_1 \cup U_2}) \sqsupseteq_{\mathrm{Ps}} [\{X^i\}, U_1 \cup U_2]$ for each $i \in \{1, \ldots, n\}$. If $X^i_1 \neq \emptyset$ then $W^i \neq \emptyset$, and we have $\mathit{occ}(\eta \circ \rho \circ \theta, w^i) = \mathit{occ}(\eta^i \circ \rho \circ \theta, w^i)$. Following the proof of Lemma C.5 with $X = X^i$, we have that $\mathit{occ}(\eta \circ \rho \circ \theta, w^i) \cap U = X^i$. When $X^i_1 = \emptyset$, we may choose $v^i \in \theta(X^i_2)$. In this case, $\mathit{occ}(\eta \circ \rho \circ \theta, v^i) \cap U = \mathit{occ}(\theta, v^i) \cap U = X^i$ as proved in Prop. C.6. □

As for Prop. C.6, in the proof of this theorem we assume that we have term symbols for each arity. However, it is possible to rewrite terms so that a constant symbol and a binary term symbol suffice.

*Theorem D.5*
$\mathbf{U}^b_{\mathrm{Sh}}$ is correct and optimal w.r.t. $\mathbf{U}^b_{\mathrm{Ps}}$.

*Proof*
Correctness immediately follows by the fact that $\mathbf{U}^b_{\mathrm{Ps}}$ is obtained by tupling and composition of correct semantic functions.

By using Theorems D.2 and D.4, it is possible to prove that

$$\mathsf{match}_{\mathrm{Sh}}([S_1, U_1], \mathsf{unif}_{\mathrm{Sh}}([S_2, U_2], \theta)) = \\ \alpha_{\mathrm{Sh}}(\mathsf{match}_{\mathrm{Ps}}(\gamma_{\mathrm{Sh}}([S_1, U_1]), \mathsf{unif}_{\mathrm{Sh}}(\gamma_{\mathrm{Sh}}([S_2, U_2]), \theta))) \enspace ,$$

i.e., that the composition of $\mathsf{match}_{\mathrm{Sh}}$ and $\mathsf{unif}_{\mathrm{Sh}}$, as used in $\mathbf{U}^b_{\mathrm{Sh}}$, is optimal.

Assume given $[S_1, U_1]$ and $[S_2, U_2] \in \mathtt{Psub}$ and $\theta \in \mathit{ISubst}$. Consider $[\{[\sigma]\}, U_2] \in \gamma_{\mathrm{Sh}}([S_2, U_2])$ obtained by Lemma D.4 such that $\mathsf{unif}_{\mathrm{Ps}}([\{[\sigma]\}, U_2]), \theta) = [\{[\delta]\}, U_2 \cup \mathrm{vars}(\theta)]$ and $\alpha_{\mathrm{Sh}}([\{[\delta]\}, U_2 \cup \mathrm{vars}(\theta)]) = \mathsf{unif}_{\mathrm{Sh}}([S_2, U_2], \theta)$. Then, we have

$$\begin{aligned} &\mathsf{match}_{\mathrm{Sh}}([S_1, U_1], \mathsf{unif}_{\mathrm{Sh}}([S_2, U_2], \theta)) \\ =&\mathsf{match}_{\mathrm{Sh}}([S_1, U_1], \alpha_{\mathrm{Sh}}(\mathsf{unif}_{\mathrm{Ps}}([\{[\sigma]\}, U_2], \theta))) \\ =&\alpha_{\mathrm{Sh}}(\mathsf{match}_{\mathrm{Ps}}(\gamma_{\mathrm{Sh}}([S_1, U_1]), \mathsf{unif}_{\mathrm{Ps}}([\{[\sigma]\}, U_2], \theta))) \end{aligned}$$

by Theorem D.2, so that, in general

$$\mathsf{match}_{\mathrm{Sh}}([S_1, U_1], \mathsf{unif}_{\mathrm{Sh}}([S_2, U_2], \theta)) \sqsubseteq_{\mathrm{Sh}} \\ \alpha_{\mathrm{Sh}}(\mathsf{match}_{\mathrm{Ps}}(\gamma_{\mathrm{Sh}}([S_1, U_1]), \mathsf{unif}_{\mathrm{Ps}}(\gamma_{\mathrm{Sh}}([S_2, U_2]), \theta))) \enspace .$$

The proof that $\mathbf{U}^b_{\mathrm{Ps}}$ is optimal follows from this result, completeness of $\pi_{\mathrm{Sh}}$ and $\gamma$-completeness of $\rho$. □